\documentclass[reprint,aps,pra,amsmath,amssymb,superscriptaddress,floatfix,longbibliography]{revtex4-2}

\usepackage{upgreek}
\usepackage{physics}
\usepackage{braket}
\usepackage{qcircuit}
\usepackage{esvect}
\usepackage{gensymb}
\usepackage{textcomp}

\usepackage[pdftex]{graphicx}
\usepackage{subcaption}
\usepackage{array}
\usepackage[colorlinks=true,linkcolor=blue,citecolor=blue]{hyperref}

\newcommand{\sref}[1]{Sec.~\ref{#1}}
\newcommand{\fref}[1]{Fig.~\ref{#1}}
\newcommand{\tref}[1]{Table~\ref{#1}}
\newcommand{\eref}[1]{Eq.~\ref{#1}}
\newcommand{\aref}[1]{Appendix~\ref{#1}}

\begin{document}

\title{\textit{Spiderweb} array: A sparse spin-qubit array}
\date{\today}

\author{Jelmer M. Boter}
\affiliation{QuTech, Delft University of Technology, Lorentzweg 1, 2628 CJ Delft, The Netherlands}
\affiliation{Kavli Institute of Nanoscience, Delft University of Technology, Lorentzweg 1, 2628 CJ Delft, The Netherlands}

\author{Juan P. Dehollain}
\affiliation{QuTech, Delft University of Technology, Lorentzweg 1, 2628 CJ Delft, The Netherlands}
\affiliation{Kavli Institute of Nanoscience, Delft University of Technology, Lorentzweg 1, 2628 CJ Delft, The Netherlands}
\affiliation{School of Mathematical and Physical Sciences, University of Technology Sydney, Ultimo NSW 2007, Australia}

\author{Jeroen P.~G. van Dijk}
\affiliation{QuTech, Delft University of Technology, Lorentzweg 1, 2628 CJ Delft, The Netherlands}
\affiliation{Kavli Institute of Nanoscience, Delft University of Technology, Lorentzweg 1, 2628 CJ Delft, The Netherlands}
\affiliation{Department of Quantum and Computer Engineering, Delft University of Technology, 2628 CJ Delft, The Netherlands}

\author{Yuanxing Xu}

\author{Toivo Hensgens}
\affiliation{QuTech, Delft University of Technology, Lorentzweg 1, 2628 CJ Delft, The Netherlands}
\affiliation{Kavli Institute of Nanoscience, Delft University of Technology, Lorentzweg 1, 2628 CJ Delft, The Netherlands}

\author{Richard Versluis}

\author{Henricus~W.~L. Naus}
\affiliation{QuTech, Delft University of Technology, Lorentzweg 1, 2628 CJ Delft, The Netherlands}
\affiliation{Netherlands Organization for Applied Scientific Research (TNO), P.O. Box 155, 2600 AD Delft, The Netherlands}

\author{James S. Clarke}
\affiliation{Components Research, Intel Corporation, 2501 NE Century Blvd, Hillsboro, OR 97124, USA}

\author{Menno Veldhorst}
\affiliation{QuTech, Delft University of Technology, Lorentzweg 1, 2628 CJ Delft, The Netherlands}
\affiliation{Kavli Institute of Nanoscience, Delft University of Technology, Lorentzweg 1, 2628 CJ Delft, The Netherlands}

\author{Fabio Sebastiano}
\affiliation{QuTech, Delft University of Technology, Lorentzweg 1, 2628 CJ Delft, The Netherlands}
\affiliation{Department of Quantum and Computer Engineering, Delft University of Technology, 2628 CJ Delft, The Netherlands}

\author{Lieven M.~K. Vandersypen}
\email{l.m.k.vandersypen@tudelft.nl}
\affiliation{QuTech, Delft University of Technology, Lorentzweg 1, 2628 CJ Delft, The Netherlands}
\affiliation{Kavli Institute of Nanoscience, Delft University of Technology, Lorentzweg 1, 2628 CJ Delft, The Netherlands}
\affiliation{Components Research, Intel Corporation, 2501 NE Century Blvd, Hillsboro, OR 97124, USA}

\begin{abstract}
One of the main bottlenecks in the pursuit of a large-scale--chip-based quantum computer is the large number of control signals needed to operate qubit systems.
As system sizes scale up, the number of terminals required to connect to off-chip control electronics quickly becomes unmanageable.
Here, we discuss a quantum-dot spin-qubit architecture that integrates on-chip control electronics, allowing for a significant reduction in the number of signal connections at the chip boundary.
By arranging the qubits in a two-dimensional (2D) array with $\sim$12 \textmu m pitch, we create space to implement locally integrated sample-and-hold circuits.
This allows to offset the inhomogeneities in the potential landscape across the array and to globally share the majority of the control signals for qubit operations.
We make use of advanced circuit modeling software to go beyond conceptual drawings of the component layout, to assess the feasibility of the scheme through a concrete floor plan, including estimates of footprints for quantum and classical electronics, as well as routing of signal lines across the chip using different interconnect layers.
We make use of local demultiplexing circuits to achieve an efficient signal-connection scaling leading to a Rent's exponent as low as $p = 0.43$.
Furthermore, we use available data from state-of-the-art spin qubit and microelectronics technology development, as well as circuit models and simulations, to estimate the operation frequencies and power consumption of a million-qubit array.
This work presents a complementary approach to previously proposed architectures, focusing on a feasible scheme to integrating quantum and classical hardware, and identifying remaining challenges for achieving full fault-tolerant quantum computation.
It thereby significantly closes the gap towards a fully CMOS-compatible quantum computer implementation.
\end{abstract}

\maketitle

\section{Introduction}
Semiconductor quantum dots~\cite{Loss1998}, particularly in silicon~\cite{Zwanenburg2013}, are attractive hosts for spin qubits in large-scale quantum computation applications, because of their assumed compatibility with conventional CMOS integration processes.
In addition, the $\sim 100$~nm spatial dimensions intrinsic to semiconductor spin qubits provide the potential to pack many millions of qubits inside a single quantum processor chip.
The last several years have seen significant progress in spin-qubit research that resulted in the demonstration of long coherence times~\cite{Veldhorst2014}, high-fidelity single-~\cite{Veldhorst2014,Kawakami2016,Yoneda2018} and two-qubit gates~\cite{Xue2019,Huang2019}, quantum algorithms~\cite{Watson2018}, quantum non-demolition measurements~\cite{Yoneda2020,Xue2020}, and electron spin~\cite{Yoneda2021} and charge~\cite{Baart2016,Seidler2021} transfer.

Scaling to millions of qubits has many technological hurdles that will need to be cleared.
Among them, the wiring bottleneck is a very challenging one for any solid-state qubit controlled by electrical signals~\cite{Vandersypen2017,Franke2019,Alexeev2019}: currently at least one wire must run from the control electronics to each and every qubit, but chips have strict interconnect wiring density limitations, making a one-wire-per-qubit scheme unfeasible in the large scale.
This was already an issue when classical processors started to scale up, leading to the development of techniques such as cross-bar addressing, that enabled the number of output pins $T$ to increase at a rate much slower than the number of transistor unit cells $U$ on the chip.
This was formally captured by Rent's rule: $T = c \, U^p$, where $c$ is the number of connections per unit cell, and $p$ is Rent's exponent, a measure of optimization in the wiring scheme~\cite{Franke2019}.

One approach to reducing Rent’s exponent in quantum-dot-based spin qubits~\cite{Veldhorst2017,Li2018a}, involves addressing 2D quantum dot arrays of $N^2$ qubits using $\sim N$ word and bit lines.
These cross-bar addressing schemes impose strong requirements on the homogeneity of the quantum dot potentials across the array, which in itself is a great technological hurdle.
Furthermore, they are inefficient in the sense that many operations can only be executed sequentially.

\begin{figure*}
    \centering
    \includegraphics[width=0.98\textwidth]{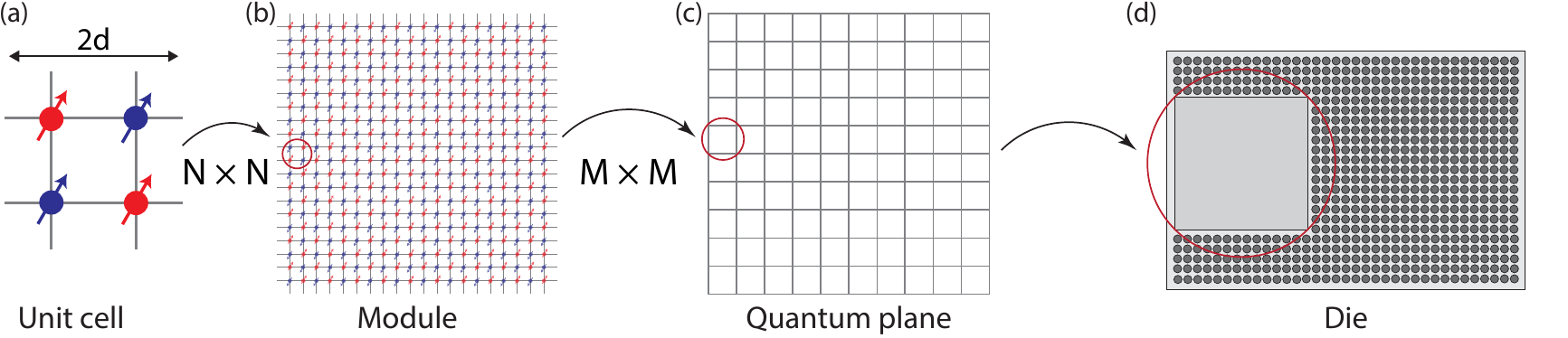}
    \caption{Overview of the \textit{spiderweb} architecture with a schematic breakdown of its components, as described in the main text. (a) A unit cell contains four qubits that are separated by the qubit pitch \textit{d}. Qubits are color coded to distinguish data qubits (blue) and SC-ancilla qubits (red), as defined in the surface code. (b) A module consists of $N \times N$ unit cells, and in turn $M \times M$ modules together form (c) the quantum plane. Generally, the quantum plane occupies only part of (d) the die and the remaining area can be used to further reduce the off-chip wire count by adding additional on-chip electronics.}
    \label{fig:overview}
\end{figure*}

A second approach is to decrease qubit density by using long-range quantum links that connect modules of qubits~\cite{Vandersypen2017,Buonacorsi2019}.
Semiconductor spin qubits provide mechanisms to control the length-scales required for qubit-qubit interactions over a wide range, allowing to configure the desired density while still maintaining very large qubit arrays on chip.
With this scheme, the distribution of the space on the chip can be customized for the integration of local, classical control electronics aimed at solving the homogeneity and operation efficiency issues.

Expanding on this potential route to solving the wiring bottleneck, we present and analyze the \textit{spiderweb array}, a sparse 2D spin qubit array with single-qubit nodes separated by gate-based shuttling channels~\cite{Boter2019}. 
In this analogy, the qubit array resembles the open structure of a spiderweb, with vacant space between spiders (spin qubits) that move along the threads of their web (shuttling channels).
We will focus on how to use the open area of the sparse array to integrate classical control electronics with quantum hardware, in order to minimize the need for off-chip interconnects, resulting in a scalable Rent’s exponent.
Different from many proposal papers, we make a rather extensive initial effort to assess the feasibility of this classical-quantum integration, by considering several relevant practical aspects, without aiming or claiming to be exhaustive in this assessment.
Another aim of this work is to identify immediate technological development opportunities that can be prioritized in order to operate a large-scale spiderweb array as a quantum processor.

We first describe in \sref{sec:array_design} the basic physical architecture of the qubit array and the implementations of the operations required to sustain the surface code. 
In \sref{sec:electronics} we describe the integration of on-chip classical electronics at different layers within the architecture.
This leads to a logarithmic reduction in the number of control and measurement lines from the qubit region to the chip boundaries, which is discussed in \sref{sec:line_scaling}.
We then consider in \sref{sec:footprint} the footprint of the local control circuits and the array sparsity required to accommodate them.
\sref{sec:heat} discusses the power dissipation in different sections of the array, assuming $\sim$1K operation~\cite{Petit2020,Petit2020a,Yang2020}, and the effects this has on the operation.
Finally, we put all these ingredients together to describe a feasible implementation of a million-qubit array in \sref{sec:example}.

\begin{figure*}[t]
    \centering
    \includegraphics[height=0.225\textwidth]{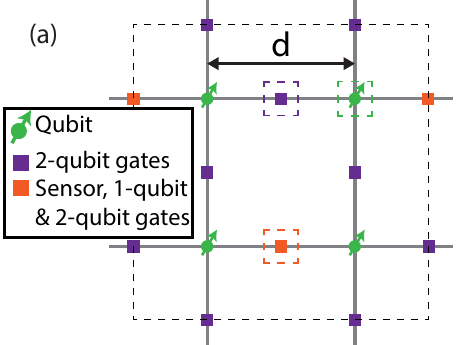}
    \includegraphics[height=0.225\textwidth]{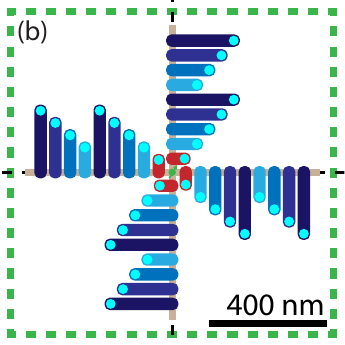}
    \includegraphics[height=0.225\textwidth]{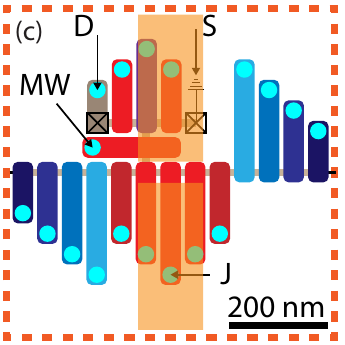}
    \includegraphics[height=0.225\textwidth]{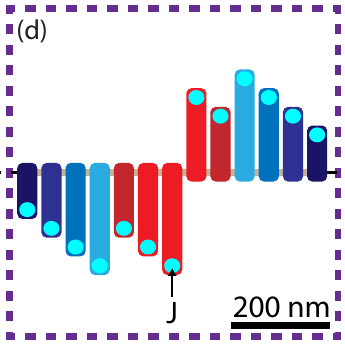}
    \caption{(a) Schematic (not to scale) of a unit cell containing four spin qubits (green), operation regions (purple/orange), connected via shuttling channels (gray lines). The dashed square indicates the repeating unit cell. (b) Qubit idling region. Four barrier gate electrodes (red) define the confinement potential and allow qubits into the shuttling channels (defined by the blue gate electrodes). Cyan circles represent vias to higher interconnect layers. The vias are arranged in a staggered configuration to enable interconnect routing. (c) Qubit operation region including control gates (red), sensing dot plunger (purple), source (S)/drain (D) Ohmic contacts (squares) and micromagnets (orange rectangles). The gate electrodes labeled MW and J are used for single- and two-qubit operations (see main text), respectively. (d) Two-qubit operation only regions. Dark red gates in (b-d) require coarse-resolution DC biasing, while the light red gates and the purple gate require a fine resolution.}
    \label{fig:unit_cell_schematics}
\end{figure*}

\section{Array design and operation}
\label{sec:array_design}
The overarching concept of the spiderweb array is presented in \fref{fig:overview}.
A large 2D square lattice of spin-qubits constitutes the \textit{quantum plane}.
The spin-qubits consist of single electrons confined in electrostatically defined quantum dots in silicon.
The large 2D array of qubits can be used to implement the surface code~\cite{Dennis2002}, by assigning qubits as data qubits or surface code (SC-) ancilla qubits in a checkerboard fashion, and allowing single-qubit operations as well as nearest-neighbor two-qubit operations.   
In contrast to most other spin-qubit architectures, this array is sparse, with a qubit pitch $d$ on the order of 10~\textmu m.
This has two major implications:
\begin{enumerate}
    \item It facilitates the local integration of classical control electronics, consisting here of sample-and-hold circuits that provide independent DC biasing of each quantum dot gate electrode, which effectively offsets inhomogeneities in the potential landscape across the array.
    Consequently, the array can be considered fully homogeneous allowing the majority of qubit control signals to be shared across the entire array, which significantly reduces the number of control lines at the quantum plane boundary.
    \item It requires a means to transfer quantum information over $\sim$10-\textmu m distances, which we implement by shuttling the qubits to operation regions--where single- and two-qubit operations are performed along with readout.
\end{enumerate}
We define a \textit{unit cell} as the the smallest operational set of elements, which can be concatenated with identical unit cells to form the large 2D array.
This unit cell (\fref{fig:overview}(a)) contains 4 qubits, and has an area of $(2d)^2$ and a perimeter of $8d$.
We define \textit{modules} (\fref{fig:overview}(b)) consisting of $N \times N$ unit cells, with an area and a perimeter of $(2dN)^2$ and $8dN$, respectively.
Modules define sections of the quantum plane where DC biasing and readout occur sequentially across the qubits in the module.
All other operations that are part of the surface code cycle occur simultaneously in all unit cells in the entire array.
Completing the array, the \textit{quantum plane} (\fref{fig:overview}(c)) consists of $M \times M$ modules ($M^2N^2$ unit cells and $4M^2N^2$ qubits), covering an area and a perimeter of $(2dNM)^2$ and $8dNM$, respectively.
As seen from \fref{fig:overview}(d), the quantum plane is designed to occupy a section of the die, with the remaining space to be used to reduce the off-chip wire count even further by adding additional on-chip electronics.
Module sizes for DC biasing and readout ($N_b$ and $N_r$, respectively) may be different, as long as $N_b M_b = N_r M_r$, where $M_b$ and $M_r$ relate to the number of DC biasing modules and readout modules in the full array, respectively.

A detailed schematic of the components of a unit cell is presented in \fref{fig:unit_cell_schematics}.
The unit cell mainly consists of large linear arrays of electrostatic gates, connecting the vertices of the qubit array to the operation regions (see \fref{fig:unit_cell_schematics}(a)).
The bulk of the gates in these linear arrays make up the shuttling channels (blue gates in \fref{fig:unit_cell_schematics}).
Four phase-shifted sinusoidal signals are applied to four consecutive gates (shades of blue in \fref{fig:unit_cell_schematics}), repeating the set of signals along the linear array to create a traveling-wave potential to trap and shuttle an electron~\cite{Taylor2005,Seidler2021}.
The sign of the phase difference between adjacent gates defines the shuttling direction.

Four gates at the vertices of the spiderweb array (red gates in \fref{fig:unit_cell_schematics}(b)), are used to control both the confinement potential that keeps the electrons at the vertex while idle, as well as the tunneling in and out of the shuttling channels.
The qubits are shuttled to and from the operation regions between the vertices (\fref{fig:unit_cell_schematics}(c,d)) in order to perform single- and two-qubit operations, as well as readout and initialization.
\fref{fig:unit_cell_schematics}(c) shows a schematic of the gate structure used in the operation regions.  

Single-qubit operations are performed via electric dipole spin resonance (EDSR) in a transverse magnetic field gradient provided by a pair of micromagnets~\cite{Pioro-Ladriere2007}.
Alternatively, the micromagnets can be omitted in qubit systems with large spin-orbit interaction~\cite{Hendrickx2020}.
After an electron is shuttled to the operation region and confined under the bottom red gates in \fref{fig:unit_cell_schematics}(c), a microwave pulse is applied to the control gate labeled MW, to drive spin rotations.
Qubit phase rotations can be achieved via geometric phase rotations~\cite{Aharonov1987} or by pulsing on the gate labeled J to temporarily detune the electron-spin energy via the Stark effect~\cite{Veldhorst2014}.

For two-qubit operations, two electrons from the vertices adjacent to the operation region are shuttled and confined under the red gates in \fref{fig:unit_cell_schematics}(c,d), and a pulsed signal on the gate labeled J activates an exchange interaction between the two electron spins~\cite{Loss1998,Divincenzo2000,Nielsen2016}.

Most state-of-the-art spin-qubit systems use one of two spin-to-charge conversion mechanisms to read out a qubit state: 1 - spin-selective tunneling to a nearby electron reservoir (Elzerman readout)~\cite{Elzerman2004}, or 2 - based on Pauli spin blockade (PSB readout)~\cite{Zwanenburg2013}, using a second quantum dot with an ancilla electron~\cite{Veldhorst2017,Seedhouse2021}.
The charge state of the single or double quantum dot is then identified using a charge sensing quantum dot connected to source/drain Ohmic contacts. 
Although both readout techniques are compatible with this architecture, we focus here on PSB readout, due to its shorter demonstrated readout times~\cite{Jones2019,Connors2020} and its compatibility with higher temperature operation~\cite{Petit2020,Petit2020a,Yang2020}.

\begin{figure*}
    \centering
    \includegraphics[width=\textwidth]{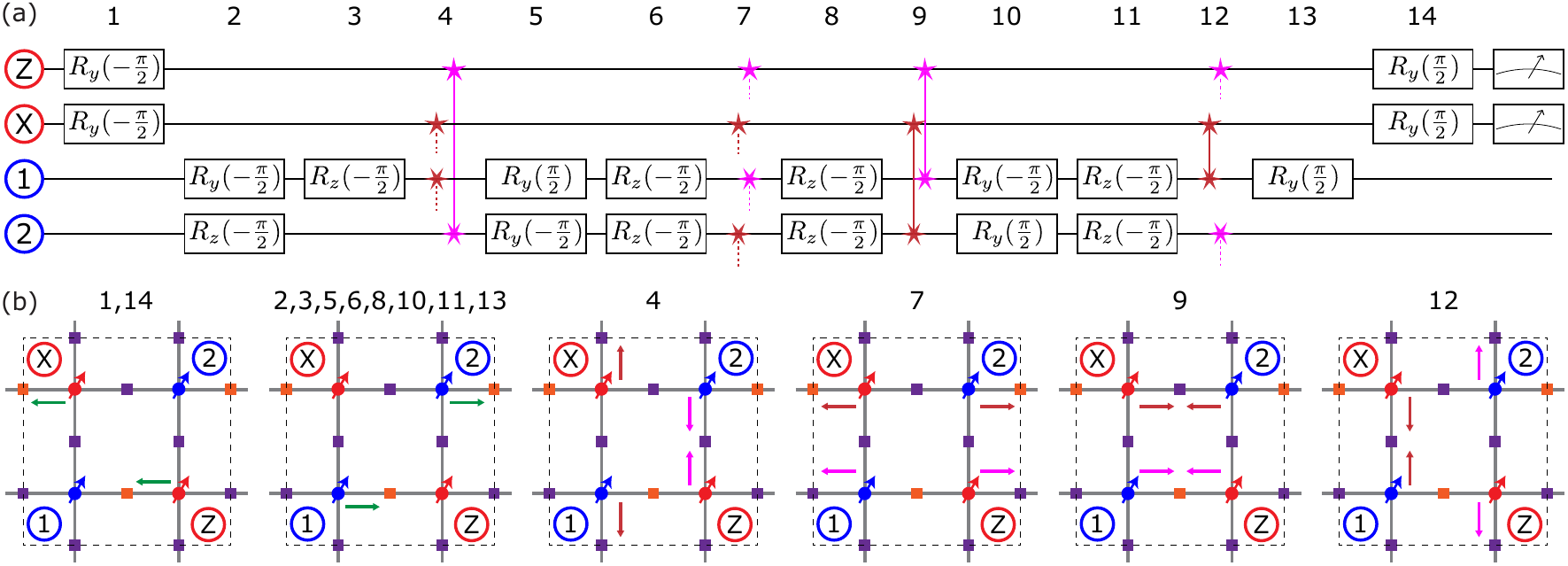}
    \caption{Surface code operation of the spiderweb array. (a) Circuit diagram showing the surface code cycle for the four-qubit unit cell. Boxes represent single-qubit gates, solid(dashed) lines represent intra(inter)-unit-cell two-qubit gates. (b) Schematics of qubit shuttling required to perform each of the steps in (a). Note that the two-qubit gates consist of three substeps (as described in the main text), and only the implementation of the $\sqrt{S_w}$ is shown. Also note that steps 3 and 13 only require shuttling of data qubit 1.}
    \label{fig:surface_code}
\end{figure*}

PSB readout can be used to measure the parity of a spin pair~\cite{Seedhouse2021}.
If prior to the measurement, one of the qubits is prepared in a known eigenstate, the parity measurement will reveal the state of the unknown spin.
Parallel spins are identified by their $(1{,}1)$ charge state while anti-parallel states are identified by their quick relaxation to the $(0{,}2)$ singlet.
The singlet can be adiabatically detuned and separated into a known product state ($\ket{\uparrow \downarrow}$ or $\ket{\downarrow \uparrow}$).
After this measurement protocol, the target qubit is left in its projected state and the RO-ancilla qubit remains in its original state.
In the surface code protocol, the measured state of the SC-ancilla qubits can be tracked to perform the required decoding of errors.
Alternatively, the target spin can be initialized after measurement by real-time feedback~\cite{Vandersypen2017} or by letting the state thermalize to the $(0{,}2)$ singlet after every measurement, then adiabatically detuning as described above.
In the spiderweb array, the target electron-spin qubit to be read out is shuttled to the operation region and confined under the bottom red gates in \fref{fig:unit_cell_schematics}(c), alongside an ancilla electron (RO-ancilla) prepared in a known state on the neighboring dot.
A PSB readout is performed to obtain a measurement of the target qubit, then the sensing dot can be tuned to store the RO-ancilla, and the two spins can be adiabatically separated with deterministic knowledge of their spin state, provided by the field gradient between the readout and sensing dot regions.

Qubits are able to share an operation region to perform single-qubit gates and only SC-ancilla qubits have to be read out.
This allows to reduce the number of gates in the unit cell by reducing some operation regions to only contain the required gates to perform two-qubit gates (see \fref{fig:unit_cell_schematics}(d)).
The arrangement of qubits and operation regions shown in \fref{fig:unit_cell_schematics}(a) constitutes a unit cell that can be tiled to obtain a fully operational surface code lattice.

Before the start of a computation, the spiderweb array needs to be initialized by loading electrons with a known spin state onto each of the qubit idling regions.
This is achieved by initializing electrons in the operation regions using the two-spin thermalization and adiabatic detuning method described above, and shuttling the electrons to the nearest idling regions.
Each operation region will initialize a data qubit first (discarding the second electron), followed by the SC- and RO-ancilla electrons.
After these two steps, the entire array is prepared to begin the surface code cycles, which we describe below.

We have designed the unit cell in the spiderweb array such that a cyclic sequence of pulsed signals can be used to perform the required operations to sustain an efficient surface code implementation~\cite{Versluis2017}, with the same sequence performed in parallel across all unit cells to sustain it in an arbitrarily large array.

\fref{fig:surface_code}(a) shows the circuit diagram of a single surface code cycle in a unit cell (see \aref{app:sc} for details of the circuit).
The circuit contains single-qubit gates denoted as $R_{\{y,z\}}(\theta)$, where $\theta$ is the angle of rotation and the subscript identifies the rotation axis on the Bloch-sphere.
The two-qubit operation in the schematic is a type of phase gate $S_p$, decomposed as: $-i S_p = \sqrt{S_w} (R_z^{[1]}(\pi) \otimes I^{[2]}) \sqrt{S_w}$, where $\sqrt{S_w}$ is the square-root-\textsc{swap} operation that is native to the exchange interaction, $I$ is the identity operator and the superscript in the single-qubit operator indicate which qubit the operation is applied to (see \aref{app:sc} for details).
Each step in the circuit is implemented by shuttling the participating qubits from their idle position to the operation region to undergo the required single- or two-qubit gate, before being shuttled back to their idle position.
Note that each two-qubit gate, as shown in the schematic, consists of three rounds of shuttling to achieve the required steps in the $S_p$ gate.
\fref{fig:surface_code}(b) shows the required qubit shuttling scheme appropriate to each step of the surface code cycle.
$R_y$ rotations are achieved by tuning the amplitude, phase and duration of the EDSR microwave pulse, while $R_z$ rotations are phase rotations achieved by using one of the methods described in \sref{sec:array_design}.
The $\sqrt{S_w}$ gate is achieved by calibrating the amplitude and duration of the pulse applied to the J gate that controls the exchange interaction between the qubits (see \fref{fig:unit_cell_schematics}(c,d)).
At the end of the cycle, the SC-ancilla qubits are measured in the operation regions.
The total operation time of a surface code cycle will be
\begin{equation}
    t_{sc} = 22\,t_{sh} + 14\,t_{1q} + 8\,t_{sw} + t_r, \label{eq:t_sc}
\end{equation}
where $t_{sh}$ is the time required to shuttle an electron from the vertex to the operation region and back, $t_{1q}$ is the single-qubit gate duration, $t_{sw}$ is the duration of a $\sqrt{S_w}$ operation and $t_r$ is the time required to readout the SC-ancilla qubits.

State-of-the-art surface code protocols follow one of two approaches to implementing logic gates.
The first, known as \textit{defect qubits}~\cite{Fowler2012}, requires disabling a subset of SC-ancilla qubits moving these defects around the lattice while the rounds of surface code error correction continue all around them.
The process of moving the defects not only involves disabling and enabling qubits in succession, but also requires measurements on individual data qubits.
The other implementation, known as \textit{lattice surgery}~\cite{Horsman2012}, divides the array into patches, separated by an idle row of interstitial data qubits.
Patches are created by splitting a section of the lattice, which requires performing a measurement of the interstitial qubits and subsequently disabling them from the surface code cycles.
Patches can also be merged, by initializing the interstitial qubits and reincorporating them into the surface code cycles.
The spiderweb array architecture can be designed to accommodate either type of logical qubit implementation.
Patches of qubits can be disabled by preventing shuttling of a subset of qubits while the rest of the surface code operations on all remaining qubits continue.
Selective qubit measurement and initialization is more complex to implement, since it will require interruptions of the regular surface code cycle.

\section{Local control electronics}\label{sec:electronics}
The spiderweb array has been designed with the main intention of providing space within the qubit plane to integrate local control electronics, with the ultimate purpose of obtaining a feasible scaling factor between number of qubits and external control and measurement signals.
In this section we describe in detail the implementation and function of these circuits, along with the corresponding routing of the signal lines between signal-generating source and gates, as summarized in \tref{tab:routing}.

\begin{table}
    \centering
    \begin{ruledtabular}
    \begin{tabular}{>{\raggedright}p{.27\columnwidth}l}
        \multicolumn{1}{c}{Gates} & \multicolumn{1}{c}{Routing} \\
        \hline
        \noalign{\vspace{3pt}}
        Shuttling (blue) & Source $\rightarrow$ gate  \\[3pt]
        Pulsed (red) & \begin{tabular}{@{}l@{}}
        DC: source $\rightarrow$ local demux \& S/H $\rightarrow$ gate \\
        AC: source $\rightarrow$ gate \end{tabular} \\
        \noalign{\vspace{3pt}}
        Sensing dot & DC: source $\rightarrow$ local demux \& S/H $\rightarrow$ gate \\
        plunger (purple) & AC: source $\rightarrow$ global demux $\rightarrow$ gate \\[3pt]
        Drain contacts & Measurement device $\leftarrow$ Ohmic \\
    \end{tabular}
    \end{ruledtabular}
    \caption{Signal routing scheme for the four different type of control lines in the array design. Demux and S/H abbreviates demultiplexer and sample-and-hold circuitry, respectively.}
    \label{tab:routing}
\end{table}

\begin{table}[b]
    \centering
    \begin{ruledtabular}
    \begin{tabular}{lcccc}
        Region & \begin{tabular}{@{}c@{}}Regions in \\ unit cell \end{tabular} & \begin{tabular}{@{}c@{}}Fine \\ (1 \textmu V) \end{tabular} & \begin{tabular}{@{}c@{}}Coarse \\ (1 mV) \end{tabular} & \begin{tabular}{@{}c@{}}Pulsed \\ gates \end{tabular} \\
        \hline
        Qubit idling & 4 & - & 4 & 4 \\
        Qubit operation & 2 & 7 & 2 & 6 \\
        Two-qubit only & 6 & 3 & 2 & 5 \\
        \hline
        \multicolumn{2}{r}{Total per unit cell} & 32 & 32 & 58 \\
    \end{tabular}
    \end{ruledtabular}
    \caption{Number of gates for each region of a unit cell with their associated type of biasing and pulsed signals.}
    \label{tab:DC_AC_count}
\end{table}

\subsection{Biasing and control signals}
In principle, the array only requires the generation of control signals for a single unit cell, and that set of signals can be replicated in parallel across the entire array to sustain the surface code.
In practice, setting a quantum dot array to a state where several qubits can be controlled simultaneously with a single set of control signals, requires careful calibration of the DC voltages of each gate in the array~\cite{Vandersypen2017}.
Inhomogeneities in the materials and fabrication properties will cause these calibration voltages to vary over a wide range across the lattice, requiring each qubit to have independent biasing.

To mitigate this effect, we design a sample-and-hold scheme to apply a local DC bias to the gate electrodes~\cite{Xu2020}.
We define two bias voltage resolutions $\Delta V$ to accommodate different gate functionalities.
For gates acting as barriers to shuttling channels (dark red gates in \fref{fig:unit_cell_schematics}), only a resolution sufficient to maintain an electron in a quantum dot is required and therefore we can afford a coarse resolution $\Delta V_c = 1$~mV.
A fine resolution $\Delta V_f = 1$~\textmu V is required for all other plunger and barrier gates~\cite{Vandersypen2017}.
The need for DC biasing of the shuttling gates is eliminated by making the traveling wave potential large enough to overcome potential landscape inhomogeneities.
As shown in \tref{tab:DC_AC_count}, there are a total of 64 gates requiring DC biasing, with an equal split between gates requiring 1~mV and 1~\textmu V resolutions.

\begin{figure}
    \centering
    \includegraphics[height=0.2\textwidth]{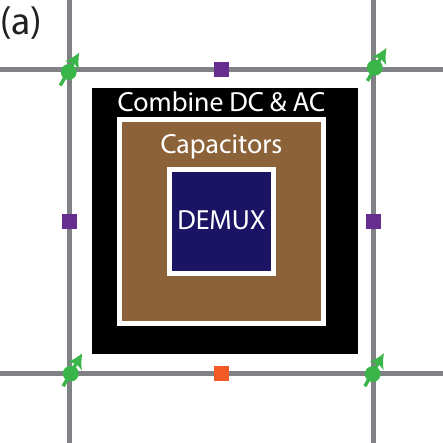}
    \includegraphics[height=0.2\textwidth]{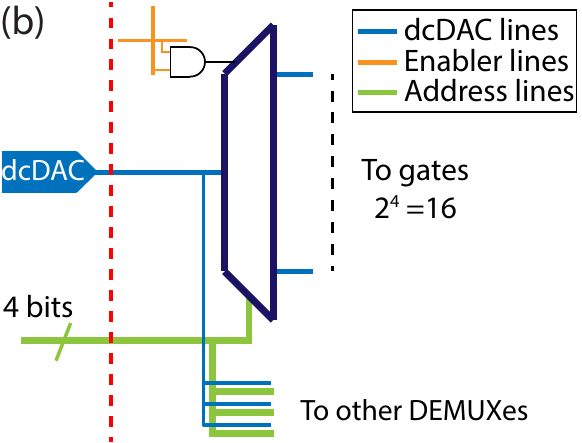}
    \caption{(a) Schematic of a unit cell containing locally integrated classical electronics. Demultiplexers in combination with capacitors form sample-and-hold-circuits to provide DC bias voltages. Additionally, circuits that combine the DC bias with AC control signals are required. (b) Input/output schematic of the demultiplexers. A demultiplexer (dark blue), once enabled via crossbar addressing (orange), ports the DC voltages coming from the dcDAC (light blue) to the output selected by the 4-bit address bus (green). The dashed red line indicates the quantum plane boundary. The color coding in (a) and the legend in (b) represent the same components in both panels as well as in \fref{fig:dc_biasing_2} and \fref{fig:dc+ac}.}
    \label{fig:dc_biasing_1}
\end{figure}

\begin{figure*}
    \centering
    \includegraphics[height=0.27\textwidth]{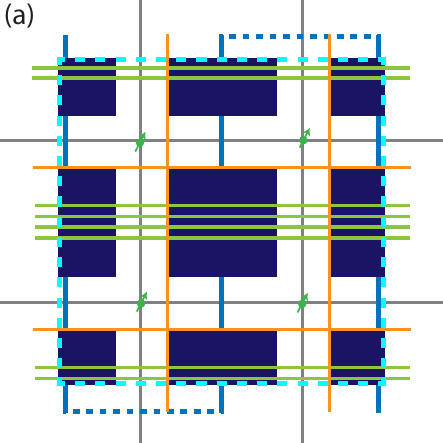}
    \includegraphics[height=0.27\textwidth]{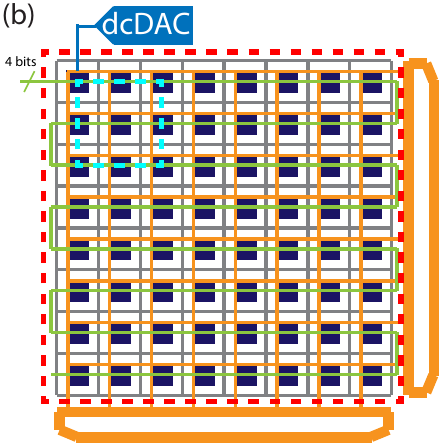}
    \includegraphics[height=0.27\textwidth]{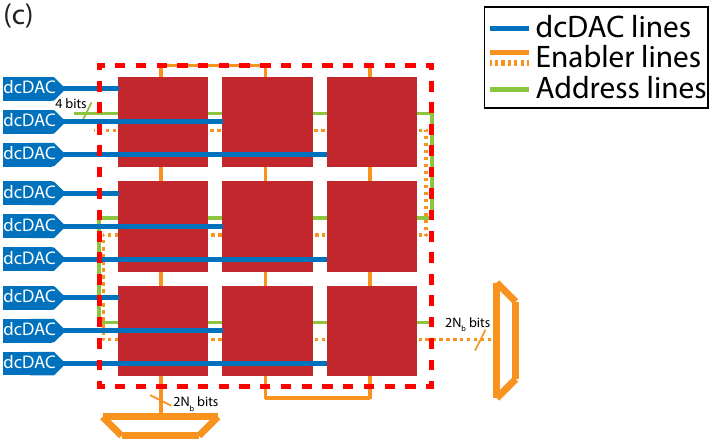}
    \caption{Schematics of (a) a unit cell and (b) a module. Demultiplexers are sequentially enabled by crossbar addressing controlled by multiplexers (orange blocks). (c) Schematic of the array of modules completing the quantum plane. Dashed red lines in (b) and (c) denote the quantum plane boundary.}
    \label{fig:dc_biasing_2}
\end{figure*}

The minimum hold capacitance required to achieve coarse ($C_c$) and fine ($C_f$) resolutions is  $C_c \approx 0.16$~fF (limited by the electron charge $e/\Delta V_c$) and $C_f \approx 14$~pF (limited by thermal noise $k_B T/\Delta V_f^2$, assuming power dissipation from the local electronics requires the operating temperature to be raised to 1 K).

The integrated electronics required to implement DC biasing consists of demultiplexers and capacitors that together form sample-and-hold circuits, as schematically depicted in \fref{fig:dc_biasing_1}.
Local demultiplexers distribute DC voltages generated remotely (i.e., outside the quantum plane) by voltage sources which we have labeled dcDAC, to local capacitors connected to the gate electrodes.
The local control electronics for a unit cell needs to be distributed within the $d^2$ area regions between the qubits, with a total of $4d^2$ open footprint available per unit cell (see \fref{fig:unit_cell_schematics}(a)). 
Therefore, in order to bias 64 gates per unit cell (as per \tref{tab:DC_AC_count}), we fit 1-to-16 demultiplexers in each open region between the qubits, which implies 4 demultiplexers per unit cell.
The demultiplexers are implemented using 4-bit digital decoders, with each output activated using the 4-bit address line.

\fref{fig:dc_biasing_2} shows the DC biasing scheme on the unit cell, module and quantum plane levels.
All demultiplexers within a module share the same input DC biasing signal (\fref{fig:dc_biasing_2}(b)), and all demultiplexers in the quantum plane share the same address bus (\fref{fig:dc_biasing_2}(c)).
The demultiplexers in a module are enabled sequentially by crossbar addressing and in turn sequentially (one by one) update each gate. This way, all modules are updated in parallel and therefore one module refresh cycle is required to refresh the entire qubit array.

\begin{figure}
    \centering
    \includegraphics[height=0.35\textwidth]{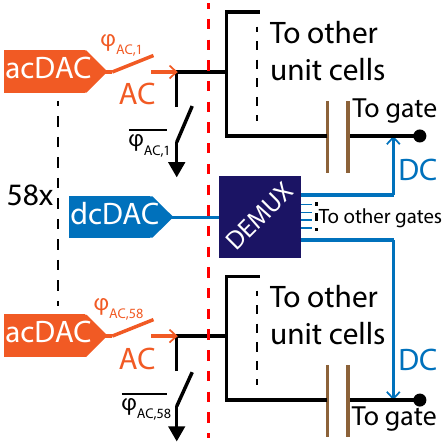}
    \caption{Circuit schematic to combine AC and DC signals as described in the main text. acDAC (dcDAC) are voltage sources for pulsed signals (DC biasing). The dashed red line indicates the quantum plane boundary.}
    \label{fig:dc+ac}
\end{figure}

AC signals (MW and pulsed) are generated remotely by sources which we have labeled acDAC.
Each signal is distributed throughout the array to their respective gates and the complementary switching circuit (see $\varphi_{AC}$ and $\overline{\varphi_{AC}}$) shown in \fref{fig:dc+ac} is used to combine the AC and DC components of the gate voltages.

\begin{figure}
    \centering
    \includegraphics[width=0.235\textwidth]{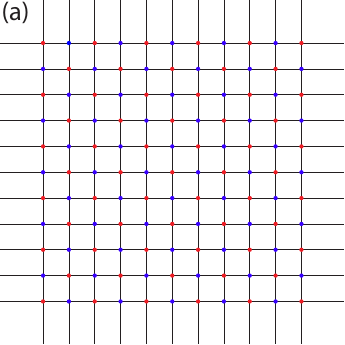}
    \includegraphics[width=0.235\textwidth]{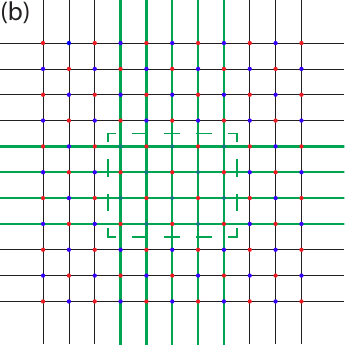}
    \caption{Crossbar scheme to control local patches of qubits. (a) Crossbars connect all columns and rows of qubits in the array. None of the crossbar lines are activated and the surface code runs on the entire array. (b) Example of the creation of a defect, where the crossbars in green are activated, disabling the qubits inside the dashed green rectangle from} tunnelling onto the shuttling channel.
    \label{fig:surface_code_2}
\end{figure}

To implement logical qubits within the surface code lattice, we have designed an additional scheme to control local patches of data and SC-ancilla qubits.
Switches connected to the barrier gates surrounding the idle regions, control the tunneling of the qubits into the shuttling channel.
When one of these switches is activated, the corresponding qubit will remain in the idle region while the rest of the operations in the array carry on.
The signals used to control these switches are arranged in a crossbar fashion across the entire quantum plane, as schematically depicted in \fref{fig:surface_code_2}.
This scheme enables the implementation of either defect qubits or lattice surgery.
Defects or interstitial boundaries can be created by disabling patches or rows of qubits, respectively.
Selective data qubit measurement or initialization can be performed by temporarily disabling the rest of the array and interrupting the surface code signal to perform the required operations.
The crossbar addressing scheme limits the shape of the defects and patches to rectangles of arbitrary size, but meets basic surface code requirements and within those limits allows for universal control.

\subsection{Readout signals}
It is difficult to realise a favourable Rent's exponent on the measurement circuit, because each SC-ancilla qubit needs to be read out independently. 
In order to use a single line outside the quantum plane to measure more than one qubit, the readout protocol will require some form of multiplexing.
With this in mind, we define a \textit{readout module} consisting of $N_r^2$ unit cells that share a single readout signal line for multiplexed measurements.

The simplest form of multiplexing is to read out qubits sequentially.
This is done by connecting the drain contacts of all sensor dots in a readout module to a single line at the quantum plane boundary
and consecutively pulsing their plungers to bring them to the low-impedance, electrostatically sensitive regime, while all other sensor dots in the readout module are in Coulomb blockade (i.e., in the high-impedance regime).
A global readout demultiplexer is used for the sequential control of the sensor plungers in a readout module. This demultiplexer can be shared between all readout modules across the entire array and can be located outside the quantum plane.
This method is technically simple to implement, but will be limited by the data qubit coherence time, since it will increase the total surface code cycle time (i.e., the last term in \eref{eq:t_sc} becomes $N_r t_r$).

Simultaneous readout of a number of qubits can also be achieved using other multiplexing techniques such as amplitude, frequency and/or phase modulation.
The plungers of the qubits to be read out in parallel can be connected to a single pulsed signal line that activates all sensors simultaneously.
The measured signal from the common Ohmic line needs to then be demodulated to extract each individual qubit measurement.

Amplitude modulation is achieved by tuning the bias voltages of the plungers from the sensors that are read out simultaneously, such that each sensor response will result in distinct current amplitudes.
The currents from all the sensors can then be added into a single output line and the total current amplitude can be used to decode the responses of all the sensors.
The number of parallel readouts using this technique will be limited by the signal-to-noise ratio of the sensor response.

Frequency and phase modulation are achieved by applying RF reflectometry~\cite{Schoelkopf1998}.
This technique requires the design of resonant circuits that connect to the readout SET Ohmics.
The challenge with this multiplexing strategy is that in order to use multiple frequencies combined into a single output line at the qubit plane boundary, the resonant circuits will need to be implemented locally, significantly increasing the footprint requirements of the control electronics.

The evolution in state-of-the-art spin-qubit measurement techniques will determine the most feasible readout multiplexing strategy that can be applied in this architecture.

\section{Line scaling}
\label{sec:line_scaling}
The local electronic circuits described above allow for significant sharing of control lines, which results in a very efficient scaling of the ratio of the number of interconnects at the quantum plane boundary to the number of lines at the unit cell level.
The line scaling factors obtained for each type of control and measurement signal are summarized in \tref{tab:wire_count}.
We note that some additional interconnects will be required at the unit cell and module level in order to route connections to adjacent cells.

\begin{table}
    \centering
    \begin{ruledtabular}
    \begin{tabular}{>{\raggedright}m{0.25\columnwidth}ccc}
        Type of line & \multicolumn{3}{c}{Connections at} \\
        & unit cell & module & quantum plane \\
        \hline
        \noalign{\vspace{3pt}}
        DC biasing & 9 & $4N_b+5$ & $M_b^2+4N_b+4$ \\[3pt]
        Shuttling & 4 & 4 & 4 \\[3pt]
        Pulsed signals \& MW & 58 & 58 & 58 \\[3pt]
        Logical operations & $4x$ & $4N_bx$ & $4N_bM_bx$ \\[3pt]
        Readout & 3 & \begin{tabular}{@{}c@{}}$2\log_2 N_r - $ \\ $\log_2 r + 1$ \end{tabular} & \begin{tabular}{@{}c@{}}$M_r^{2} + 2\log_2 N_r -$  \\ $\log_2 r$ \end{tabular} \\
        \hline
        \noalign{\vspace{3pt}}
        Total & $74+4x$ & \begin{tabular}{@{}c@{}}$4N_b(1+x)+$ \\ $2\log_2 N_r - $ \\ $\log_2 r + 68$ \end{tabular} & \begin{tabular}{@{}c@{}}$M_b^{2} + M_r^{2} + $ \\ $4N_b(1+M_bx)$ \\ $+2\log_2 N_r -$ \\ $\log_2 r + 66$ \end{tabular}
    \end{tabular}
    \end{ruledtabular}
    \caption{Scaling of the number of local connections at different levels of the array.}
    \label{tab:wire_count}
\end{table}

The sample-and-hold circuits for independent DC biasing of the gates require $O(M_b^2+N_b)$ lines at the quantum plane boundary.
Concretely, at the unit cell level, four digital address lines (green lines in \fref{fig:dc_biasing_2}(a)) and four enabler lines (orange lines) select a specific output of one demultiplexer, which is set to the correct voltage via a single connection to a voltage source (i.e., dcDAC) outside the quantum plane.
This makes a total of 9 lines required for DC biasing.
At the module level only the enabler lines scale with the number of unit cells as $4N_b$.
Meanwhile, the 4 digital address lines and the connection to the voltage source is shared between all unit cells in a module.
Every module is served by one dcDAC, so at the quantum plane level $M_b^2$ connections to dcDACs are needed, while both the digital address lines and the enabler lines are shared between all modules.

Shuttling of electrons across the array requires four signals that can be fully shared over the entire array.

Since all pulsed and microwave control signals can be shared across all unit cells in the array, a constant number of 58 of these lines (see \tref{tab:DC_AC_count}) are required at the quantum plane boundary to sustain the surface code, irrespective of the total number of qubits.

To control the switches that disable shuttling of qubits from the idle region, we propose to use $x$ crossbars, in order to allow for $x$ qubit patches to be simultaneously created and manipulated, therefore accommodating approximately $x$ logical qubits.
To obtain an estimate of $x$, we first consider that the number of physical qubits required to implement a logical qubit will depend on the maximum error correction distance $d_c$, measured as the number of neighboring physical qubits that can have errors before the logical qubit produces an error.
The maximum number of logical qubits that will fit in the array will then depend on the chosen logical qubit implementation~\cite{Devitt2009}:
\begin{equation}
\begin{split}
x_{\textrm{max,dq}} &\approx \frac{2U}{3d_c^2} \qquad \textrm{(defect qubits)} \\
x_{\textrm{max,ls}} &\approx \frac{U}{d_c^2} \qquad \textrm{(lattice surgery)}, \label{eq:logical_crossbars}
\end{split}
\end{equation}
where $U = M_b^2N_b^2$ is the number of unit cells in the array.
In this line-counting exercise, we consider that each of the crossbars span the entire array, which require two horizontal and two vertical lines per unit cell.
The number of crossbar lines then scales with $N_b$ and $N_bM_b$ at the module and quantum plane level, respectively.
This is a worst-case estimate, since we envision that it will be good enough to define crossbars over partial sections of the array, therefore reducing the total line count.
In addition, the number of lines required per crossbar for lattice surgery will likely be significantly less than for defect qubits, since only the interstitial qubits need to be locally addressed.

For the readout signals, a unit cell contains two readout SETs that both require a connection to their plunger and share a single Ohmic line.
A module requires a single Ohmic line, and the plunger line scaling will depend on the multiplexing technique used.
Performing $q$ sequential readouts will require $\log_2 q$ lines to address the readout decoders.
Simultaneous readout of $r$ unit cells will just require a single line that connects all the plungers together.
Therefore, a readout module consisting of $N_r^2 = q r$ unit cells, will require $2\log_2 N_r - \log_2 r + 1$ lines in total for the readout scheme.
To read out the full quantum plane, all $M_r^2$ modules require their own Ohmic line, while the address lines for the readout demultiplexers are shared.

\subsection{Module size considerations}
From the total achievable signal connections at the quantum plane boundary (see \tref{tab:wire_count}), it is clear that Rent's exponent will be minimized if the module size is made as large as possible (i.e., maximize $N$, thus minimize $M$).

The main consideration for the DC bias module size $N_b$ is the required refresh rate of the sample-and-hold circuits.
Leakage current from the hold capacitors will cause a drift in the DC bias voltage on the gates $dV/dt$~\cite{Puddy2015,Xu2020}.
The ratio between the voltage drift and the required voltage stability (which we assume equal to the voltage resolution $\Delta V$) will determine the minimum refresh rate
\begin{equation}
    f_c = \frac{\frac{dV}{dt}}{\Delta V}.
\end{equation}
The leakage current is very technology dependent and values are not readily available for state-of-the-art integrated capacitor technology.
To date, there have only been two proof-of-principle demonstrations of sample-and-hold circuits integrated with qubit device gates, with reported $dV/dt$ values ranging from 2~\textmu V/s~\cite{Puddy2015} to 0.1~V/s~\cite{Xu2020}.
Using these values, we obtain minimum refresh rates ranging from $f_c = 2$~Hz to $100$~kHz, limited by the sample-and-hold circuits with fine resolution $\Delta V_f = 1$~\textmu V.
The module size will then set the minimum clock frequency required to run the DC biasing demultiplexers $f_b = 64 N_b^2 f_c$.
Therefore, the maximum module size will be limited by the feasibility of distributing the dcDAC signals across the array, which becomes more difficult as $f_b$ increases.
This issue will be discussed in more detail below.

The readout module size has two limiting factors. The number of parallel readouts $r$ through a single line will depend on the feasibility of applying readout multiplexing techniques.
As readout is generally the operation that takes the most amount of time, $q$ will need to be kept to a minimum in order to restrain $t_{sc}$ to within the appropriate bounds of coherence that will achieve a good quantum memory.

\section{Footprint}
\label{sec:footprint}
We now consider the footprint requirements of the control electronics that need to be locally integrated in the quantum plane.
This is the minimum area that needs to be available adjacent to the qubits, and therefore sets the minimum qubit pitch $d$.
The most significant contribution to the footprint comes from the capacitors that are required for the sample-and-hold scheme.
As summarized in \tref{tab:DC_AC_count}, 32 gate electrodes per unit cell require a fine voltage resolution and another 32 gates require coarse resolution, which comprise a total capacitance per unit cell of $\sim$450~pF.
Assuming $\sim$1~pF/\textmu m\textsuperscript{2} (using state-of-the-art deep-trench capacitor technology~\cite{Hou2019}), we estimate a total capacitor footprint $A_c \approx 450$~\textmu m\textsuperscript{2}.
In addition, we modeled a decoder circuit using 40-nm technology (see \aref{app:demux} for details) and obtained an estimate of the total footprint of the required demultiplexers $A_d \approx 180$~\textmu m\textsuperscript{2} per unit cell.
This adds to a total footprint per unit cell of $A_u \approx 630$~\textmu m\textsuperscript{2}, which allows to set the qubit pitch to $d \gtrsim 13$~\textmu m.
Assuming a 50-nm pitch between gate electrodes, this would require linear arrays of 260 gate electrodes per lattice arm and would set the unit cell area to $4\,d^2\approx676$~\textmu m\textsuperscript{2}.

The required unit cell footprint sets the qubit pitch and consequently the perimeter through which the required interconnects have to enter and leave the unit cell.
To evaluate the feasibility of implementing the spiderweb array gate structure integrated with the local control electronics, we have constructed a circuit model in \textsc{cadence} (see \aref{app:cadence} for details) with realistic parameters that includes all the essential components and connection routing scheme described here, using a total of four metal layers. To obtain an estimate of the maximum number $x_{\textrm{max,fab}}$ of crossbars that can be added for the implementation of logical qubits, we need to first estimate the maximum number of interconnect lines that can be routed across the perimeter of a unit cell $N_{\textrm{lines}} = 8\,d\,N_{\textrm{layers}} / \Delta_\textrm{i}$, where $N_{\textrm{layers}}$ is the number of metallic layers that can be used for routing and $\Delta_\textrm{i}$ is the pitch of the interconnect lines.
Then $x_{\textrm{max,fab}} = N_{\textrm{lines}} / 8$, since each of the 4 crossbar lines will cross the unit cell perimeter twice.
Assuming current numbers from the latest device roadmap report~\cite{IRDS2020} $N_{\textrm{layers}} = 12$ and $\Delta_\textrm{i} = 80$~nm, we estimate $x_{\textrm{max,fab}} \approx 2000$.

\section{Heat dissipation}
\label{sec:heat}
As with all quantum processors that will operate at cryogenic temperatures, it is necessary to ensure that the heat dissipated during the operation is kept to within the stringent requirements set by the cooling power available~\cite{Vandersypen2017}.
In this section we will discuss some of the main sources of heat dissipation in the spiderweb array.

Routing of the signals lines across the chip requires a high-density of metallic lines at the different interconnect levels, which results in parasitic capacitances between the lines.
Any oscillating or pulsed signal on these lines will dissipate energy due to the charging and discharging of these parasitic capacitances.
Using the interconnect circuit model drawn in \textsc{cadence} as a guide, we consider the two layers with the highest line density as a regular grid of metal lines and use this to obtain an estimate of $C_p = 700$~fF for the total parasitic capacitance of the signal lines in a unit cell (see \aref{app:heat}.1 for details).
When a capacitor $C_p$ is charged by a voltage source $v_p$, the energy stored in the capacitor ($0.5C_pv_p^2$) is half the energy supplied by the source ($Qv_p = C_pv_p^2$).
The other half is dissipated by the circuit as heat by the parasitic resistance between the voltage source and the capacitor, independent of the resistance value.
This is known as the \emph{dynamic power} dissipation from the parasitic capacitance and can be expressed as:
\begin{equation}
    P_p = \tfrac{1}{2} C_p v_p^2 f_p, \label{eq:power_cap}
\end{equation}
where $v_p$ and $f_p$ are the amplitude and frequency of the pulses applied to the lines.
We note that this estimate includes power dissipated both by the signal lines and the driving circuit--which will be outside the quantum plane--so it should be taken as a worst case estimate.

Next, we estimate the dissipation for the sample-and-hold circuits, which comes mainly from the dynamic power consumption of the network of transistor switches driving the hold capacitors.
We use the decoder model introduced in \sref{sec:footprint} and described in \aref{app:demux} to estimate a maximum energy dissipation of 0.35~pJ for a 4-bit decoder cycling through all 16 outputs.
Using the worst-case $100$~kHz gate refresh rate previously estimated, we obtain the transient power dissipation $P_d < 140$~nW for 4 demultiplexers in a unit cell, noting that capacitor leakage rates in state-of-the-art technologies will likely make this dissipation orders of magnitude lower.
Additional capacitive-load power from charging and discharging the array of hold capacitors is negligible ($\sim$fW per unit cell).

Another important consideration is the power dissipated due to the finite resistance of lengthy lines carrying AC signals.
We model the signal lines as transmission lines with finite resistance and parasitic capacitance to ground (see \aref{app:heat}.2 for details), to obtain the following expression for the power dissipation of a line running above a unit cell of length $2d = 26$~$\mu$m:
\begin{equation}
P_t = 1.1 \tfrac{\textrm{nW\,ns$^2$}}{\textrm{V$^2$}} \, \left( v_t f_t \right)^2, \label{eq:power_tl}
\end{equation}
where $v_t$ and $f_t$ are the amplitude and frequency of the signal, respectively.

The total power that will be dissipated by the operation of the entire array will then be
\begin{equation}
    P_T = U (P_p + P_d + P_t). \label{eq:power_total}
\end{equation}

\section{Example: a million-qubit array}
\label{sec:example}
To illustrate the advantages of implementing this architecture on a large scale, let us consider a spiderweb array of $2^{20}$ ($\approx10^6$) qubits, or $U = 2^{18}$ unit cells.

To make a concrete assessment of the line scaling, we assume a DC biasing module size $N_b^2 = 1024$, with $M_b^2 = 256$ modules to complete the array.
This sets the maximum clock frequency of the biasing demultiplexers to $f_b \approx 6$~GHz.
We assume a measurement multiplexing strategy with $q = 4$ sequential readouts of sets of 8 SC-ancilla qubits read out in parallel using amplitude, frequency and/or phase multiplexing ($r = 4$).
This implies a readout module size $N_r^2 = 16$ and $M_r^2 = 16384$. 
Omitting for the moment the crossbars required to implement logical qubits, and using the result in \tref{tab:wire_count}, the array contains a total of $c = 74$ connections per unit cell and $T = 16,836$ connections at the quantum plane boundary.
Using the formula for Rent's rule $T = c \, U^p$, we extract a Rent's exponent $p = 0.43$.
Adding crossbar circuits will increase Rent's exponent to a maximum $p=0.5$ for $x \gtrsim 200$ (see \aref{app:rent} for further discussion).

Considering that state-of-the-art qubit fidelities, require a code distance $d_c \approx 16$ to perform complex quantum algorithms~\cite{Devitt2009}, we can estimate upper bounds of $x$ from \eref{eq:logical_crossbars}, $x_{\textrm{max,dq}} \approx 700$ and $x_{\textrm{max,ls}} \approx 1000$.

The total area covered by the quantum plane of this million qubit array with local control electronics, will be $(2dNM)^2\approx$ 177 mm\textsuperscript{2}.
The remaining area on, for example, a 22 mm $\times$ 33 mm (726 mm\textsuperscript{2}) die is $\sim$550 mm\textsuperscript{2}, and can be used to implement classical control circuits, i.e., among others the pulsed voltage sources we have described.
In addition, additional levels of multiplexing can be employed to bring the off-chip wire count, typically being the real bottleneck for Rent’s rule, to well below the wire count at the quantum plane boundary.

We can evaluate the total duration of a surface code cycle by estimating the operation times in \eref{eq:t_sc}.
From recent modeling of a shuttling protocol similar to the one used here~\cite{Buonacorsi2020}, we estimate that with a 50~nm gate pitch we can achieve $t_{sh} \approx 50$~ns, maintaining $>$99.9\% fidelity.
We note that a recent experimental demonstration of this shuttling protocol~\cite{Seidler2021} showed shuttling fidelities $>$90~\% for distances $\sim$420~nm, an encouraging initial result towards the feasibility of this protocol at larger scales.
State-of-the-art quantum dot qubit systems, with operation mechanisms similar to the ones proposed here, have demonstrated Rabi frequencies and exchange couplings $\sim$10~MHz~\cite{Yoneda2018,Watson2018,Leon2020}, which correspond in our system to $t_{1q},t_{sw} \approx 25$~ns.
In addition, those same systems exhibit dephasing times as long as $T_2^* \approx 20$~\textmu s.
High-fidelity readout using Pauli spin blockade has been achieved on timescales $t_r \approx 1$~\textmu s~\cite{Jones2019,Connors2020}.
Assuming these operation times, we estimate an entire surface code cycle will take $t_{sc} \approx 6$~\textmu s, more than 3$\times$ shorter than the coherence time.

Finally, we calculate the power dissipation from the sources described in \sref{sec:heat}.
To estimate the power dissipated from the parasitic capacitance of the interconnects, we first note from \tref{tab:wire_count} that the bulk of lines that need to be routed above the unit cell correspond to pulsed signals for qubit operations.
These signal lines will need to be activated on average $\sim 6$ times per surface code cycle, from which we estimate $f_p \approx 1$~MHz.
Assuming $v_p \approx 1$~V of pulse amplitude on these lines, we can use \eref{eq:power_cap} to estimate $U P_p \lesssim 90$~mW for these lines.
The biasing multiplexers will dissipate $U P_d \lesssim 40$~mW for the entire array.
We use \eref{eq:power_tl} to calculate the power dissipation of the lossy transmission lines carrying the higher-frequency signals.
The MW signals used for single-qubit control are relatively low amplitude and pulsed with very short duty cycle, so their power dissipation will be negligible.
A larger contribution will come from the lines that carry the signals for the shuttling channels, since they will be nearly continuously activated.
Assuming a voltage amplitude $v_t = 1$~V and frequency $f_t = 1$~GHz, we estimate $U P_t \lesssim 0.3$~mW.
These estimates suggest that the parasitic capacitance and the local demultiplexers will be the main contributors to the total power dissipation in the array, with a million-qubit spiderweb array expected to dissipate on the order of $100$~mW of power. This is well within the cooling power capabilities for 1~K operation using state-of-the-art cryogenic technologies.
We can also estimate how well this power can be transferred from the chip to the cryogenic environment with currently used heat sinking techniques.
Recent work~\cite{vanDijk2020} studying a CMOS integrated chip in a cryogenic environment (3~K) found that a circuit dissipating 200~mW increased the chip temperature by 7~K over an area of $\sim 4$~mm$^2$.
Considering that the qubit plane area is 40$\times$ larger than that, we can roughly estimate $\sim 0.2$~K of self heating in the qubit plane of the spiderweb array.

\section{Discussion}
The goal of this work is to provide insight into the feasibility of implementing a quantum hardware architecture with integrated local control electronics.
We have shown that, with reasonable assumptions regarding quantum and classical fabrication and measurement techniques, a million qubit array made from spins in quantum dots is achievable.
We have used the state-of-the-art in quantum-dot qubit development to consider a range of technical implementation aspects, and will now discuss some of the open questions that will need to be addressed based on how the technology evolves.
As with most other electrically controlled solid-state qubit architectures, the distribution of high-frequency signals is not trivial, with potential issues such as standing waves and signal synchronization across the array.
Some additional local circuitry will likely be needed to solve these issues.

Furthermore, qubit operation performance could be significantly improved with significant restructuring of the array topology, to place micromagnets at the qubit idle regions and add a fast gate to perform single-qubit gates.
This would reduce the number of shuttles per surface code cycle, allow for dynamical decoupling during idle times to extend coherence times and correct phase errors that may arise, e.g., from the shuttling process.

If the number of cross-bars becomes the limiting factor to Rent's exponent, it would be beneficial to move the DC-biasing and readout demultiplexers outside the quantum plane.
This would reduce the footprint and in turn relax the shuttling performance requirements.

This work builds on previous large-scale qubit architectures based on quantum dots, focussing on the implementation of integrated classical and quantum electronics.
We believe it will help identify the key areas of technological development required to shortcut the path towards building the future generation of quantum computers.

\begin{acknowledgments}
We would like acknowledge Simon Devitt and Ramon W.J. Overwater for useful discussions.

This work was supported by Intel Corporation and the Early Research Programme of the Netherlands Organisation for Applied Scientific
Research (TNO) with additional support from the Top Sector High Tech Systems and Materials.
\end{acknowledgments}

\appendix	
\section{Demultiplexer design} \label{app:demux}
In order to estimate the footprint and the power consumption of the demultiplexer used to route the DC  biasing to the gates in a unit cell, we model a 1-to-16 demultiplexer using a 4-bit digital decoder driving an array of 16 switches.
As explained in the main text, each unit cell accommodates 4 4-bit demultiplexers to bias 64 gates, so that the demultiplexers can be easily placed in the 4 free areas within the quantum-dot grid.

\begin{figure*}
    \centering
    \includegraphics[width=0.9\textwidth]{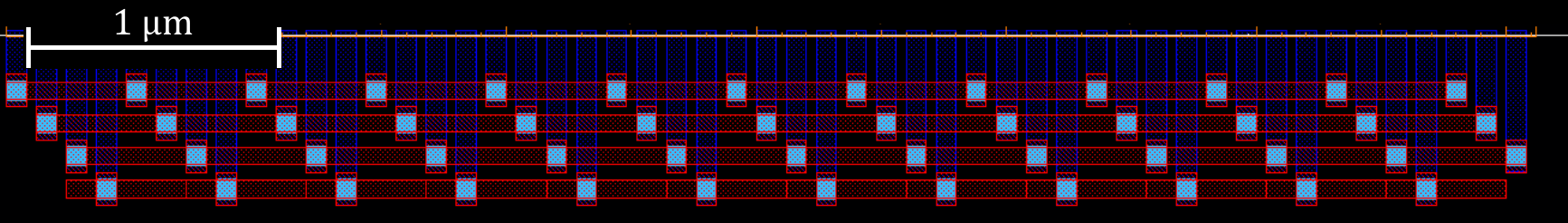}
    \caption{Shuttling arm of the spiderweb array ($>$6 \textmu m) consisting of concatenated sets of four-gate electrodes residing in the bottom metal layer (blue). Every fifth gate electrode is connected together via the second metal layer (red). Vias (cyan) connect the two metal layers and are arranged in a staggered configuration to enable interconnect routing. All gate electrodes carry a sinusoidal signal with equal frequency and amplitude, with a phase-shift of $\frac{\pi}{2}$ between neighboring gates in order to create a traveling-wave potential.}
    \label{fig:supp_cadence_1}
\end{figure*}

\begin{figure*}
    \centering
    \includegraphics[height=0.35\textwidth]{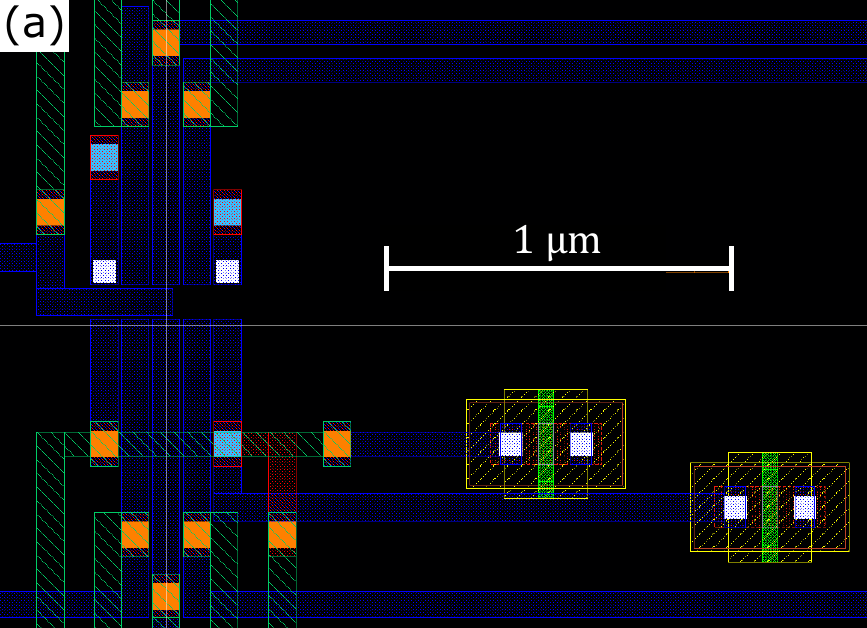}
    \includegraphics[height=0.35\textwidth]{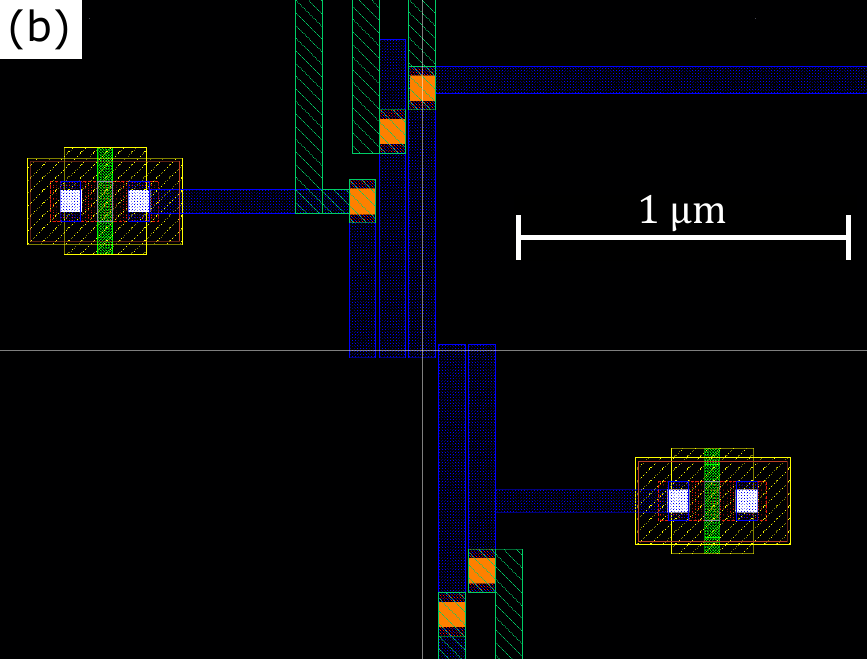}
    \caption{The two types of qubit operation regions, showing gate configurations for (a) single- and two-qubit operations, as well as readout, and (b) two-qubit operations only. The rulers indicate a length of 1 \textmu m.}
    \label{fig:supp_cadence_2}
\end{figure*}

The switches can be implemented as a simple transistor switch (NMOS or PMOS) or by a pass gate (NMOS and PMOS in parallel) depending on the voltage levels expected by the gates, the demultiplexer supply and the threshold voltage of the transistors in the employed technology. 
Due to the increased threshold voltage at cryogenic temperatures with respect to room temperature, a dead zone for voltage biasing around mid supply may appear, in which even the pass-gate impedance could be too high to allow proper biasing of the gates in the unit cell~\cite{vanDijk2020a}.
Possible alternatives could then be the well-known bootstrapped switches or the use for thick-oxide transistors driven by level shifters (or eventually driven by a thick-oxide decoder).
Noting that the above-mentioned design choices strongly depend on the required DC biasing levels, we assume $\sim 1$~V outputs and design a standard thin-oxide decoder in a nanometer CMOS technology driving single-transistor  switches.
Independent of the design choice, we expect the effective footprint and energy to be well within an order of magnitude of the reported estimates.

A 4-bit decoder enabled by the combination of a row and column address (see \fref{fig:dc_biasing_1}) has been designed in \textsc{verilog} and synthesized using a commercial TSMC 40-nm CMOS process. After place-and-route, the decoder occupies an area of 36~$\mu\textrm{m}^2$. By budgeting 25\% extra area for the switch array, each demultiplexer would occupy 45~$\mu\textrm{m}^2$ in 40-nm CMOS, and the total footprint required to fit 4 demultiplexers in a unit cell will be $A_d = 180$~$\mu\textrm{m}^2$.

To estimate the power dissipation, the decoder binary input has been swept from 0000 to 1111, leading to an energy dissipation between 0.2~pJ and 0.35~pJ from a 1-V supply for the whole 16-phase cycle for a decoder load ranging from 1~fF to 10~fF to emulate the switch input capacitance.
A unit cell containing 4 demultiplexers will then dissipate a maximum energy of 1.4~pJ.
This estimate uses the standard room-temperature device models and includes the parasitics extracted from the layout but excludes the power required to drive the demultiplexer inputs.

\section{\textsc{Cadence} drawings} \label{app:cadence}
In order assess more concretely the feasibility of implementing the local classical control electronics described in the spiderweb array, we have drawn a unit cell with all its elements using \textsc{cadence} circuit design software.
Although we have used realistic gate pitch and width dimensions (80~nm) for the densest part of the array, the drawings do not strictly follow design rules and are not optimized, as they are intended as an illustrative feasibility exercise.
Additionally, we have not implemented the connection that will be required for readout (these will depend on the exact implementation) or the lines connecting the dcDACs outside of the quantum plane to the local demultiplexers.
Footprint is reserved for the local demultiplexers themselves based on the design described above, but they are not implemented explicitly.
We have not drawn the two-dimensional electron gas (2DEG) channels that run underneath the end of the gate electrodes, which are electrostatically depleted to create quantum dots.

For all the drawings presented we use the following color convention: metal layers from lowest to highest are blue, red, green and pink, while vias connecting different layers are represented by squares colored cyan, orange and purple, connecting to the second, third and fourth layers, respectively.
White areas represent doped regions.

\begin{figure*}
    \centering
    \includegraphics[width=\textwidth]{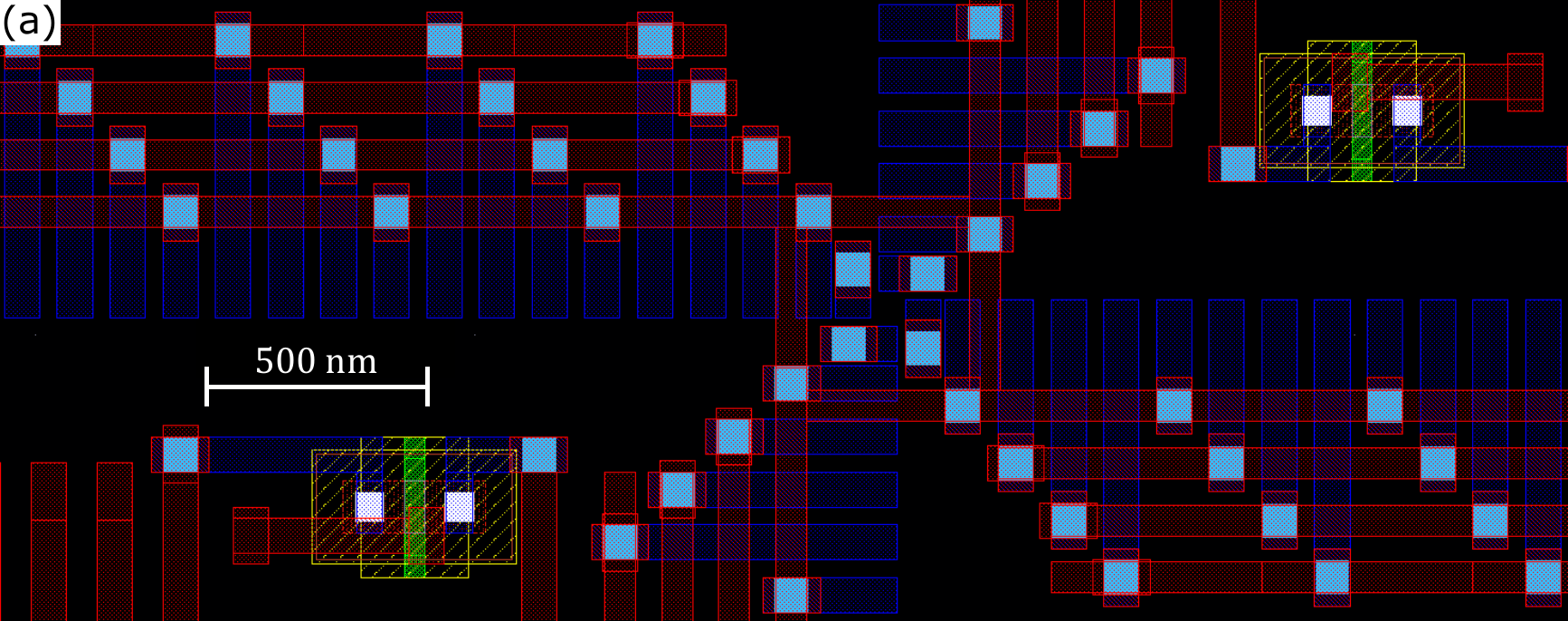} \\
    \includegraphics[width=\textwidth]{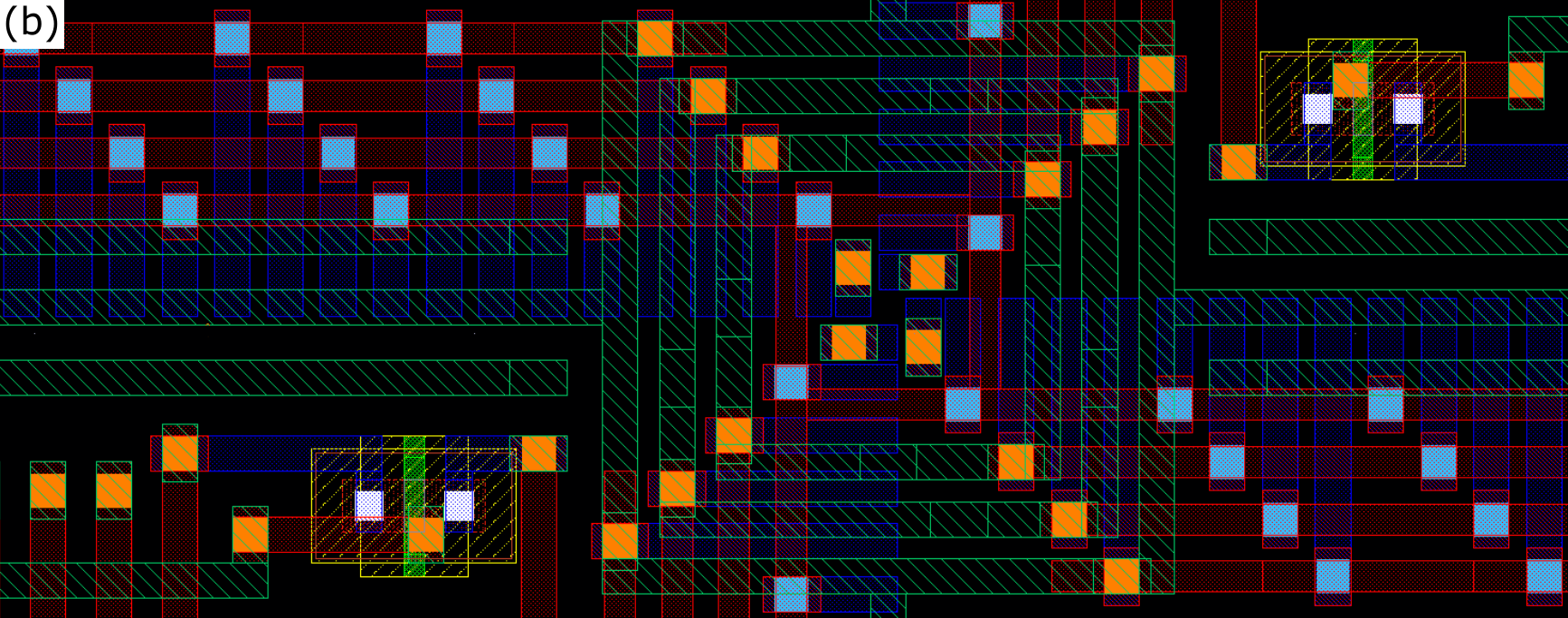} \\
    \includegraphics[width=\textwidth]{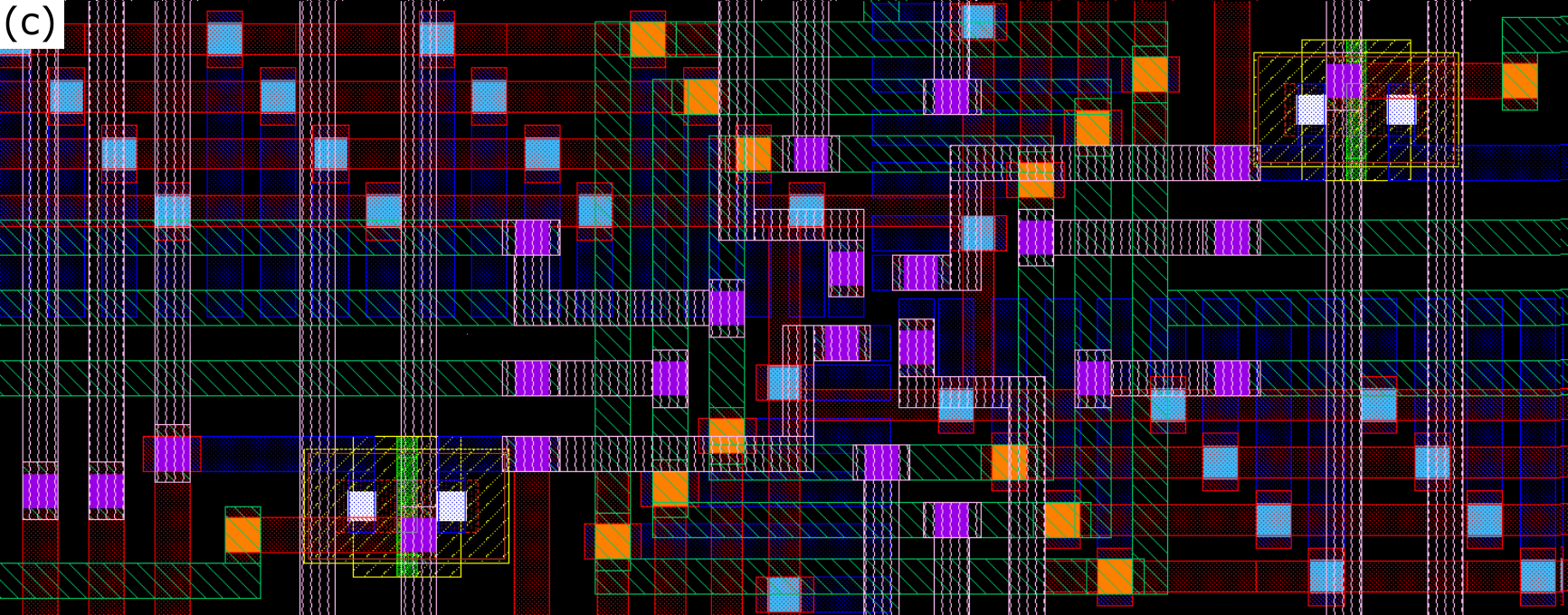}
    \caption{The qubit idling region, showing (a) the first two, (b) three and (c) four metal layers.}
    \label{fig:supp_cadence_3}
\end{figure*}

\begin{figure*}
    \centering
    \includegraphics[height=0.39\textwidth]{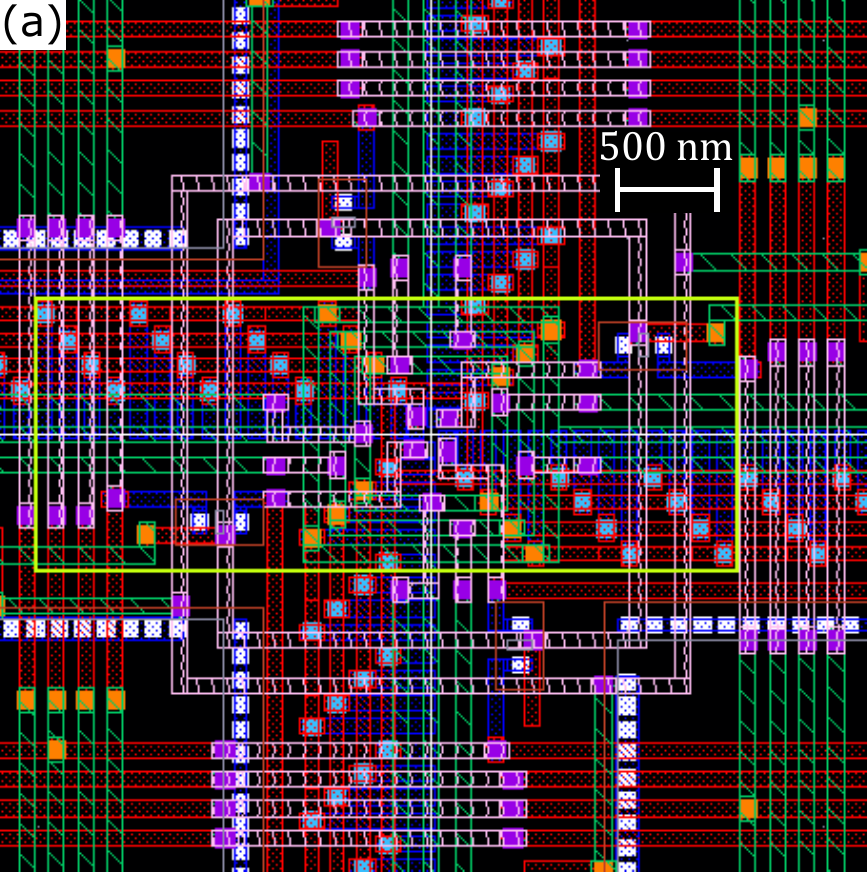}
    \includegraphics[height=0.39\textwidth]{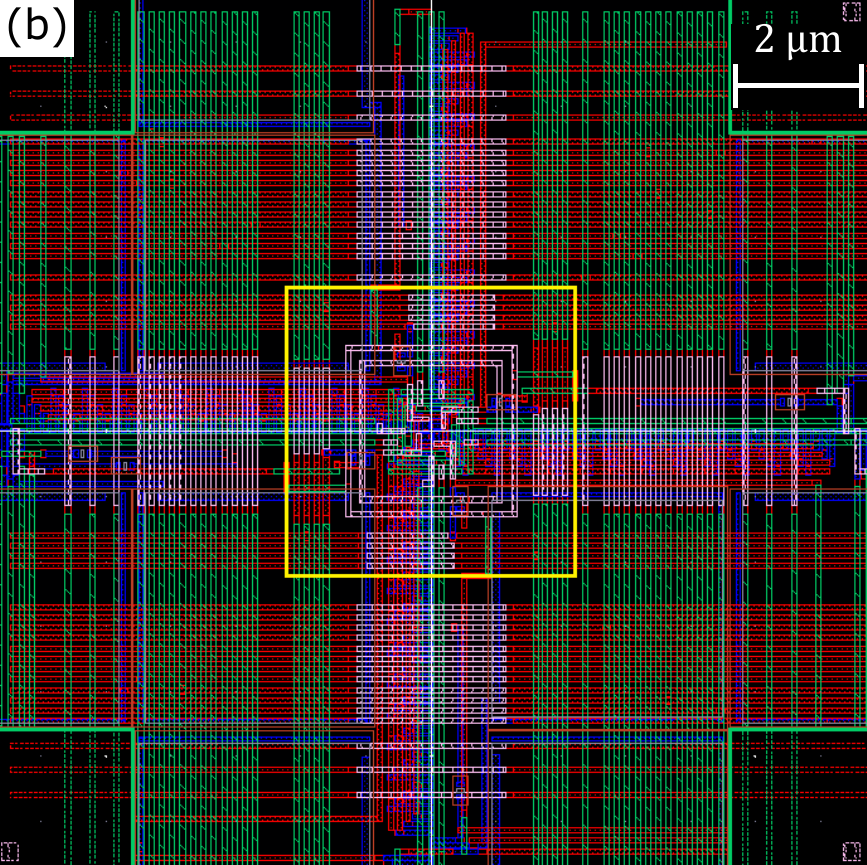}
    \caption{Zoomed out images of the qubit idle region. The light green rectangle in (a) corresponds to the area shown in \fref{fig:supp_cadence_3}. The yellow square in (b) has a dimension of $\sim$4.4 \textmu m and indicates the area shown in (a).}
    \label{fig:supp_cadence_4}
\end{figure*}

\begin{figure*}
    \centering
    \includegraphics[height=0.78\textwidth]{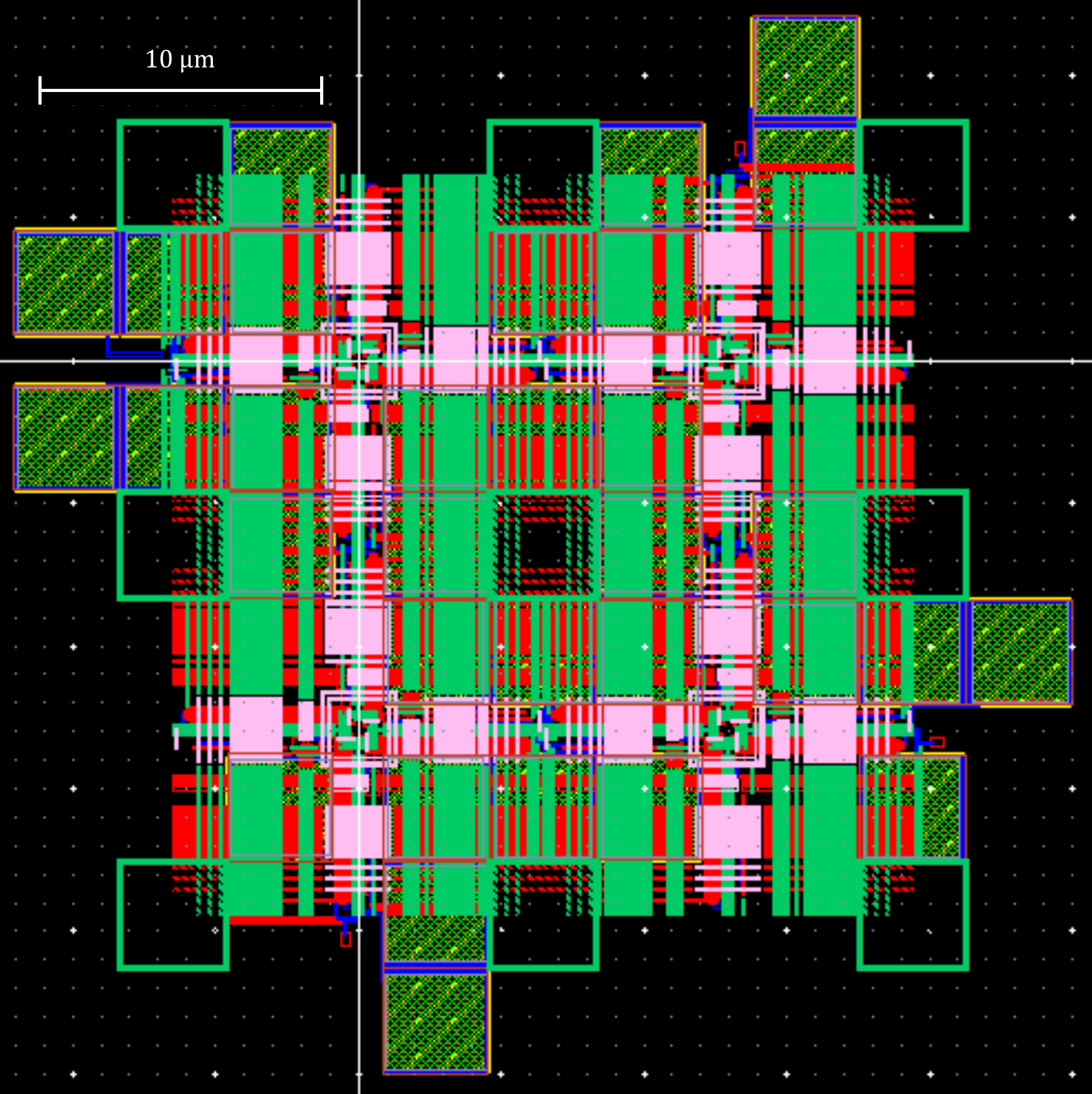}
    \caption{Full unit cell of the spiderweb array. The distance between two qubit idle regions is $\sim$13 \textmu m.}
    \label{fig:supp_cadence_5}
\end{figure*}

\fref{fig:supp_cadence_1} shows the drawing of a $\sim$6.5~\textmu m shuttling arm.
Identical copies of these shuttling arms link every qubit idling and operation regions to form the spiderweb array.
The 2DEG channel runs horizontally in the top of the image below the tips of the blue gate electrodes.

\fref{fig:supp_cadence_2} show the two types of operation regions, as shown schematically in \fref{fig:unit_cell_schematics} of the main text. In these images, the 2DEG channels run below the tips of the blue gate electrodes, along the horizontal white line in the center.
To combine DC biasing signals with AC pulses, the quantum dot gates  are connected to DC biasing capacitors in a sample-and-hold scheme, as explained in detail in the main text. Coarse-resolution capacitors are visible in \fref{fig:supp_cadence_2} (yellowish structures). For simplicity, we have represented these capacitors here by parallel plate capacitors, although we envision to use more advanced capacitor types that have a higher capacitance per unit area. The capacitor footprint used in these drawings (and not the parallel plate capacitance that it represents) is in agreement with our estimates in the main text.
The gate electrodes for which their capacitor is not seen in the image, connect to larger, fine-resolution capacitors that are not shown here, but are visible (as larger greenish hatched squares) in \fref{fig:supp_cadence_5}.
The green wires from the third metal layer carry the AC signals and are routed to the corresponding gates in neighboring unit cells.
In the top part of \fref{fig:supp_cadence_2}(a) show two doped square regions in white that serve as source and drain reservoirs for the sensing dot, as well as electron reservoirs for initialization of the spiderweb array.
The other doped regions in \fref{fig:supp_cadence_2} are part of the capacitor structures.
Connections to the boundary of the quantum plane for any of these doped regions are not implemented.
In the bottom part of \fref{fig:supp_cadence_2}(a) a short connection via the second metal layer (red) is required to by-pass another metal line before connecting to the third metal layer (green).

The qubit idle region is surrounded by four shuttling arms, as shown in \fref{fig:supp_cadence_3}.
The four gates directly adjacent to the qubit idle region act as gate keepers that control the transfer of electrons into the shuttling arms.
They are activated through a transistor structure arranged in an AND configuration.
Two of these transistors are visible in the top right and bottom left of the images, while the remaining two transistors are outside the field of view.
The fourth metal layer (pink) connects the transistors to the gate electrodes, as is visible in \fref{fig:supp_cadence_3}(c).
The second ring of gate electrodes around the qubit idling region (i.e., the first set of shuttling gates) connect to a line in the second metal layer (red), visible in \fref{fig:supp_cadence_3}(a).
The subsequent three rings of gate electrodes that make up the rest of the shuttling set connect to three lines from the third metal layer (green), visible in \fref{fig:supp_cadence_3}(b).
The horizontal green wires that are visible in \fref{fig:supp_cadence_3}(b) connect the shuttling signals to the next qubit idle regions, and the vertical pink wires in \fref{fig:supp_cadence_3}(c) connect the AC signals in adjacent unit cells together.
The idle regions for ancilla and data qubits are very similar, with minor differences to allow for the crossbar addressing to prevent data qubits from being shuttled in the surface code operation. The images shown here correspond to the ancilla qubits.

\fref{fig:supp_cadence_4} progressively shows larger areas of the \textsc{cadence} drawing, leading to the drawing of the full unit cell in \fref{fig:supp_cadence_5}. 
All layers are included in both figures.
Both panels of \fref{fig:supp_cadence_4} are centered on the qubit idling region.
The horizontal (red/pink) and vertical (green/pink) wires connect the AC signals in adjacent unit cells together.
The open areas visible in the corners of \fref{fig:supp_cadence_4}(b) and as nine green open squares in \fref{fig:supp_cadence_5}, are reserved footprint for the local demultiplexers used for DC biasing.
The greenish hatched squares on the perimeter of the full unit cell image are fine-resolution capacitors for the sample-and-hold scheme.

We have drawn the unit cell, with a qubit pitch of $\sim$13 \textmu m, in such a way that it can be tiled horizontally and vertically to form modules, with lines ending at each edge of the unit cell aligned to connect to their respective counterparts at the opposite edge of the adjacent unit cells, and the ancillary structures (such as the fine-resolution capacitors) neatly complement each other.

\section{Heat dissipation estimates} \label{app:heat}
\subsection{Parasitic capacitance}
\begin{figure}
    \centering
    \includegraphics[width=\linewidth]{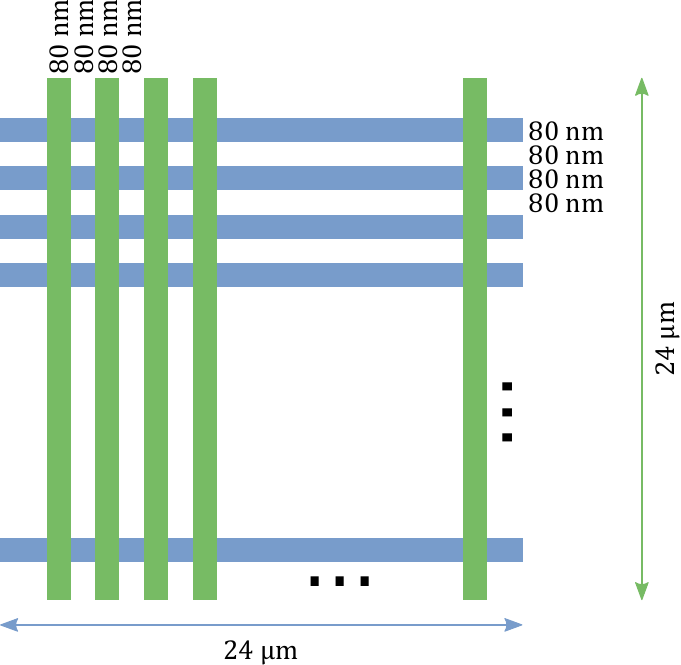}
    \caption{Schematic of signal lines in a 24 $\mu$m by 24 $\mu$m unit cell. Line width and lateral distance in between are both 80 nm. The first layer (blue) and second layer (green) are separated by presumably 500 nm SiO$_2$ (not shown in this figure of top view).}
    \label{fig:Interconnect}
\end{figure}

To assess the heat dissipation of the signal lines, we consider the dominant contribution from the parasitic capacitance caused by having metallic lines next to each other, as well as overlapping across layers.
Here we will estimate this capacitance based on simplified models, noting that a more accurate estimate can be obtained using capacitance simulation software.
We consider a grid of metallic lines such as that shown schematically in \fref{fig:Interconnect}, consisting of a set of parallel lines in one layer, overlapped by another set of orthogonally placed lines placed on a layer above.
The two main contributors to the total parasitic capacitance will be the capacitance between parallel lines in the same layer $C_1$ and the capacitance caused by overlapping metal regions across two layers $C_2$.
The total parasitic capacitance of a metallic grid with $N_l$ lines per layer is then $C_p = 2 N_l C_1 + N_l^2 C_2$.
We consider both the parallel plate capacitances and the fringe capacitance between metal edges, to obtain a capacitance model~\cite{Paul2008,Wong1998}:
\begin{align*}
C_1 &= \epsilon L \left\{ \frac{H}{d_1} + \left[ 377 \pi v_0 \ln \left( -2 \frac{\sqrt{\alpha_1+1}}{\sqrt{\alpha_1-1}} \right) \right]^{-1} \right\} \\
C_2 &= \epsilon W \left( 3.285 \frac{W}{d_2} + 9.01 \alpha_2 - 8.696 \alpha_2^2 \right)\\
\textrm{with } &\alpha_1 = \frac{d_1}{d_1+2W} \textrm{ and } \alpha_2 = \frac {H}{H+0.2 d_2} , 
\end{align*}
where $L$/$W$/$H$ is the length/width/height of a metal line, $d_1$ is the distance between neighboring metal lines in the same layer, $d_2$ is the thickness of the dielectric separating two layers, $\epsilon$ is the permittivity of the dielectric between metals and $v_0$ is the speed of light in vacuum.

To estimate the $C_p$ of a unit cell, we consider the bottom two layers as the main contributors, since they contain the highest density of signal lines.
Metal layers higher up would have signal lines separated widely, thus bringing in less parasitic capacitance.
Vias that make connections between layers introduce a parasitic capacitance as well, while it would be reduced through proper layout design that avoids placing vias close together.
Using the drawings in \aref{app:cadence} as a guide, we assume $N_l = 150$, $L = 24$~\textmu m, $W = 80$~nm, $H = 50$~nm, $d1 = 80$~nm and $d_2 = 500$~nm. 
The dielectric is assumed to be SiO$_2$ making $\epsilon = 3.9 \epsilon_0$, where $\epsilon_0$ is the vacuum permittivity. 
From this we obtain the estimate $C_p \approx 700$~fF, used to calculate this portion of the power dissipation in the main text.

\subsection{AC transmission lines}
To estimate the power dissipated by lossy transmission lines, we model a segment of transmission line running above a unit cell, as a resistor $R_t$ in series with a capacitor $C_t$ to ground. The amplitude of the current drawn by the capacitor while sustaining a signal of amplitude $v_t$ and frequency $f_t$ is $i_c = 2 \pi v_t f_t C_t$.
This current flowing through the finite resistance will cause a power dissipation $P_t = i_c^2 R_t / 2 = 2 R_t \left( \pi v_t f_t C_t \right)^2$.
Assuming transmission lines with capacitance to ground $\sim 0.2$~fF/$\mu$m and sheet resistance $\sim 0.1$~$\Omega$/$\square$\cite{Razavi2016} with width $\sim 1$~$\mu$m, we obtain the expression in \eref{eq:power_tl} used in the calculation in the main text.
This is a conservative estimate for our application, since the resistance of the lines is expected to reduce at cryogenic temperatures~\cite{Patra2020}.

\section{Square-root-SWAP as two-qubit gate in the surface code for spin qubits} \label{app:sc}
A scalable scheme for implementing error correction cycles in a surface code for
superconducting qubits has recently been presented~\cite{Versluis2017}, where an effective \textsc{cnot} gate is applied by combining a conditional phase gate with appropriate single-qubit Hadamard ($\boldsymbol{H}$) operations.
Here we implement a similar scheme, adapted to efficiently accommodate the square-root-\textsc{swap} ($\sqrt{S_w}$) operation, i.e., the native two-qubit gate of the spiderweb array, defined as
\begin{equation*}
\begin{split}
    \sqrt{S_w} &= \frac{1}{2} \begin{pmatrix}
        2 & 0 & 0 & 0 \\
        0 & 1+i & 1-i & 0 \\
        0 & 1-i & 1+i & 0 \\
        0 & 0 & 0 & 2 
      \end{pmatrix}\\
    &= \frac{\sqrt{2}}{2} \begin{pmatrix}
        \sqrt{2} & 0 & 0 & 0 \\
        0 & e^{i\frac{\pi}{4}} & e^{-i\frac{\pi}{4}} & 0 \\ 
        0 & e^{-i\frac{\pi}{4}} & e^{i\frac{\pi}{4}} & 0 \\ 
        0 & 0 & 0 & \sqrt{2} 
      \end{pmatrix}.
\end{split}
\end{equation*}
We propose $\sqrt{S_w}$ modifications of the $X$- and $Z$-plaquettes 
which are the building blocks of larger error-correction schemes.

\subsection{Entangling phase gate}
\begin{figure}
    \centering
    \includegraphics[width=\linewidth]{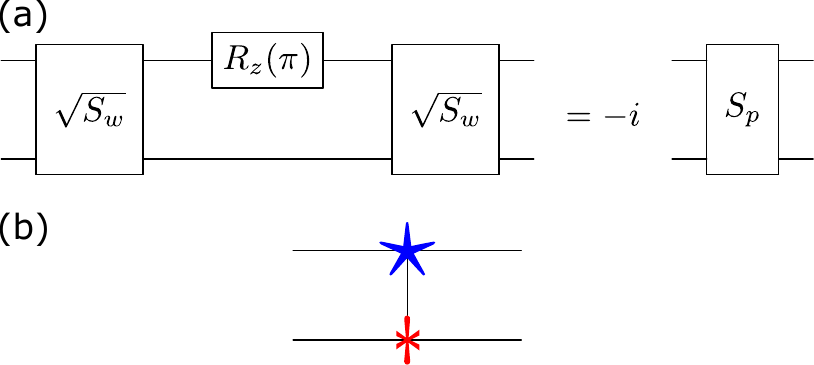}
    \caption{The $S_p$ gate. (a) Decomposition in quantum circuit diagram. (b) Simplified circuit diagram representation.}\label{fig:Sp}
\end{figure}

An explicit calculation of an operator product proposed in~\cite{Loss1998,Schuch2003} yields
\begin{equation}
    \sqrt{S_w} \left(R_z^{[1]}(\pi) \otimes I^{[2]}\right) \sqrt{S_w} = -i S_p,
    \label{eq:S_p}
\end{equation}
where we have defined the phase gate as
\begin{equation*}
    S_p = 
     \begin{pmatrix}
        1 & 0 & 0 & 0 \\
        0 & i & 0 & 0 \\
        0 & 0 & -i &  0  \\
        0 & 0 & 0 & -1 
      \end{pmatrix}.
\end{equation*}
This relation is pictorially shown in \fref{fig:Sp}a and in what follows the $S_p$ gate will be depicted as in \fref{fig:Sp}b.

Note that $S_p$ is an entangling gate, in the sense that it can map a product state into a an entangled state. This is most easily verified by using the concurrence of a two-qubit state.
Its Hermitian conjugate $S_p^\dagger$ can be written as $-i S_p^\dagger = \sqrt{S_w} \left( I^{[1]} \otimes R_z^{[2]}(\pi) \right) \sqrt{S_w}$.

The controlled-phase gate $C_Z$ may be used as primitive data-ancilla interaction in
a surface code~\cite{Versluis2017}. 
With only two single-qubit rotations, this $C_Z$ gate is constructed from $S_p$ as
\begin{equation}
\begin{split}
C_Z&=
 \left(R_z^{[1]}(\tfrac{\pi}{2}) \otimes I^{[2]}\right)
\left( I^{[1]} \otimes R_z^{[2]}(-\tfrac{\pi}{2})\right) S_p\\
&=
 \left(R_z^{[1]}(\tfrac{\pi}{2}) \otimes R_z^{[2]}(-\tfrac{\pi}{2})\right) S_p.
\label{eq:CZ-S}
\end{split}
\end{equation}
Since the three matrices involved are diagonal, they commute and the order of applying the 
operations is immaterial. This can be advantageous in actually implementing
error-correction cycles. To this end, the following relation with the
Hermitian conjugate is convenient as well 
\begin{equation}
\begin{split}
C_Z&=
 \left(R_z^{[1]}(-\tfrac{\pi}{2}) \otimes I^{[2]}\right)
\left( I^{[1]} \otimes R_z^{[2]}(\tfrac{\pi}{2})\right) S_p^\dagger\\
&=
 \left(R_z^{[1]}(-\tfrac{\pi}{2}) \otimes R_z^{[2]}(\tfrac{\pi}{2})\right) S_p^\dagger.
\label{eq:CZ-S2}
\end{split}
\end{equation}

If one alternatively desires the \textsc{cnot} gate as primitive data-ancilla interaction~\cite{Versluis2017}, two additional Hadamard gates are necessary~\cite{Nielsen2016}.
The result can be written as
\begin{equation}
\begin{split}
    C_N=
    & i \left(R_z^{[1]}(\tfrac{\pi}{2}) \otimes I^{[2]}\right)
    \left( I^{[1]} \otimes R_z^{[2]}(\tfrac{\pi}{2})\right)\\
    &\times \left( I^{[1]} \otimes R_x^{[2]}(\tfrac{\pi}{2})\right)
    S_p
    \left( I^{[1]} \otimes \boldsymbol{H}^{[2]}\right).\\
    \label{eq:CN-Sp}
\end{split}
\end{equation}
Here, non-commuting operators have to be applied in the right order. A similar relation involving $S_p^\dagger$ can be readily be derived. 

\subsection{Surface code cycles}
We adapt the pipeline-based error-correction scheme~\cite{Versluis2017} to use it with the $S_p$ operator described above.
To achieve this, we are only required to re-analyze the $X$-type and $Z$-type plaquettes using the $C_Z$ ancilla-data interaction.

\begin{figure}
\centering
\includegraphics[width=\linewidth]{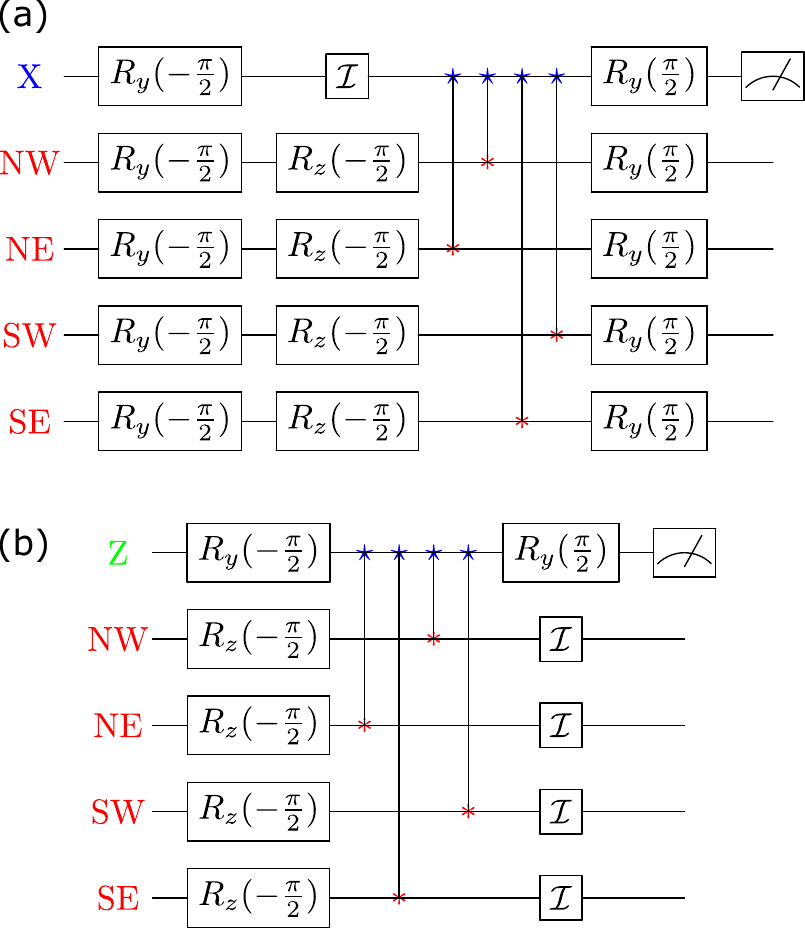}
\caption{Quantum circuits for surface code cycles. (a) $X$-plaquette. (b) $Z$-plaquette.}\label{fig:circuits}
\end{figure}

For the $X$-type plaquette, we insert 
$\boldsymbol{H} = R_y(\tfrac{\pi}{2}) Z = Z R_y(-\tfrac{\pi}{2})$
for all Hadamard operations.
The $C_Z$ gates are to be implemented as \eref{eq:CZ-S}; here the actual realization of $S_p$ as \eref{eq:S_p} is implicitly assumed but irrelevant for the remaining analysis.
Because of the earlier mentioned commutativity of the diagonal operators, the circuit can readily be simplified.
In particular, the $Z$-operators stemming form the initial and final Hadamard gates combine to the identity. The $Z$-rotations can be combined as well.
All final Hadamard gates are replaced by $R_y(\tfrac{\pi}{2})$.
The $C_Z$ ancilla-data operations are replaced by $S_p$ gates without changing the time-ordering.
The initial Hadamard operation on the ancilla is replaced by 
$[R_z(\tfrac{\pi}{2})]^{4}R_y(-\tfrac{\pi}{2})= -R_y(-\tfrac{\pi}{2}),$
where the minus sign can be omitted.
The initial Hadamard data-qubit gates are replaced by the product $R_z(-\tfrac{\pi}{2}) R_y(-\tfrac{\pi}{2})$.
The resulting circuit diagram for the $X$-plaquette cycle is depicted in \fref{fig:circuits}a.

The $Z$-plaquette is modified analogously.
The final Hadamard gate on the ancilla is again replaced by $R_y(\tfrac{\pi}{2})$, whereas the initial Hadamard operation is replaced by $-R_y(-\tfrac{\pi}{2})$.
Once more, $C_Z$ ancilla-data operations are replaced by $S_p$ gates at the same instant of time.
For the data-qubits, rotations $R_z(-\tfrac{\pi}{2})$ have to be added at some instant of time.
In the circuit shown in \fref{fig:circuits}b, it is done at the onset.

These building blocks can be used to construct depth-nine quantum circuits
for parallel as well as pipeline-based, quantum-error-correction cycle of a surface code; the proposed scheme is scalable.

\begin{figure}
    \centering
    \includegraphics[width=\linewidth]{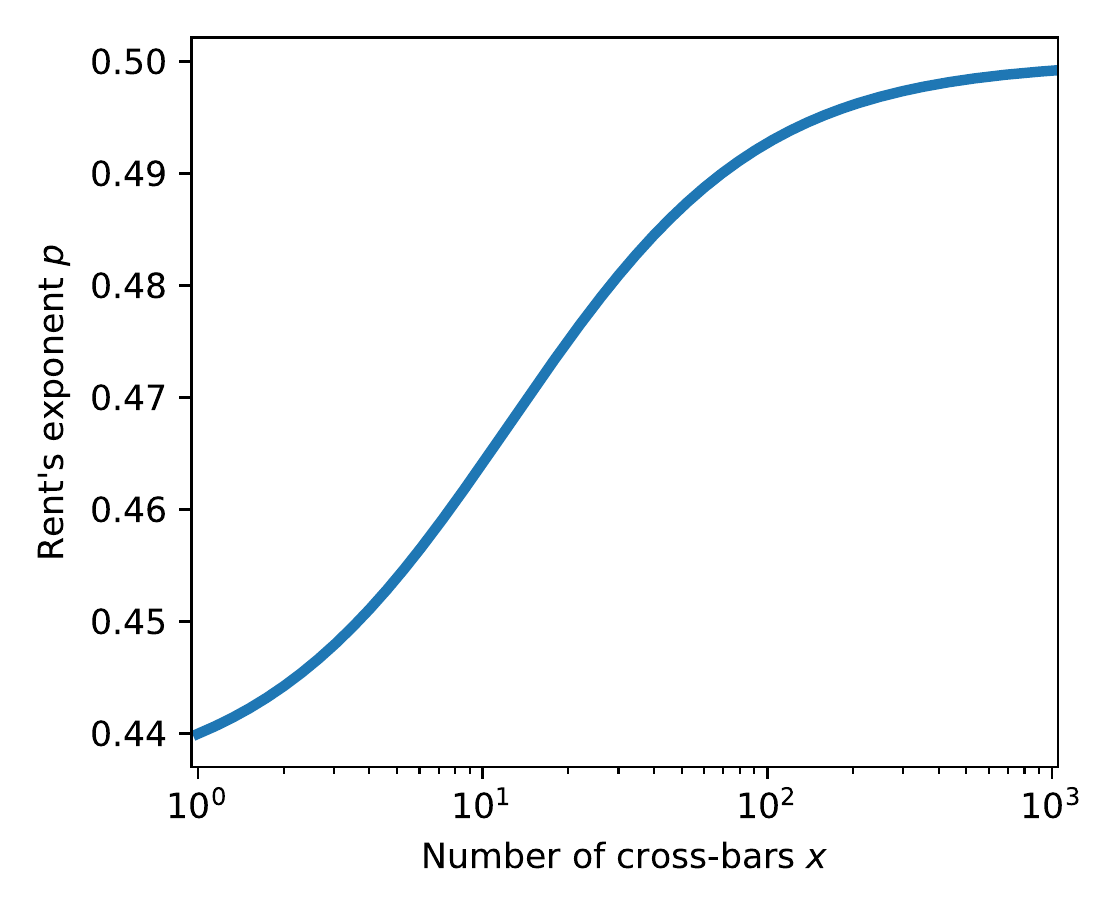}
    \caption{Rent's exponent as a function of the number of cross-bars used for logical operations in the spiderweb array. All other counts in parameters that affect Rent's exponent are as per the example in \sref{sec:example}.}
    \label{fig:rent}
\end{figure}

\section{Rent's exponent considerations}
\label{app:rent}
We calculate Rent's exponent using
\begin{equation*}
    p = \log_U \left( \frac{T}{c} \right) \textrm{ ,}
\end{equation*}
where, as per \tref{tab:wire_count}, $T = M_b^{2} + M_r^{2} + 4N_b(1+M_bx) + 2\log_2 N_r - \log_2 r + 66$ is the total number of connections on the quantum plane, $c = 74 + 4x$ is the number of connection in a unit cell and $U$ is the number of unit cells.

Using the numbers presented in the example in \sref{sec:example}, we obtain the quoted value $p = 0.43$ for an array without any cross bar circuits ($x = 0$, i.e. a single qubit memory). As we add cross-bar circuits for the implementation of logical operations, Rent's exponent will increase as per \fref{fig:rent} and saturate at a value of 0.5.

\bibliography{spiderweb}

\begin{thebibliography}{52}%
\makeatletter
\providecommand \@ifxundefined [1]{%
 \@ifx{#1\undefined}
}%
\providecommand \@ifnum [1]{%
 \ifnum #1\expandafter \@firstoftwo
 \else \expandafter \@secondoftwo
 \fi
}%
\providecommand \@ifx [1]{%
 \ifx #1\expandafter \@firstoftwo
 \else \expandafter \@secondoftwo
 \fi
}%
\providecommand \natexlab [1]{#1}%
\providecommand \enquote  [1]{``#1''}%
\providecommand \bibnamefont  [1]{#1}%
\providecommand \bibfnamefont [1]{#1}%
\providecommand \citenamefont [1]{#1}%
\providecommand \href@noop [0]{\@secondoftwo}%
\providecommand \href [0]{\begingroup \@sanitize@url \@href}%
\providecommand \@href[1]{\@@startlink{#1}\@@href}%
\providecommand \@@href[1]{\endgroup#1\@@endlink}%
\providecommand \@sanitize@url [0]{\catcode `\\12\catcode `\$12\catcode
  `\&12\catcode `\#12\catcode `\^12\catcode `\_12\catcode `\%12\relax}%
\providecommand \@@startlink[1]{}%
\providecommand \@@endlink[0]{}%
\providecommand \url  [0]{\begingroup\@sanitize@url \@url }%
\providecommand \@url [1]{\endgroup\@href {#1}{\urlprefix }}%
\providecommand \urlprefix  [0]{URL }%
\providecommand \Eprint [0]{\href }%
\providecommand \doibase [0]{https://doi.org/}%
\providecommand \selectlanguage [0]{\@gobble}%
\providecommand \bibinfo  [0]{\@secondoftwo}%
\providecommand \bibfield  [0]{\@secondoftwo}%
\providecommand \translation [1]{[#1]}%
\providecommand \BibitemOpen [0]{}%
\providecommand \bibitemStop [0]{}%
\providecommand \bibitemNoStop [0]{.\EOS\space}%
\providecommand \EOS [0]{\spacefactor3000\relax}%
\providecommand \BibitemShut  [1]{\csname bibitem#1\endcsname}%
\let\auto@bib@innerbib\@empty
\bibitem [{\citenamefont {Loss}\ and\ \citenamefont
  {DiVincenzo}(1998)}]{Loss1998}%
  \BibitemOpen
  \bibfield  {author} {\bibinfo {author} {\bibfnamefont {D.}~\bibnamefont
  {Loss}}\ and\ \bibinfo {author} {\bibfnamefont {D.~P.}\ \bibnamefont
  {DiVincenzo}},\ }\bibfield  {title} {\bibinfo {title} {{Quantum computation
  with quantum dots}},\ }\href {https://doi.org/10.1103/PhysRevA.57.120}
  {\bibfield  {journal} {\bibinfo  {journal} {Phys. Rev. A}\ }\textbf {\bibinfo
  {volume} {57}},\ \bibinfo {pages} {120} (\bibinfo {year} {1998})},\ \Eprint
  {https://arxiv.org/abs/cond-mat/9701055} {cond-mat/9701055} \BibitemShut
  {NoStop}%
\bibitem [{\citenamefont {Zwanenburg}\ \emph {et~al.}(2013)\citenamefont
  {Zwanenburg}, \citenamefont {Dzurak}, \citenamefont {Morello}, \citenamefont
  {Simmons}, \citenamefont {Hollenberg}, \citenamefont {Klimeck}, \citenamefont
  {Rogge}, \citenamefont {Coppersmith},\ and\ \citenamefont
  {Eriksson}}]{Zwanenburg2013}%
  \BibitemOpen
  \bibfield  {author} {\bibinfo {author} {\bibfnamefont {F.~A.}\ \bibnamefont
  {Zwanenburg}}, \bibinfo {author} {\bibfnamefont {A.~S.}\ \bibnamefont
  {Dzurak}}, \bibinfo {author} {\bibfnamefont {A.}~\bibnamefont {Morello}},
  \bibinfo {author} {\bibfnamefont {M.~Y.}\ \bibnamefont {Simmons}}, \bibinfo
  {author} {\bibfnamefont {L.~C.~L.}\ \bibnamefont {Hollenberg}}, \bibinfo
  {author} {\bibfnamefont {G.}~\bibnamefont {Klimeck}}, \bibinfo {author}
  {\bibfnamefont {S.}~\bibnamefont {Rogge}}, \bibinfo {author} {\bibfnamefont
  {S.~N.}\ \bibnamefont {Coppersmith}},\ and\ \bibinfo {author} {\bibfnamefont
  {M.~A.}\ \bibnamefont {Eriksson}},\ }\bibfield  {title} {\bibinfo {title}
  {{Silicon quantum electronics}},\ }\href
  {https://doi.org/10.1103/RevModPhys.85.961} {\bibfield  {journal} {\bibinfo
  {journal} {Rev. Mod. Phys.}\ }\textbf {\bibinfo {volume} {85}},\ \bibinfo
  {pages} {961} (\bibinfo {year} {2013})}\BibitemShut {NoStop}%
\bibitem [{\citenamefont {Veldhorst}\ \emph {et~al.}(2014)\citenamefont
  {Veldhorst}, \citenamefont {Hwang}, \citenamefont {Yang}, \citenamefont
  {Leenstra}, \citenamefont {de~Ronde}, \citenamefont {Dehollain},
  \citenamefont {Muhonen}, \citenamefont {Hudson}, \citenamefont {Itoh},
  \citenamefont {Morello},\ and\ \citenamefont {Dzurak}}]{Veldhorst2014}%
  \BibitemOpen
  \bibfield  {author} {\bibinfo {author} {\bibfnamefont {M.}~\bibnamefont
  {Veldhorst}}, \bibinfo {author} {\bibfnamefont {J.~C.~C.}\ \bibnamefont
  {Hwang}}, \bibinfo {author} {\bibfnamefont {C.~H.}\ \bibnamefont {Yang}},
  \bibinfo {author} {\bibfnamefont {A.~W.}\ \bibnamefont {Leenstra}}, \bibinfo
  {author} {\bibfnamefont {B.}~\bibnamefont {de~Ronde}}, \bibinfo {author}
  {\bibfnamefont {J.~P.}\ \bibnamefont {Dehollain}}, \bibinfo {author}
  {\bibfnamefont {J.~T.}\ \bibnamefont {Muhonen}}, \bibinfo {author}
  {\bibfnamefont {F.~E.}\ \bibnamefont {Hudson}}, \bibinfo {author}
  {\bibfnamefont {K.~M.}\ \bibnamefont {Itoh}}, \bibinfo {author}
  {\bibfnamefont {A.}~\bibnamefont {Morello}},\ and\ \bibinfo {author}
  {\bibfnamefont {A.~S.}\ \bibnamefont {Dzurak}},\ }\bibfield  {title}
  {\bibinfo {title} {{An addressable quantum dot qubit with fault-tolerant
  control fidelity}},\ }\href {https://doi.org/10.1038/nnano.2014.216}
  {\bibfield  {journal} {\bibinfo  {journal} {Nat. Nanotechnology}\ }\textbf
  {\bibinfo {volume} {9}},\ \bibinfo {pages} {981} (\bibinfo {year}
  {2014})}\BibitemShut {NoStop}%
\bibitem [{\citenamefont {Kawakami}\ \emph {et~al.}(2016)\citenamefont
  {Kawakami}, \citenamefont {Jullien}, \citenamefont {Scarlino}, \citenamefont
  {Ward}, \citenamefont {Savage}, \citenamefont {Lagally}, \citenamefont
  {Dobrovitski}, \citenamefont {Friesen}, \citenamefont {Coppersmith},
  \citenamefont {Eriksson},\ and\ \citenamefont {Vandersypen}}]{Kawakami2016}%
  \BibitemOpen
  \bibfield  {author} {\bibinfo {author} {\bibfnamefont {E.}~\bibnamefont
  {Kawakami}}, \bibinfo {author} {\bibfnamefont {T.}~\bibnamefont {Jullien}},
  \bibinfo {author} {\bibfnamefont {P.}~\bibnamefont {Scarlino}}, \bibinfo
  {author} {\bibfnamefont {D.~R.}\ \bibnamefont {Ward}}, \bibinfo {author}
  {\bibfnamefont {D.~E.}\ \bibnamefont {Savage}}, \bibinfo {author}
  {\bibfnamefont {M.~G.}\ \bibnamefont {Lagally}}, \bibinfo {author}
  {\bibfnamefont {V.~V.}\ \bibnamefont {Dobrovitski}}, \bibinfo {author}
  {\bibfnamefont {M.}~\bibnamefont {Friesen}}, \bibinfo {author} {\bibfnamefont
  {S.~N.}\ \bibnamefont {Coppersmith}}, \bibinfo {author} {\bibfnamefont
  {M.~A.}\ \bibnamefont {Eriksson}},\ and\ \bibinfo {author} {\bibfnamefont
  {L.~M.~K.}\ \bibnamefont {Vandersypen}},\ }\bibfield  {title} {\bibinfo
  {title} {{Gate fidelity and coherence of an electron spin in a Si/SiGe
  quantum dot with micromagnet}},\ }\href
  {https://doi.org/10.1073/pnas.1603251113} {\bibfield  {journal} {\bibinfo
  {journal} {PNAS}\ }\textbf {\bibinfo {volume} {113}},\ \bibinfo {pages}
  {11738} (\bibinfo {year} {2016})}\BibitemShut {NoStop}%
\bibitem [{\citenamefont {Yoneda}\ \emph {et~al.}(2018)\citenamefont {Yoneda},
  \citenamefont {Takeda}, \citenamefont {Otsuka}, \citenamefont {Nakajima},
  \citenamefont {Delbecq}, \citenamefont {Allison}, \citenamefont {Honda},
  \citenamefont {Kodera}, \citenamefont {Oda}, \citenamefont {Hoshi},
  \citenamefont {Usami}, \citenamefont {Itoh},\ and\ \citenamefont
  {Tarucha}}]{Yoneda2018}%
  \BibitemOpen
  \bibfield  {author} {\bibinfo {author} {\bibfnamefont {J.}~\bibnamefont
  {Yoneda}}, \bibinfo {author} {\bibfnamefont {K.}~\bibnamefont {Takeda}},
  \bibinfo {author} {\bibfnamefont {T.}~\bibnamefont {Otsuka}}, \bibinfo
  {author} {\bibfnamefont {T.}~\bibnamefont {Nakajima}}, \bibinfo {author}
  {\bibfnamefont {M.~R.}\ \bibnamefont {Delbecq}}, \bibinfo {author}
  {\bibfnamefont {G.}~\bibnamefont {Allison}}, \bibinfo {author} {\bibfnamefont
  {T.}~\bibnamefont {Honda}}, \bibinfo {author} {\bibfnamefont
  {T.}~\bibnamefont {Kodera}}, \bibinfo {author} {\bibfnamefont
  {S.}~\bibnamefont {Oda}}, \bibinfo {author} {\bibfnamefont {Y.}~\bibnamefont
  {Hoshi}}, \bibinfo {author} {\bibfnamefont {N.}~\bibnamefont {Usami}},
  \bibinfo {author} {\bibfnamefont {K.~M.}\ \bibnamefont {Itoh}},\ and\
  \bibinfo {author} {\bibfnamefont {S.}~\bibnamefont {Tarucha}},\ }\bibfield
  {title} {\bibinfo {title} {{A quantum-dot spin qubit with coherence limited
  by charge noise and fidelity higher than 99.9{\%}}},\ }\href
  {https://doi.org/10.1038/s41565-017-0014-x} {\bibfield  {journal} {\bibinfo
  {journal} {Nat. Nanotechnology}\ }\textbf {\bibinfo {volume} {13}},\ \bibinfo
  {pages} {102} (\bibinfo {year} {2018})}\BibitemShut {NoStop}%
\bibitem [{\citenamefont {Xue}\ \emph {et~al.}(2019)\citenamefont {Xue},
  \citenamefont {Watson}, \citenamefont {Helsen}, \citenamefont {Ward},
  \citenamefont {Savage}, \citenamefont {Lagally}, \citenamefont {Coppersmith},
  \citenamefont {Eriksson}, \citenamefont {Wehner},\ and\ \citenamefont
  {Vandersypen}}]{Xue2019}%
  \BibitemOpen
  \bibfield  {author} {\bibinfo {author} {\bibfnamefont {X.}~\bibnamefont
  {Xue}}, \bibinfo {author} {\bibfnamefont {T.~F.}\ \bibnamefont {Watson}},
  \bibinfo {author} {\bibfnamefont {J.}~\bibnamefont {Helsen}}, \bibinfo
  {author} {\bibfnamefont {D.~R.}\ \bibnamefont {Ward}}, \bibinfo {author}
  {\bibfnamefont {D.~E.}\ \bibnamefont {Savage}}, \bibinfo {author}
  {\bibfnamefont {M.~G.}\ \bibnamefont {Lagally}}, \bibinfo {author}
  {\bibfnamefont {S.~N.}\ \bibnamefont {Coppersmith}}, \bibinfo {author}
  {\bibfnamefont {M.~A.}\ \bibnamefont {Eriksson}}, \bibinfo {author}
  {\bibfnamefont {S.}~\bibnamefont {Wehner}},\ and\ \bibinfo {author}
  {\bibfnamefont {L.~M.~K.}\ \bibnamefont {Vandersypen}},\ }\bibfield  {title}
  {\bibinfo {title} {{Benchmarking Gate Fidelities in a Si/SiGe Two-Qubit
  Device}},\ }\href {https://doi.org/10.1103/PhysRevX.9.021011} {\bibfield
  {journal} {\bibinfo  {journal} {Phys. Rev. X}\ }\textbf {\bibinfo {volume}
  {9}},\ \bibinfo {pages} {021011} (\bibinfo {year} {2019})}\BibitemShut
  {NoStop}%
\bibitem [{\citenamefont {Huang}\ \emph {et~al.}(2019)\citenamefont {Huang},
  \citenamefont {Yang}, \citenamefont {Chan}, \citenamefont {Tanttu},
  \citenamefont {Hensen}, \citenamefont {Leon}, \citenamefont {Fogarty},
  \citenamefont {Hwang}, \citenamefont {Hudson}, \citenamefont {Itoh},
  \citenamefont {Morello}, \citenamefont {Laucht},\ and\ \citenamefont
  {Dzurak}}]{Huang2019}%
  \BibitemOpen
  \bibfield  {author} {\bibinfo {author} {\bibfnamefont {W.}~\bibnamefont
  {Huang}}, \bibinfo {author} {\bibfnamefont {C.~H.}\ \bibnamefont {Yang}},
  \bibinfo {author} {\bibfnamefont {K.~W.}\ \bibnamefont {Chan}}, \bibinfo
  {author} {\bibfnamefont {T.}~\bibnamefont {Tanttu}}, \bibinfo {author}
  {\bibfnamefont {B.}~\bibnamefont {Hensen}}, \bibinfo {author} {\bibfnamefont
  {R.~C.~C.}\ \bibnamefont {Leon}}, \bibinfo {author} {\bibfnamefont {M.~A.}\
  \bibnamefont {Fogarty}}, \bibinfo {author} {\bibfnamefont {J.~C.~C.}\
  \bibnamefont {Hwang}}, \bibinfo {author} {\bibfnamefont {F.~E.}\ \bibnamefont
  {Hudson}}, \bibinfo {author} {\bibfnamefont {K.~M.}\ \bibnamefont {Itoh}},
  \bibinfo {author} {\bibfnamefont {A.}~\bibnamefont {Morello}}, \bibinfo
  {author} {\bibfnamefont {A.}~\bibnamefont {Laucht}},\ and\ \bibinfo {author}
  {\bibfnamefont {A.~S.}\ \bibnamefont {Dzurak}},\ }\bibfield  {title}
  {\bibinfo {title} {{Fidelity benchmarks for two-qubit gates in silicon}},\
  }\href {https://doi.org/10.1038/s41586-019-1197-0} {\bibfield  {journal}
  {\bibinfo  {journal} {Nature}\ }\textbf {\bibinfo {volume} {569}},\ \bibinfo
  {pages} {532} (\bibinfo {year} {2019})}\BibitemShut {NoStop}%
\bibitem [{\citenamefont {Watson}\ \emph {et~al.}(2018)\citenamefont {Watson},
  \citenamefont {Philips}, \citenamefont {Kawakami}, \citenamefont {Ward},
  \citenamefont {Scarlino}, \citenamefont {Veldhorst}, \citenamefont {Savage},
  \citenamefont {Lagally}, \citenamefont {Friesen}, \citenamefont
  {Coppersmith}, \citenamefont {Eriksson},\ and\ \citenamefont
  {Vandersypen}}]{Watson2018}%
  \BibitemOpen
  \bibfield  {author} {\bibinfo {author} {\bibfnamefont {T.~F.}\ \bibnamefont
  {Watson}}, \bibinfo {author} {\bibfnamefont {S.~G.~J.}\ \bibnamefont
  {Philips}}, \bibinfo {author} {\bibfnamefont {E.}~\bibnamefont {Kawakami}},
  \bibinfo {author} {\bibfnamefont {D.~R.}\ \bibnamefont {Ward}}, \bibinfo
  {author} {\bibfnamefont {P.}~\bibnamefont {Scarlino}}, \bibinfo {author}
  {\bibfnamefont {M.}~\bibnamefont {Veldhorst}}, \bibinfo {author}
  {\bibfnamefont {D.~E.}\ \bibnamefont {Savage}}, \bibinfo {author}
  {\bibfnamefont {M.~G.}\ \bibnamefont {Lagally}}, \bibinfo {author}
  {\bibfnamefont {M.}~\bibnamefont {Friesen}}, \bibinfo {author} {\bibfnamefont
  {S.~N.}\ \bibnamefont {Coppersmith}}, \bibinfo {author} {\bibfnamefont
  {M.~A.}\ \bibnamefont {Eriksson}},\ and\ \bibinfo {author} {\bibfnamefont
  {L.~M.~K.}\ \bibnamefont {Vandersypen}},\ }\bibfield  {title} {\bibinfo
  {title} {{A programmable two-qubit quantum processor in silicon}},\ }\href
  {https://doi.org/10.1038/nature25766} {\bibfield  {journal} {\bibinfo
  {journal} {Nature}\ }\textbf {\bibinfo {volume} {555}},\ \bibinfo {pages}
  {633} (\bibinfo {year} {2018})}\BibitemShut {NoStop}%
\bibitem [{\citenamefont {Yoneda}\ \emph {et~al.}(2020)\citenamefont {Yoneda},
  \citenamefont {Takeda}, \citenamefont {Noiri}, \citenamefont {Nakajima},
  \citenamefont {Li}, \citenamefont {Kamioka}, \citenamefont {Kodera},\ and\
  \citenamefont {Tarucha}}]{Yoneda2020}%
  \BibitemOpen
  \bibfield  {author} {\bibinfo {author} {\bibfnamefont {J.}~\bibnamefont
  {Yoneda}}, \bibinfo {author} {\bibfnamefont {K.}~\bibnamefont {Takeda}},
  \bibinfo {author} {\bibfnamefont {A.}~\bibnamefont {Noiri}}, \bibinfo
  {author} {\bibfnamefont {T.}~\bibnamefont {Nakajima}}, \bibinfo {author}
  {\bibfnamefont {S.}~\bibnamefont {Li}}, \bibinfo {author} {\bibfnamefont
  {J.}~\bibnamefont {Kamioka}}, \bibinfo {author} {\bibfnamefont
  {T.}~\bibnamefont {Kodera}},\ and\ \bibinfo {author} {\bibfnamefont
  {S.}~\bibnamefont {Tarucha}},\ }\bibfield  {title} {\bibinfo {title}
  {{Quantum non-demolition readout of an electron spin in silicon}},\ }\href
  {https://doi.org/10.1038/s41467-020-14818-8} {\bibfield  {journal} {\bibinfo
  {journal} {Nat. Commun.}\ }\textbf {\bibinfo {volume} {11}},\ \bibinfo
  {pages} {1144} (\bibinfo {year} {2020})}\BibitemShut {NoStop}%
\bibitem [{\citenamefont {Xue}\ \emph {et~al.}(2020)\citenamefont {Xue},
  \citenamefont {D'Anjou}, \citenamefont {Watson}, \citenamefont {Ward},
  \citenamefont {Savage}, \citenamefont {Lagally}, \citenamefont {Friesen},
  \citenamefont {Coppersmith}, \citenamefont {Eriksson}, \citenamefont
  {Coish},\ and\ \citenamefont {Vandersypen}}]{Xue2020}%
  \BibitemOpen
  \bibfield  {author} {\bibinfo {author} {\bibfnamefont {X.}~\bibnamefont
  {Xue}}, \bibinfo {author} {\bibfnamefont {B.}~\bibnamefont {D'Anjou}},
  \bibinfo {author} {\bibfnamefont {T.~F.}\ \bibnamefont {Watson}}, \bibinfo
  {author} {\bibfnamefont {D.~R.}\ \bibnamefont {Ward}}, \bibinfo {author}
  {\bibfnamefont {D.~E.}\ \bibnamefont {Savage}}, \bibinfo {author}
  {\bibfnamefont {M.~G.}\ \bibnamefont {Lagally}}, \bibinfo {author}
  {\bibfnamefont {M.}~\bibnamefont {Friesen}}, \bibinfo {author} {\bibfnamefont
  {S.~N.}\ \bibnamefont {Coppersmith}}, \bibinfo {author} {\bibfnamefont
  {M.~A.}\ \bibnamefont {Eriksson}}, \bibinfo {author} {\bibfnamefont {W.~A.}\
  \bibnamefont {Coish}},\ and\ \bibinfo {author} {\bibfnamefont {L.~M.~K.}\
  \bibnamefont {Vandersypen}},\ }\bibfield  {title} {\bibinfo {title}
  {{Repetitive Quantum Nondemolition Measurement and Soft Decoding of a Silicon
  Spin Qubit}},\ }\href {https://doi.org/10.1103/PhysRevX.10.021006} {\bibfield
   {journal} {\bibinfo  {journal} {Phys. Rev. X}\ }\textbf {\bibinfo {volume}
  {10}},\ \bibinfo {pages} {021006} (\bibinfo {year} {2020})}\BibitemShut
  {NoStop}%
\bibitem [{\citenamefont {Yoneda}\ \emph {et~al.}(2021)\citenamefont {Yoneda},
  \citenamefont {Huang}, \citenamefont {Feng}, \citenamefont {Yang},
  \citenamefont {Chan}, \citenamefont {Tanttu}, \citenamefont {Gilbert},
  \citenamefont {Leon}, \citenamefont {Hudson}, \citenamefont {Itoh},
  \citenamefont {Morello}, \citenamefont {Bartlett}, \citenamefont {Laucht},
  \citenamefont {Saraiva},\ and\ \citenamefont {Dzurak}}]{Yoneda2021}%
  \BibitemOpen
  \bibfield  {author} {\bibinfo {author} {\bibfnamefont {J.}~\bibnamefont
  {Yoneda}}, \bibinfo {author} {\bibfnamefont {W.}~\bibnamefont {Huang}},
  \bibinfo {author} {\bibfnamefont {M.}~\bibnamefont {Feng}}, \bibinfo {author}
  {\bibfnamefont {C.~H.}\ \bibnamefont {Yang}}, \bibinfo {author}
  {\bibfnamefont {K.~W.}\ \bibnamefont {Chan}}, \bibinfo {author}
  {\bibfnamefont {T.}~\bibnamefont {Tanttu}}, \bibinfo {author} {\bibfnamefont
  {W.}~\bibnamefont {Gilbert}}, \bibinfo {author} {\bibfnamefont {R.~C.~C.}\
  \bibnamefont {Leon}}, \bibinfo {author} {\bibfnamefont {F.~E.}\ \bibnamefont
  {Hudson}}, \bibinfo {author} {\bibfnamefont {K.~M.}\ \bibnamefont {Itoh}},
  \bibinfo {author} {\bibfnamefont {A.}~\bibnamefont {Morello}}, \bibinfo
  {author} {\bibfnamefont {S.~D.}\ \bibnamefont {Bartlett}}, \bibinfo {author}
  {\bibfnamefont {A.}~\bibnamefont {Laucht}}, \bibinfo {author} {\bibfnamefont
  {A.}~\bibnamefont {Saraiva}},\ and\ \bibinfo {author} {\bibfnamefont {A.~S.}\
  \bibnamefont {Dzurak}},\ }\bibfield  {title} {\bibinfo {title} {Coherent spin
  qubit transport in silicon},\ }\href
  {https://doi.org/10.1038/s41467-021-24371-7} {\bibfield  {journal} {\bibinfo
  {journal} {Nat. Commun.}\ }\textbf {\bibinfo {volume} {12}},\ \bibinfo
  {pages} {4114} (\bibinfo {year} {2021})}\BibitemShut {NoStop}%
\bibitem [{\citenamefont {Baart}\ \emph {et~al.}(2016)\citenamefont {Baart},
  \citenamefont {Shafiei}, \citenamefont {Fujita}, \citenamefont {Reichl},
  \citenamefont {Wegscheider},\ and\ \citenamefont {Vandersypen}}]{Baart2016}%
  \BibitemOpen
  \bibfield  {author} {\bibinfo {author} {\bibfnamefont {T.~A.}\ \bibnamefont
  {Baart}}, \bibinfo {author} {\bibfnamefont {M.}~\bibnamefont {Shafiei}},
  \bibinfo {author} {\bibfnamefont {T.}~\bibnamefont {Fujita}}, \bibinfo
  {author} {\bibfnamefont {C.}~\bibnamefont {Reichl}}, \bibinfo {author}
  {\bibfnamefont {W.}~\bibnamefont {Wegscheider}},\ and\ \bibinfo {author}
  {\bibfnamefont {L.~M.~K.}\ \bibnamefont {Vandersypen}},\ }\bibfield  {title}
  {\bibinfo {title} {Single-spin {CCD}},\ }\bibfield  {journal} {\bibinfo
  {journal} {Nature Nanotechnology}\ }\href
  {https://doi.org/10.1038/nnano.2015.291} {10.1038/nnano.2015.291} (\bibinfo
  {year} {2016})\BibitemShut {NoStop}%
\bibitem [{\citenamefont {Seidler}\ \emph {et~al.}(2021)\citenamefont
  {Seidler}, \citenamefont {Struck}, \citenamefont {Xue}, \citenamefont
  {Focke}, \citenamefont {Trellenkamp}, \citenamefont {Bluhm},\ and\
  \citenamefont {Schreiber}}]{Seidler2021}%
  \BibitemOpen
  \bibfield  {author} {\bibinfo {author} {\bibfnamefont {I.}~\bibnamefont
  {Seidler}}, \bibinfo {author} {\bibfnamefont {T.}~\bibnamefont {Struck}},
  \bibinfo {author} {\bibfnamefont {R.}~\bibnamefont {Xue}}, \bibinfo {author}
  {\bibfnamefont {N.}~\bibnamefont {Focke}}, \bibinfo {author} {\bibfnamefont
  {S.}~\bibnamefont {Trellenkamp}}, \bibinfo {author} {\bibfnamefont
  {H.}~\bibnamefont {Bluhm}},\ and\ \bibinfo {author} {\bibfnamefont {L.~R.}\
  \bibnamefont {Schreiber}},\ }\href@noop {} {\bibinfo {title} {Conveyor-mode
  single-electron shuttling in si/sige for a scalable quantum computing
  architecture}} (\bibinfo {year} {2021}),\ \Eprint
  {https://arxiv.org/abs/2108.00879} {arXiv:2108.00879 [cond-mat.mes-hall]}
  \BibitemShut {NoStop}%
\bibitem [{\citenamefont {Vandersypen}\ \emph {et~al.}(2017)\citenamefont
  {Vandersypen}, \citenamefont {Bluhm}, \citenamefont {Clarke}, \citenamefont
  {Dzurak}, \citenamefont {Ishihara}, \citenamefont {Morello}, \citenamefont
  {Reilly}, \citenamefont {Schreiber},\ and\ \citenamefont
  {Veldhorst}}]{Vandersypen2017}%
  \BibitemOpen
  \bibfield  {author} {\bibinfo {author} {\bibfnamefont {L.~M.~K.}\
  \bibnamefont {Vandersypen}}, \bibinfo {author} {\bibfnamefont
  {H.}~\bibnamefont {Bluhm}}, \bibinfo {author} {\bibfnamefont {J.~S.}\
  \bibnamefont {Clarke}}, \bibinfo {author} {\bibfnamefont {A.~S.}\
  \bibnamefont {Dzurak}}, \bibinfo {author} {\bibfnamefont {R.}~\bibnamefont
  {Ishihara}}, \bibinfo {author} {\bibfnamefont {A.}~\bibnamefont {Morello}},
  \bibinfo {author} {\bibfnamefont {D.~J.}\ \bibnamefont {Reilly}}, \bibinfo
  {author} {\bibfnamefont {L.~R.}\ \bibnamefont {Schreiber}},\ and\ \bibinfo
  {author} {\bibfnamefont {M.}~\bibnamefont {Veldhorst}},\ }\bibfield  {title}
  {\bibinfo {title} {{Interfacing spin qubits in quantum dots and donors--hot,
  dense and coherent}},\ }\href {https://doi.org/10.1038/s41534-017-0038-y}
  {\bibfield  {journal} {\bibinfo  {journal} {npj Quantum Inf.}\ }\textbf
  {\bibinfo {volume} {3}},\ \bibinfo {pages} {34} (\bibinfo {year}
  {2017})}\BibitemShut {NoStop}%
\bibitem [{\citenamefont {Franke}\ \emph {et~al.}(2019)\citenamefont {Franke},
  \citenamefont {Clarke}, \citenamefont {Vandersypen},\ and\ \citenamefont
  {Veldhorst}}]{Franke2019}%
  \BibitemOpen
  \bibfield  {author} {\bibinfo {author} {\bibfnamefont {D.~P.}\ \bibnamefont
  {Franke}}, \bibinfo {author} {\bibfnamefont {J.~S.}\ \bibnamefont {Clarke}},
  \bibinfo {author} {\bibfnamefont {L.~M.~K.}\ \bibnamefont {Vandersypen}},\
  and\ \bibinfo {author} {\bibfnamefont {M.}~\bibnamefont {Veldhorst}},\
  }\bibfield  {title} {\bibinfo {title} {{Rent's rule and extensibility in
  quantum computing}},\ }\href {https://doi.org/10.1016/j.micpro.2019.02.006}
  {\bibfield  {journal} {\bibinfo  {journal} {Microprocess. Microsy.}\ }\textbf
  {\bibinfo {volume} {67}},\ \bibinfo {pages} {1} (\bibinfo {year}
  {2019})}\BibitemShut {NoStop}%
\bibitem [{\citenamefont {Alexeev}\ \emph {et~al.}(2021)\citenamefont
  {Alexeev}, \citenamefont {Bacon}, \citenamefont {Brown}, \citenamefont
  {Calderbank}, \citenamefont {Carr}, \citenamefont {Chong}, \citenamefont
  {DeMarco}, \citenamefont {Englund}, \citenamefont {Farhi}, \citenamefont
  {Fefferman}, \citenamefont {Gorshkov}, \citenamefont {Houck}, \citenamefont
  {Kim}, \citenamefont {Kimmel}, \citenamefont {Lange}, \citenamefont {Lloyd},
  \citenamefont {Lukin}, \citenamefont {Maslov}, \citenamefont {Maunz},
  \citenamefont {Monroe}, \citenamefont {Preskill}, \citenamefont {Roetteler},
  \citenamefont {Savage},\ and\ \citenamefont {Thompson}}]{Alexeev2019}%
  \BibitemOpen
  \bibfield  {author} {\bibinfo {author} {\bibfnamefont {Y.}~\bibnamefont
  {Alexeev}}, \bibinfo {author} {\bibfnamefont {D.}~\bibnamefont {Bacon}},
  \bibinfo {author} {\bibfnamefont {K.~R.}\ \bibnamefont {Brown}}, \bibinfo
  {author} {\bibfnamefont {R.}~\bibnamefont {Calderbank}}, \bibinfo {author}
  {\bibfnamefont {L.~D.}\ \bibnamefont {Carr}}, \bibinfo {author}
  {\bibfnamefont {F.~T.}\ \bibnamefont {Chong}}, \bibinfo {author}
  {\bibfnamefont {B.}~\bibnamefont {DeMarco}}, \bibinfo {author} {\bibfnamefont
  {D.}~\bibnamefont {Englund}}, \bibinfo {author} {\bibfnamefont
  {E.}~\bibnamefont {Farhi}}, \bibinfo {author} {\bibfnamefont
  {B.}~\bibnamefont {Fefferman}}, \bibinfo {author} {\bibfnamefont {A.~V.}\
  \bibnamefont {Gorshkov}}, \bibinfo {author} {\bibfnamefont {A.}~\bibnamefont
  {Houck}}, \bibinfo {author} {\bibfnamefont {J.}~\bibnamefont {Kim}}, \bibinfo
  {author} {\bibfnamefont {S.}~\bibnamefont {Kimmel}}, \bibinfo {author}
  {\bibfnamefont {M.}~\bibnamefont {Lange}}, \bibinfo {author} {\bibfnamefont
  {S.}~\bibnamefont {Lloyd}}, \bibinfo {author} {\bibfnamefont {M.~D.}\
  \bibnamefont {Lukin}}, \bibinfo {author} {\bibfnamefont {D.}~\bibnamefont
  {Maslov}}, \bibinfo {author} {\bibfnamefont {P.}~\bibnamefont {Maunz}},
  \bibinfo {author} {\bibfnamefont {C.}~\bibnamefont {Monroe}}, \bibinfo
  {author} {\bibfnamefont {J.}~\bibnamefont {Preskill}}, \bibinfo {author}
  {\bibfnamefont {M.}~\bibnamefont {Roetteler}}, \bibinfo {author}
  {\bibfnamefont {M.}~\bibnamefont {Savage}},\ and\ \bibinfo {author}
  {\bibfnamefont {J.}~\bibnamefont {Thompson}},\ }\bibfield  {title} {\bibinfo
  {title} {{Quantum Computer Systems for Scientific Discovery}},\ }\href
  {https://doi.org/10.1103/PRXQuantum.2.017001} {\bibfield  {journal} {\bibinfo
   {journal} {PRX Quantum}\ }\textbf {\bibinfo {volume} {2}},\ \bibinfo {pages}
  {017001} (\bibinfo {year} {2021})},\ \Eprint
  {https://arxiv.org/abs/1912.07577} {1912.07577} \BibitemShut {NoStop}%
\bibitem [{\citenamefont {Veldhorst}\ \emph {et~al.}(2017)\citenamefont
  {Veldhorst}, \citenamefont {Eenink}, \citenamefont {Yang},\ and\
  \citenamefont {Dzurak}}]{Veldhorst2017}%
  \BibitemOpen
  \bibfield  {author} {\bibinfo {author} {\bibfnamefont {M.}~\bibnamefont
  {Veldhorst}}, \bibinfo {author} {\bibfnamefont {H.~G.~J.}\ \bibnamefont
  {Eenink}}, \bibinfo {author} {\bibfnamefont {C.~H.}\ \bibnamefont {Yang}},\
  and\ \bibinfo {author} {\bibfnamefont {A.~S.}\ \bibnamefont {Dzurak}},\
  }\bibfield  {title} {\bibinfo {title} {{Silicon CMOS architecture for a
  spin-based quantum computer}},\ }\href
  {https://doi.org/10.1038/s41467-017-01905-6} {\bibfield  {journal} {\bibinfo
  {journal} {Nat. Commun.}\ }\textbf {\bibinfo {volume} {8}},\ \bibinfo {pages}
  {1766} (\bibinfo {year} {2017})}\BibitemShut {NoStop}%
\bibitem [{\citenamefont {Li}\ \emph {et~al.}(2018)\citenamefont {Li},
  \citenamefont {Petit}, \citenamefont {Franke}, \citenamefont {Dehollain},
  \citenamefont {Helsen}, \citenamefont {Steudtner}, \citenamefont {Thomas},
  \citenamefont {Yoscovits}, \citenamefont {Singh}, \citenamefont {Wehner},
  \citenamefont {Vandersypen}, \citenamefont {Clarke},\ and\ \citenamefont
  {Veldhorst}}]{Li2018a}%
  \BibitemOpen
  \bibfield  {author} {\bibinfo {author} {\bibfnamefont {R.}~\bibnamefont
  {Li}}, \bibinfo {author} {\bibfnamefont {L.}~\bibnamefont {Petit}}, \bibinfo
  {author} {\bibfnamefont {D.~P.}\ \bibnamefont {Franke}}, \bibinfo {author}
  {\bibfnamefont {J.~P.}\ \bibnamefont {Dehollain}}, \bibinfo {author}
  {\bibfnamefont {J.}~\bibnamefont {Helsen}}, \bibinfo {author} {\bibfnamefont
  {M.}~\bibnamefont {Steudtner}}, \bibinfo {author} {\bibfnamefont {N.~K.}\
  \bibnamefont {Thomas}}, \bibinfo {author} {\bibfnamefont {Z.~R.}\
  \bibnamefont {Yoscovits}}, \bibinfo {author} {\bibfnamefont {K.~J.}\
  \bibnamefont {Singh}}, \bibinfo {author} {\bibfnamefont {S.}~\bibnamefont
  {Wehner}}, \bibinfo {author} {\bibfnamefont {L.~M.~K.}\ \bibnamefont
  {Vandersypen}}, \bibinfo {author} {\bibfnamefont {J.~S.}\ \bibnamefont
  {Clarke}},\ and\ \bibinfo {author} {\bibfnamefont {M.}~\bibnamefont
  {Veldhorst}},\ }\bibfield  {title} {\bibinfo {title} {{A crossbar network for
  silicon quantum dot qubits}},\ }\href
  {https://doi.org/10.1126/sciadv.aar3960} {\bibfield  {journal} {\bibinfo
  {journal} {Sci. Adv.}\ }\textbf {\bibinfo {volume} {4}},\ \bibinfo {pages}
  {eaar3960} (\bibinfo {year} {2018})}\BibitemShut {NoStop}%
\bibitem [{\citenamefont {Buonacorsi}\ \emph {et~al.}(2019)\citenamefont
  {Buonacorsi}, \citenamefont {Cai}, \citenamefont {Ramirez}, \citenamefont
  {Willick}, \citenamefont {Walker}, \citenamefont {Li}, \citenamefont {Shaw},
  \citenamefont {Xu}, \citenamefont {Benjamin},\ and\ \citenamefont
  {Baugh}}]{Buonacorsi2019}%
  \BibitemOpen
  \bibfield  {author} {\bibinfo {author} {\bibfnamefont {B.}~\bibnamefont
  {Buonacorsi}}, \bibinfo {author} {\bibfnamefont {Z.}~\bibnamefont {Cai}},
  \bibinfo {author} {\bibfnamefont {E.~B.}\ \bibnamefont {Ramirez}}, \bibinfo
  {author} {\bibfnamefont {K.~S.}\ \bibnamefont {Willick}}, \bibinfo {author}
  {\bibfnamefont {S.~M.}\ \bibnamefont {Walker}}, \bibinfo {author}
  {\bibfnamefont {J.}~\bibnamefont {Li}}, \bibinfo {author} {\bibfnamefont
  {B.~D.}\ \bibnamefont {Shaw}}, \bibinfo {author} {\bibfnamefont
  {X.}~\bibnamefont {Xu}}, \bibinfo {author} {\bibfnamefont {S.~C.}\
  \bibnamefont {Benjamin}},\ and\ \bibinfo {author} {\bibfnamefont
  {J.}~\bibnamefont {Baugh}},\ }\bibfield  {title} {\bibinfo {title} {{Network
  architecture for a topological quantum computer in silicon}},\ }\href
  {https://doi.org/10.1088/2058-9565/aaf3c4} {\bibfield  {journal} {\bibinfo
  {journal} {Quantum Sci. Technol.}\ }\textbf {\bibinfo {volume} {4}},\
  \bibinfo {pages} {025003} (\bibinfo {year} {2019})}\BibitemShut {NoStop}%
\bibitem [{\citenamefont {Boter}\ \emph {et~al.}(2019)\citenamefont {Boter},
  \citenamefont {Dehollain}, \citenamefont {van Dijk}, \citenamefont
  {Hensgens}, \citenamefont {Versluis}, \citenamefont {Clarke}, \citenamefont
  {Veldhorst}, \citenamefont {Sebastiano},\ and\ \citenamefont
  {Vandersypen}}]{Boter2019}%
  \BibitemOpen
  \bibfield  {author} {\bibinfo {author} {\bibfnamefont {J.~M.}\ \bibnamefont
  {Boter}}, \bibinfo {author} {\bibfnamefont {J.~P.}\ \bibnamefont
  {Dehollain}}, \bibinfo {author} {\bibfnamefont {J.~P.~G.}\ \bibnamefont {van
  Dijk}}, \bibinfo {author} {\bibfnamefont {T.}~\bibnamefont {Hensgens}},
  \bibinfo {author} {\bibfnamefont {R.}~\bibnamefont {Versluis}}, \bibinfo
  {author} {\bibfnamefont {J.~S.}\ \bibnamefont {Clarke}}, \bibinfo {author}
  {\bibfnamefont {M.}~\bibnamefont {Veldhorst}}, \bibinfo {author}
  {\bibfnamefont {F.}~\bibnamefont {Sebastiano}},\ and\ \bibinfo {author}
  {\bibfnamefont {L.~M.~K.}\ \bibnamefont {Vandersypen}},\ }\bibfield  {title}
  {\bibinfo {title} {{A sparse spin qubit array with integrated control
  electronics}},\ }in\ \href {https://doi.org/10.1109/IEDM19573.2019.8993570}
  {\emph {\bibinfo {booktitle} {2019 IEEE International Electron Devices
  Meeting (IEDM)}}}\ (\bibinfo  {publisher} {IEEE},\ \bibinfo {address} {New
  York},\ \bibinfo {year} {2019})\ pp.\ \bibinfo {pages} {1--31}\BibitemShut
  {NoStop}%
\bibitem [{\citenamefont {Petit}\ \emph
  {et~al.}(2020{\natexlab{a}})\citenamefont {Petit}, \citenamefont {Eenink},
  \citenamefont {Russ}, \citenamefont {Lawrie}, \citenamefont {Hendrickx},
  \citenamefont {Philips}, \citenamefont {Clarke}, \citenamefont
  {Vandersypen},\ and\ \citenamefont {Veldhorst}}]{Petit2020}%
  \BibitemOpen
  \bibfield  {author} {\bibinfo {author} {\bibfnamefont {L.}~\bibnamefont
  {Petit}}, \bibinfo {author} {\bibfnamefont {H.~G.~J.}\ \bibnamefont
  {Eenink}}, \bibinfo {author} {\bibfnamefont {M.}~\bibnamefont {Russ}},
  \bibinfo {author} {\bibfnamefont {W.~I.~L.}\ \bibnamefont {Lawrie}}, \bibinfo
  {author} {\bibfnamefont {N.~W.}\ \bibnamefont {Hendrickx}}, \bibinfo {author}
  {\bibfnamefont {S.~G.~J.}\ \bibnamefont {Philips}}, \bibinfo {author}
  {\bibfnamefont {J.~S.}\ \bibnamefont {Clarke}}, \bibinfo {author}
  {\bibfnamefont {L.~M.~K.}\ \bibnamefont {Vandersypen}},\ and\ \bibinfo
  {author} {\bibfnamefont {M.}~\bibnamefont {Veldhorst}},\ }\bibfield  {title}
  {\bibinfo {title} {{Universal quantum logic in hot silicon qubits}},\ }\href
  {https://doi.org/10.1038/s41586-020-2170-7} {\bibfield  {journal} {\bibinfo
  {journal} {Nature}\ }\textbf {\bibinfo {volume} {580}},\ \bibinfo {pages}
  {355} (\bibinfo {year} {2020}{\natexlab{a}})}\BibitemShut {NoStop}%
\bibitem [{\citenamefont {Petit}\ \emph
  {et~al.}(2020{\natexlab{b}})\citenamefont {Petit}, \citenamefont {Russ},
  \citenamefont {Eenink}, \citenamefont {Lawrie}, \citenamefont {Clarke},
  \citenamefont {Vandersypen},\ and\ \citenamefont {Veldhorst}}]{Petit2020a}%
  \BibitemOpen
  \bibfield  {author} {\bibinfo {author} {\bibfnamefont {L.}~\bibnamefont
  {Petit}}, \bibinfo {author} {\bibfnamefont {M.}~\bibnamefont {Russ}},
  \bibinfo {author} {\bibfnamefont {H.~G.~J.}\ \bibnamefont {Eenink}}, \bibinfo
  {author} {\bibfnamefont {W.~I.~L.}\ \bibnamefont {Lawrie}}, \bibinfo {author}
  {\bibfnamefont {J.~S.}\ \bibnamefont {Clarke}}, \bibinfo {author}
  {\bibfnamefont {L.~M.~K.}\ \bibnamefont {Vandersypen}},\ and\ \bibinfo
  {author} {\bibfnamefont {M.}~\bibnamefont {Veldhorst}},\ }\href@noop {}
  {\bibinfo {title} {High-fidelity two-qubit gates in silicon above one
  {K}elvin}} (\bibinfo {year} {2020}{\natexlab{b}}),\ \Eprint
  {https://arxiv.org/abs/2007.09034} {arXiv:2007.09034 [cond-mat.mes-hall]}
  \BibitemShut {NoStop}%
\bibitem [{\citenamefont {Yang}\ \emph {et~al.}(2020)\citenamefont {Yang},
  \citenamefont {Leon}, \citenamefont {Hwang}, \citenamefont {Saraiva},
  \citenamefont {Tanttu}, \citenamefont {Huang}, \citenamefont
  {Camirand~Lemyre}, \citenamefont {Chan}, \citenamefont {Tan}, \citenamefont
  {Hudson}, \citenamefont {Itoh}, \citenamefont {Morello}, \citenamefont
  {Pioro-Ladri{\`{e}}re}, \citenamefont {Laucht},\ and\ \citenamefont
  {Dzurak}}]{Yang2020}%
  \BibitemOpen
  \bibfield  {author} {\bibinfo {author} {\bibfnamefont {C.~H.}\ \bibnamefont
  {Yang}}, \bibinfo {author} {\bibfnamefont {R.~C.~C.}\ \bibnamefont {Leon}},
  \bibinfo {author} {\bibfnamefont {J.~C.~C.}\ \bibnamefont {Hwang}}, \bibinfo
  {author} {\bibfnamefont {A.}~\bibnamefont {Saraiva}}, \bibinfo {author}
  {\bibfnamefont {T.}~\bibnamefont {Tanttu}}, \bibinfo {author} {\bibfnamefont
  {W.}~\bibnamefont {Huang}}, \bibinfo {author} {\bibfnamefont
  {J.}~\bibnamefont {Camirand~Lemyre}}, \bibinfo {author} {\bibfnamefont
  {K.~W.}\ \bibnamefont {Chan}}, \bibinfo {author} {\bibfnamefont {K.~Y.}\
  \bibnamefont {Tan}}, \bibinfo {author} {\bibfnamefont {F.~E.}\ \bibnamefont
  {Hudson}}, \bibinfo {author} {\bibfnamefont {K.~M.}\ \bibnamefont {Itoh}},
  \bibinfo {author} {\bibfnamefont {A.}~\bibnamefont {Morello}}, \bibinfo
  {author} {\bibfnamefont {M.}~\bibnamefont {Pioro-Ladri{\`{e}}re}}, \bibinfo
  {author} {\bibfnamefont {A.}~\bibnamefont {Laucht}},\ and\ \bibinfo {author}
  {\bibfnamefont {A.~S.}\ \bibnamefont {Dzurak}},\ }\bibfield  {title}
  {\bibinfo {title} {{Operation of a silicon quantum processor unit cell above
  one kelvin}},\ }\href {https://doi.org/10.1038/s41586-020-2171-6} {\bibfield
  {journal} {\bibinfo  {journal} {Nature}\ }\textbf {\bibinfo {volume} {580}},\
  \bibinfo {pages} {350} (\bibinfo {year} {2020})}\BibitemShut {NoStop}%
\bibitem [{\citenamefont {Dennis}\ \emph {et~al.}(2002)\citenamefont {Dennis},
  \citenamefont {Kitaev}, \citenamefont {Landahl},\ and\ \citenamefont
  {Preskill}}]{Dennis2002}%
  \BibitemOpen
  \bibfield  {author} {\bibinfo {author} {\bibfnamefont {E.}~\bibnamefont
  {Dennis}}, \bibinfo {author} {\bibfnamefont {A.}~\bibnamefont {Kitaev}},
  \bibinfo {author} {\bibfnamefont {A.}~\bibnamefont {Landahl}},\ and\ \bibinfo
  {author} {\bibfnamefont {J.}~\bibnamefont {Preskill}},\ }\bibfield  {title}
  {\bibinfo {title} {{Topological quantum memory}},\ }\href
  {https://doi.org/10.1063/1.1499754} {\bibfield  {journal} {\bibinfo
  {journal} {J. Math. Phys.}\ }\textbf {\bibinfo {volume} {43}},\ \bibinfo
  {pages} {4452} (\bibinfo {year} {2002})}\BibitemShut {NoStop}%
\bibitem [{\citenamefont {Taylor}\ \emph {et~al.}(2005)\citenamefont {Taylor},
  \citenamefont {Engel}, \citenamefont {D{\"{u}}r}, \citenamefont {Yacoby},
  \citenamefont {Marcus}, \citenamefont {Zoller},\ and\ \citenamefont
  {Lukin}}]{Taylor2005}%
  \BibitemOpen
  \bibfield  {author} {\bibinfo {author} {\bibfnamefont {J.~M.}\ \bibnamefont
  {Taylor}}, \bibinfo {author} {\bibfnamefont {H.-A.}\ \bibnamefont {Engel}},
  \bibinfo {author} {\bibfnamefont {W.}~\bibnamefont {D{\"{u}}r}}, \bibinfo
  {author} {\bibfnamefont {A.}~\bibnamefont {Yacoby}}, \bibinfo {author}
  {\bibfnamefont {C.~M.}\ \bibnamefont {Marcus}}, \bibinfo {author}
  {\bibfnamefont {P.}~\bibnamefont {Zoller}},\ and\ \bibinfo {author}
  {\bibfnamefont {M.~D.}\ \bibnamefont {Lukin}},\ }\bibfield  {title} {\bibinfo
  {title} {{Fault-tolerant architecture for quantum computation using
  electrically controlled semiconductor spins}},\ }\href
  {https://doi.org/10.1038/nphys174} {\bibfield  {journal} {\bibinfo  {journal}
  {Nat. Phys.}\ }\textbf {\bibinfo {volume} {1}},\ \bibinfo {pages} {177}
  (\bibinfo {year} {2005})}\BibitemShut {NoStop}%
\bibitem [{\citenamefont {Pioro-Ladri{\`{e}}re}\ \emph
  {et~al.}(2007)\citenamefont {Pioro-Ladri{\`{e}}re}, \citenamefont {Tokura},
  \citenamefont {Obata}, \citenamefont {Kubo},\ and\ \citenamefont
  {Tarucha}}]{Pioro-Ladriere2007}%
  \BibitemOpen
  \bibfield  {author} {\bibinfo {author} {\bibfnamefont {M.}~\bibnamefont
  {Pioro-Ladri{\`{e}}re}}, \bibinfo {author} {\bibfnamefont {Y.}~\bibnamefont
  {Tokura}}, \bibinfo {author} {\bibfnamefont {T.}~\bibnamefont {Obata}},
  \bibinfo {author} {\bibfnamefont {T.}~\bibnamefont {Kubo}},\ and\ \bibinfo
  {author} {\bibfnamefont {S.}~\bibnamefont {Tarucha}},\ }\bibfield  {title}
  {\bibinfo {title} {{Micromagnets for coherent control of spin-charge qubit in
  lateral quantum dots}},\ }\href {https://doi.org/10.1063/1.2430906}
  {\bibfield  {journal} {\bibinfo  {journal} {Appl. Phys. Lett.}\ }\textbf
  {\bibinfo {volume} {90}},\ \bibinfo {pages} {024105} (\bibinfo {year}
  {2007})}\BibitemShut {NoStop}%
\bibitem [{\citenamefont {Hendrickx}\ \emph {et~al.}(2020)\citenamefont
  {Hendrickx}, \citenamefont {Lawrie}, \citenamefont {Petit}, \citenamefont
  {Sammak}, \citenamefont {Scappucci},\ and\ \citenamefont
  {Veldhorst}}]{Hendrickx2020}%
  \BibitemOpen
  \bibfield  {author} {\bibinfo {author} {\bibfnamefont {N.~W.}\ \bibnamefont
  {Hendrickx}}, \bibinfo {author} {\bibfnamefont {W.~I.~L.}\ \bibnamefont
  {Lawrie}}, \bibinfo {author} {\bibfnamefont {L.}~\bibnamefont {Petit}},
  \bibinfo {author} {\bibfnamefont {A.}~\bibnamefont {Sammak}}, \bibinfo
  {author} {\bibfnamefont {G.}~\bibnamefont {Scappucci}},\ and\ \bibinfo
  {author} {\bibfnamefont {M.}~\bibnamefont {Veldhorst}},\ }\bibfield  {title}
  {\bibinfo {title} {{A single-hole spin qubit}},\ }\href
  {https://doi.org/10.1038/s41467-020-17211-7} {\bibfield  {journal} {\bibinfo
  {journal} {Nat. Commun.}\ }\textbf {\bibinfo {volume} {11}},\ \bibinfo
  {pages} {3478} (\bibinfo {year} {2020})}\BibitemShut {NoStop}%
\bibitem [{\citenamefont {Aharonov}\ and\ \citenamefont
  {Anandan}(1987)}]{Aharonov1987}%
  \BibitemOpen
  \bibfield  {author} {\bibinfo {author} {\bibfnamefont {Y.}~\bibnamefont
  {Aharonov}}\ and\ \bibinfo {author} {\bibfnamefont {J.}~\bibnamefont
  {Anandan}},\ }\bibfield  {title} {\bibinfo {title} {{Phase change during a
  cyclic quantum evolution}},\ }\href
  {https://doi.org/10.1103/PhysRevLett.58.1593} {\bibfield  {journal} {\bibinfo
   {journal} {Phys. Rev. Lett.}\ }\textbf {\bibinfo {volume} {58}},\ \bibinfo
  {pages} {1593} (\bibinfo {year} {1987})}\BibitemShut {NoStop}%
\bibitem [{\citenamefont {DiVincenzo}(2000)}]{Divincenzo2000}%
  \BibitemOpen
  \bibfield  {author} {\bibinfo {author} {\bibfnamefont {D.~P.}\ \bibnamefont
  {DiVincenzo}},\ }\bibfield  {title} {\bibinfo {title} {{The Physical
  Implementation of Quantum Computation}},\ }\href
  {https://doi.org/10.1002/1521-3978(200009)48:9/11<771::AID-PROP771>3.0.CO;2-E}
  {\bibfield  {journal} {\bibinfo  {journal} {Fortschr. Phys.}\ }\textbf
  {\bibinfo {volume} {48}},\ \bibinfo {pages} {771} (\bibinfo {year}
  {2000})}\BibitemShut {NoStop}%
\bibitem [{\citenamefont {Michael A.~Nielsen}(2010)}]{Nielsen2016}%
  \BibitemOpen
  \bibfield  {author} {\bibinfo {author} {\bibfnamefont {I.~L.~C.}\
  \bibnamefont {Michael A.~Nielsen}},\ }\href
  {https://www.ebook.de/de/product/13055864/michael_a_nielsen_isaac_l_chuang_quantum_computation_and_quantum_information.html}
  {\emph {\bibinfo {title} {{Quantum Computation and Quantum Information}}}}\
  (\bibinfo  {publisher} {Cambridge University Pr.},\ \bibinfo {year}
  {2010})\BibitemShut {NoStop}%
\bibitem [{\citenamefont {Elzerman}\ \emph {et~al.}(2004)\citenamefont
  {Elzerman}, \citenamefont {Hanson}, \citenamefont {Willems~van Beveren},
  \citenamefont {Witkamp}, \citenamefont {Vandersypen},\ and\ \citenamefont
  {Kouwenhoven}}]{Elzerman2004}%
  \BibitemOpen
  \bibfield  {author} {\bibinfo {author} {\bibfnamefont {J.~M.}\ \bibnamefont
  {Elzerman}}, \bibinfo {author} {\bibfnamefont {R.}~\bibnamefont {Hanson}},
  \bibinfo {author} {\bibfnamefont {L.~H.}\ \bibnamefont {Willems~van
  Beveren}}, \bibinfo {author} {\bibfnamefont {B.}~\bibnamefont {Witkamp}},
  \bibinfo {author} {\bibfnamefont {L.~M.~K.}\ \bibnamefont {Vandersypen}},\
  and\ \bibinfo {author} {\bibfnamefont {L.~P.}\ \bibnamefont {Kouwenhoven}},\
  }\bibfield  {title} {\bibinfo {title} {{Single-shot read-out of an individual
  electron spin in a quantum dot}},\ }\href
  {https://doi.org/10.1038/nature02693} {\bibfield  {journal} {\bibinfo
  {journal} {Nature}\ }\textbf {\bibinfo {volume} {430}},\ \bibinfo {pages}
  {431} (\bibinfo {year} {2004})}\BibitemShut {NoStop}%
\bibitem [{\citenamefont {Seedhouse}\ \emph {et~al.}(2021)\citenamefont
  {Seedhouse}, \citenamefont {Tanttu}, \citenamefont {Leon}, \citenamefont
  {Zhao}, \citenamefont {Tan}, \citenamefont {Hensen}, \citenamefont {Hudson},
  \citenamefont {Itoh}, \citenamefont {Yoneda}, \citenamefont {Yang},
  \citenamefont {Morello}, \citenamefont {Laucht}, \citenamefont {Coppersmith},
  \citenamefont {Saraiva},\ and\ \citenamefont {Dzurak}}]{Seedhouse2021}%
  \BibitemOpen
  \bibfield  {author} {\bibinfo {author} {\bibfnamefont {A.~E.}\ \bibnamefont
  {Seedhouse}}, \bibinfo {author} {\bibfnamefont {T.}~\bibnamefont {Tanttu}},
  \bibinfo {author} {\bibfnamefont {R.~C.~C.}\ \bibnamefont {Leon}}, \bibinfo
  {author} {\bibfnamefont {R.}~\bibnamefont {Zhao}}, \bibinfo {author}
  {\bibfnamefont {K.~Y.}\ \bibnamefont {Tan}}, \bibinfo {author} {\bibfnamefont
  {B.}~\bibnamefont {Hensen}}, \bibinfo {author} {\bibfnamefont {F.~E.}\
  \bibnamefont {Hudson}}, \bibinfo {author} {\bibfnamefont {K.~M.}\
  \bibnamefont {Itoh}}, \bibinfo {author} {\bibfnamefont {J.}~\bibnamefont
  {Yoneda}}, \bibinfo {author} {\bibfnamefont {C.~H.}\ \bibnamefont {Yang}},
  \bibinfo {author} {\bibfnamefont {A.}~\bibnamefont {Morello}}, \bibinfo
  {author} {\bibfnamefont {A.}~\bibnamefont {Laucht}}, \bibinfo {author}
  {\bibfnamefont {S.~N.}\ \bibnamefont {Coppersmith}}, \bibinfo {author}
  {\bibfnamefont {A.}~\bibnamefont {Saraiva}},\ and\ \bibinfo {author}
  {\bibfnamefont {A.~S.}\ \bibnamefont {Dzurak}},\ }\bibfield  {title}
  {\bibinfo {title} {Pauli blockade in silicon quantum dots with spin-orbit
  control},\ }\href {https://doi.org/10.1103/PRXQuantum.2.010303} {\bibfield
  {journal} {\bibinfo  {journal} {PRX Quantum}\ }\textbf {\bibinfo {volume}
  {2}},\ \bibinfo {pages} {010303} (\bibinfo {year} {2021})}\BibitemShut
  {NoStop}%
\bibitem [{\citenamefont {Jones}\ \emph {et~al.}(2019)\citenamefont {Jones},
  \citenamefont {Pritchett}, \citenamefont {Chen}, \citenamefont {Keating},
  \citenamefont {Andrews}, \citenamefont {Blumoff}, \citenamefont {De~Lorenzo},
  \citenamefont {Eng}, \citenamefont {Ha}, \citenamefont {Kiselev},
  \citenamefont {Meenehan}, \citenamefont {Merkel}, \citenamefont {Wright},
  \citenamefont {Edge}, \citenamefont {Ross}, \citenamefont {Rakher},
  \citenamefont {Borselli},\ and\ \citenamefont {Hunter}}]{Jones2019}%
  \BibitemOpen
  \bibfield  {author} {\bibinfo {author} {\bibfnamefont {A.~M.}\ \bibnamefont
  {Jones}}, \bibinfo {author} {\bibfnamefont {E.~J.}\ \bibnamefont
  {Pritchett}}, \bibinfo {author} {\bibfnamefont {E.~H.}\ \bibnamefont {Chen}},
  \bibinfo {author} {\bibfnamefont {T.~E.}\ \bibnamefont {Keating}}, \bibinfo
  {author} {\bibfnamefont {R.~W.}\ \bibnamefont {Andrews}}, \bibinfo {author}
  {\bibfnamefont {J.~Z.}\ \bibnamefont {Blumoff}}, \bibinfo {author}
  {\bibfnamefont {L.~A.}\ \bibnamefont {De~Lorenzo}}, \bibinfo {author}
  {\bibfnamefont {K.}~\bibnamefont {Eng}}, \bibinfo {author} {\bibfnamefont
  {S.~D.}\ \bibnamefont {Ha}}, \bibinfo {author} {\bibfnamefont {A.~A.}\
  \bibnamefont {Kiselev}}, \bibinfo {author} {\bibfnamefont {S.~M.}\
  \bibnamefont {Meenehan}}, \bibinfo {author} {\bibfnamefont {S.~T.}\
  \bibnamefont {Merkel}}, \bibinfo {author} {\bibfnamefont {J.~A.}\
  \bibnamefont {Wright}}, \bibinfo {author} {\bibfnamefont {L.~F.}\
  \bibnamefont {Edge}}, \bibinfo {author} {\bibfnamefont {R.~S.}\ \bibnamefont
  {Ross}}, \bibinfo {author} {\bibfnamefont {M.~T.}\ \bibnamefont {Rakher}},
  \bibinfo {author} {\bibfnamefont {M.~G.}\ \bibnamefont {Borselli}},\ and\
  \bibinfo {author} {\bibfnamefont {A.}~\bibnamefont {Hunter}},\ }\bibfield
  {title} {\bibinfo {title} {{Spin-Blockade Spectroscopy of Si/Si-Ge Quantum
  Dots}},\ }\href {https://doi.org/10.1103/PhysRevApplied.12.014026} {\bibfield
   {journal} {\bibinfo  {journal} {Phys. Rev. Applied}\ }\textbf {\bibinfo
  {volume} {12}},\ \bibinfo {pages} {014026} (\bibinfo {year}
  {2019})}\BibitemShut {NoStop}%
\bibitem [{\citenamefont {Connors}\ \emph {et~al.}(2020)\citenamefont
  {Connors}, \citenamefont {Nelson},\ and\ \citenamefont
  {Nichol}}]{Connors2020}%
  \BibitemOpen
  \bibfield  {author} {\bibinfo {author} {\bibfnamefont {E.~J.}\ \bibnamefont
  {Connors}}, \bibinfo {author} {\bibfnamefont {J.~J.}\ \bibnamefont
  {Nelson}},\ and\ \bibinfo {author} {\bibfnamefont {J.~M.}\ \bibnamefont
  {Nichol}},\ }\bibfield  {title} {\bibinfo {title} {{Rapid High-Fidelity
  Spin-State Readout in Si/Si-Ge Quantum Dots via rf Reflectometry}},\ }\href
  {https://doi.org/10.1103/PhysRevApplied.13.024019} {\bibfield  {journal}
  {\bibinfo  {journal} {Phys. Rev. Applied}\ }\textbf {\bibinfo {volume}
  {13}},\ \bibinfo {pages} {024019} (\bibinfo {year} {2020})}\BibitemShut
  {NoStop}%
\bibitem [{\citenamefont {Versluis}\ \emph {et~al.}(2017)\citenamefont
  {Versluis}, \citenamefont {Poletto}, \citenamefont {Khammassi}, \citenamefont
  {Tarasinski}, \citenamefont {Haider}, \citenamefont {Michalak}, \citenamefont
  {Bruno}, \citenamefont {Bertels}, \citenamefont {DiCarlo},\ and\
  \citenamefont {Bruno}}]{Versluis2017}%
  \BibitemOpen
  \bibfield  {author} {\bibinfo {author} {\bibfnamefont {R.}~\bibnamefont
  {Versluis}}, \bibinfo {author} {\bibfnamefont {S.}~\bibnamefont {Poletto}},
  \bibinfo {author} {\bibfnamefont {N.}~\bibnamefont {Khammassi}}, \bibinfo
  {author} {\bibfnamefont {B.}~\bibnamefont {Tarasinski}}, \bibinfo {author}
  {\bibfnamefont {N.}~\bibnamefont {Haider}}, \bibinfo {author} {\bibfnamefont
  {D.~J.}\ \bibnamefont {Michalak}}, \bibinfo {author} {\bibfnamefont
  {A.}~\bibnamefont {Bruno}}, \bibinfo {author} {\bibfnamefont
  {K.}~\bibnamefont {Bertels}}, \bibinfo {author} {\bibfnamefont
  {L.}~\bibnamefont {DiCarlo}},\ and\ \bibinfo {author} {\bibfnamefont
  {A.}~\bibnamefont {Bruno}},\ }\bibfield  {title} {\bibinfo {title} {{Scalable
  Quantum Circuit and Control for a Superconducting Surface Code}},\ }\href
  {https://doi.org/10.1103/PhysRevApplied.8.034021} {\bibfield  {journal}
  {\bibinfo  {journal} {Phys. Rev. Applied}\ }\textbf {\bibinfo {volume} {8}},\
  \bibinfo {pages} {034021} (\bibinfo {year} {2017})}\BibitemShut {NoStop}%
\bibitem [{\citenamefont {Fowler}\ \emph {et~al.}(2012)\citenamefont {Fowler},
  \citenamefont {Mariantoni}, \citenamefont {Martinis},\ and\ \citenamefont
  {Cleland}}]{Fowler2012}%
  \BibitemOpen
  \bibfield  {author} {\bibinfo {author} {\bibfnamefont {A.~G.}\ \bibnamefont
  {Fowler}}, \bibinfo {author} {\bibfnamefont {M.}~\bibnamefont {Mariantoni}},
  \bibinfo {author} {\bibfnamefont {J.~M.}\ \bibnamefont {Martinis}},\ and\
  \bibinfo {author} {\bibfnamefont {A.~N.}\ \bibnamefont {Cleland}},\
  }\bibfield  {title} {\bibinfo {title} {{Surface codes: Towards practical
  large-scale quantum computation}},\ }\href
  {https://doi.org/10.1103/PhysRevA.86.032324} {\bibfield  {journal} {\bibinfo
  {journal} {Phys. Rev. A}\ }\textbf {\bibinfo {volume} {86}},\ \bibinfo
  {pages} {032324} (\bibinfo {year} {2012})}\BibitemShut {NoStop}%
\bibitem [{\citenamefont {Horsman}\ \emph {et~al.}(2012)\citenamefont
  {Horsman}, \citenamefont {Fowler}, \citenamefont {Devitt},\ and\
  \citenamefont {Van~Meter}}]{Horsman2012}%
  \BibitemOpen
  \bibfield  {author} {\bibinfo {author} {\bibfnamefont {C.}~\bibnamefont
  {Horsman}}, \bibinfo {author} {\bibfnamefont {A.~G.}\ \bibnamefont {Fowler}},
  \bibinfo {author} {\bibfnamefont {S.}~\bibnamefont {Devitt}},\ and\ \bibinfo
  {author} {\bibfnamefont {R.}~\bibnamefont {Van~Meter}},\ }\bibfield  {title}
  {\bibinfo {title} {{Surface code quantum computing by lattice surgery}},\
  }\href {https://doi.org/10.1088/1367-2630/14/12/123011} {\bibfield  {journal}
  {\bibinfo  {journal} {New J. Phys.}\ }\textbf {\bibinfo {volume} {14}},\
  \bibinfo {pages} {123011} (\bibinfo {year} {2012})}\BibitemShut {NoStop}%
\bibitem [{\citenamefont {Xu}\ \emph {et~al.}(2020)\citenamefont {Xu},
  \citenamefont {Unseld}, \citenamefont {Corna}, \citenamefont {Zwerver},
  \citenamefont {Sammak}, \citenamefont {Brousse}, \citenamefont {Samkharadze},
  \citenamefont {Amitonov}, \citenamefont {Veldhorst}, \citenamefont
  {Scappucci}, \citenamefont {Ishihara},\ and\ \citenamefont
  {Vandersypen}}]{Xu2020}%
  \BibitemOpen
  \bibfield  {author} {\bibinfo {author} {\bibfnamefont {Y.}~\bibnamefont
  {Xu}}, \bibinfo {author} {\bibfnamefont {F.~K.}\ \bibnamefont {Unseld}},
  \bibinfo {author} {\bibfnamefont {A.}~\bibnamefont {Corna}}, \bibinfo
  {author} {\bibfnamefont {A.~M.~J.}\ \bibnamefont {Zwerver}}, \bibinfo
  {author} {\bibfnamefont {A.}~\bibnamefont {Sammak}}, \bibinfo {author}
  {\bibfnamefont {D.}~\bibnamefont {Brousse}}, \bibinfo {author} {\bibfnamefont
  {N.}~\bibnamefont {Samkharadze}}, \bibinfo {author} {\bibfnamefont {S.~V.}\
  \bibnamefont {Amitonov}}, \bibinfo {author} {\bibfnamefont {M.}~\bibnamefont
  {Veldhorst}}, \bibinfo {author} {\bibfnamefont {G.}~\bibnamefont
  {Scappucci}}, \bibinfo {author} {\bibfnamefont {R.}~\bibnamefont
  {Ishihara}},\ and\ \bibinfo {author} {\bibfnamefont {L.~M.~K.}\ \bibnamefont
  {Vandersypen}},\ }\bibfield  {title} {\bibinfo {title} {{On-chip integration
  of Si/SiGe-based quantum dots and switched-capacitor circuits}},\ }\href
  {https://doi.org/10.1063/5.0012883} {\bibfield  {journal} {\bibinfo
  {journal} {Appl. Phys. Lett.}\ }\textbf {\bibinfo {volume} {117}},\ \bibinfo
  {pages} {144002} (\bibinfo {year} {2020})}\BibitemShut {NoStop}%
\bibitem [{\citenamefont {Schoelkopf}\ \emph {et~al.}(1998)\citenamefont
  {Schoelkopf}, \citenamefont {Wahlgren}, \citenamefont {Kozhevnikov},
  \citenamefont {Delsing},\ and\ \citenamefont {Prober}}]{Schoelkopf1998}%
  \BibitemOpen
  \bibfield  {author} {\bibinfo {author} {\bibfnamefont {R.~J.}\ \bibnamefont
  {Schoelkopf}}, \bibinfo {author} {\bibfnamefont {P.}~\bibnamefont
  {Wahlgren}}, \bibinfo {author} {\bibfnamefont {A.~A.}\ \bibnamefont
  {Kozhevnikov}}, \bibinfo {author} {\bibfnamefont {P.}~\bibnamefont
  {Delsing}},\ and\ \bibinfo {author} {\bibfnamefont {D.~E.}\ \bibnamefont
  {Prober}},\ }\bibfield  {title} {\bibinfo {title} {{The Radio-Frequency
  Single-Electron Transistor (RF-SET): A Fast and Ultrasensitive
  Electrometer}},\ }\href {https://doi.org/10.1126/science.280.5367.1238}
  {\bibfield  {journal} {\bibinfo  {journal} {Science}\ }\textbf {\bibinfo
  {volume} {280}},\ \bibinfo {pages} {1238} (\bibinfo {year}
  {1998})}\BibitemShut {NoStop}%
\bibitem [{\citenamefont {Devitt}\ \emph {et~al.}(2009)\citenamefont {Devitt},
  \citenamefont {Fowler}, \citenamefont {Stephens}, \citenamefont {Greentree},
  \citenamefont {Hollenberg}, \citenamefont {Munro},\ and\ \citenamefont
  {Nemoto}}]{Devitt2009}%
  \BibitemOpen
  \bibfield  {author} {\bibinfo {author} {\bibfnamefont {S.~J.}\ \bibnamefont
  {Devitt}}, \bibinfo {author} {\bibfnamefont {A.~G.}\ \bibnamefont {Fowler}},
  \bibinfo {author} {\bibfnamefont {A.~M.}\ \bibnamefont {Stephens}}, \bibinfo
  {author} {\bibfnamefont {A.~D.}\ \bibnamefont {Greentree}}, \bibinfo {author}
  {\bibfnamefont {L.~C.~L.}\ \bibnamefont {Hollenberg}}, \bibinfo {author}
  {\bibfnamefont {W.~J.}\ \bibnamefont {Munro}},\ and\ \bibinfo {author}
  {\bibfnamefont {K.}~\bibnamefont {Nemoto}},\ }\bibfield  {title} {\bibinfo
  {title} {{Architectural design for a topological cluster state quantum
  computer}},\ }\href {https://doi.org/10.1088/1367-2630/11/8/083032}
  {\bibfield  {journal} {\bibinfo  {journal} {New J. Phys.}\ }\textbf {\bibinfo
  {volume} {11}},\ \bibinfo {pages} {083032} (\bibinfo {year}
  {2009})}\BibitemShut {NoStop}%
\bibitem [{\citenamefont {Puddy}\ \emph {et~al.}(2015)\citenamefont {Puddy},
  \citenamefont {Smith}, \citenamefont {Al-Taie}, \citenamefont {Chong},
  \citenamefont {Farrer}, \citenamefont {Griffiths}, \citenamefont {Ritchie},
  \citenamefont {Kelly}, \citenamefont {Pepper},\ and\ \citenamefont
  {Smith}}]{Puddy2015}%
  \BibitemOpen
  \bibfield  {author} {\bibinfo {author} {\bibfnamefont {R.~K.}\ \bibnamefont
  {Puddy}}, \bibinfo {author} {\bibfnamefont {L.~W.}\ \bibnamefont {Smith}},
  \bibinfo {author} {\bibfnamefont {H.}~\bibnamefont {Al-Taie}}, \bibinfo
  {author} {\bibfnamefont {C.~H.}\ \bibnamefont {Chong}}, \bibinfo {author}
  {\bibfnamefont {I.}~\bibnamefont {Farrer}}, \bibinfo {author} {\bibfnamefont
  {J.~P.}\ \bibnamefont {Griffiths}}, \bibinfo {author} {\bibfnamefont {D.~A.}\
  \bibnamefont {Ritchie}}, \bibinfo {author} {\bibfnamefont {M.~J.}\
  \bibnamefont {Kelly}}, \bibinfo {author} {\bibfnamefont {M.}~\bibnamefont
  {Pepper}},\ and\ \bibinfo {author} {\bibfnamefont {C.~G.}\ \bibnamefont
  {Smith}},\ }\bibfield  {title} {\bibinfo {title} {{Multiplexed charge-locking
  device for large arrays of quantum devices}},\ }\href
  {https://doi.org/10.1063/1.4932012} {\bibfield  {journal} {\bibinfo
  {journal} {Appl. Phys. Lett.}\ }\textbf {\bibinfo {volume} {107}},\ \bibinfo
  {pages} {143501} (\bibinfo {year} {2015})}\BibitemShut {NoStop}%
\bibitem [{\citenamefont {Hou}\ \emph {et~al.}(2019)\citenamefont {Hou},
  \citenamefont {Wu}, \citenamefont {Yu}, \citenamefont {Hsia}, \citenamefont
  {Tsai}, \citenamefont {Ting}, \citenamefont {Yu}, \citenamefont {Lee},
  \citenamefont {Chen}, \citenamefont {Chiou},\ and\ \citenamefont
  {Wang}}]{Hou2019}%
  \BibitemOpen
  \bibfield  {author} {\bibinfo {author} {\bibfnamefont {S.~Y.}\ \bibnamefont
  {Hou}}, \bibinfo {author} {\bibfnamefont {C.~H.}\ \bibnamefont {Wu}},
  \bibinfo {author} {\bibfnamefont {D.}~\bibnamefont {Yu}}, \bibinfo {author}
  {\bibfnamefont {H.}~\bibnamefont {Hsia}}, \bibinfo {author} {\bibfnamefont
  {C.~H.}\ \bibnamefont {Tsai}}, \bibinfo {author} {\bibfnamefont {K.~C.}\
  \bibnamefont {Ting}}, \bibinfo {author} {\bibfnamefont {T.~H.}\ \bibnamefont
  {Yu}}, \bibinfo {author} {\bibfnamefont {Y.~W.}\ \bibnamefont {Lee}},
  \bibinfo {author} {\bibfnamefont {F.~C.}\ \bibnamefont {Chen}}, \bibinfo
  {author} {\bibfnamefont {W.~C.}\ \bibnamefont {Chiou}},\ and\ \bibinfo
  {author} {\bibfnamefont {C.~T.}\ \bibnamefont {Wang}},\ }\bibfield  {title}
  {\bibinfo {title} {{Integrated Deep Trench Capacitor in Si Interposer for
  CoWoS Heterogeneous Integration}},\ }in\ \href
  {https://doi.org/10.1109/IEDM19573.2019.8993498} {\emph {\bibinfo {booktitle}
  {2019 IEEE International Electron Devices Meeting (IEDM)}}}\ (\bibinfo
  {publisher} {IEEE},\ \bibinfo {year} {2019})\BibitemShut {NoStop}%
\bibitem [{IRD(2020)}]{IRDS2020}%
  \BibitemOpen
  \href {https://irds.ieee.org/editions/2020} {\bibinfo {title} {{International
  Roadmap for Devices and Systems (IRDS\textsuperscript{TM}) 2020 Edition -
  More Moore}}} (\bibinfo {year} {2020})\BibitemShut {NoStop}%
\bibitem [{\citenamefont {Buonacorsi}\ \emph {et~al.}(2020)\citenamefont
  {Buonacorsi}, \citenamefont {Shaw},\ and\ \citenamefont
  {Baugh}}]{Buonacorsi2020}%
  \BibitemOpen
  \bibfield  {author} {\bibinfo {author} {\bibfnamefont {B.}~\bibnamefont
  {Buonacorsi}}, \bibinfo {author} {\bibfnamefont {B.}~\bibnamefont {Shaw}},\
  and\ \bibinfo {author} {\bibfnamefont {J.}~\bibnamefont {Baugh}},\ }\bibfield
   {title} {\bibinfo {title} {{Simulated coherent electron shuttling in silicon
  quantum dots}},\ }\href {https://doi.org/10.1103/PhysRevB.102.125406}
  {\bibfield  {journal} {\bibinfo  {journal} {Phys. Rev. B}\ }\textbf {\bibinfo
  {volume} {102}},\ \bibinfo {pages} {125406} (\bibinfo {year}
  {2020})}\BibitemShut {NoStop}%
\bibitem [{\citenamefont {Leon}\ \emph {et~al.}(2020)\citenamefont {Leon},
  \citenamefont {Yang}, \citenamefont {Hwang}, \citenamefont {Lemyre},
  \citenamefont {Tanttu}, \citenamefont {Huang}, \citenamefont {Chan},
  \citenamefont {Tan}, \citenamefont {Hudson}, \citenamefont {Itoh},
  \citenamefont {Morello}, \citenamefont {Laucht}, \citenamefont
  {Pioro-Ladri{\`{e}}re}, \citenamefont {Saraiva},\ and\ \citenamefont
  {Dzurak}}]{Leon2020}%
  \BibitemOpen
  \bibfield  {author} {\bibinfo {author} {\bibfnamefont {R.~C.~C.}\
  \bibnamefont {Leon}}, \bibinfo {author} {\bibfnamefont {C.~H.}\ \bibnamefont
  {Yang}}, \bibinfo {author} {\bibfnamefont {J.~C.~C.}\ \bibnamefont {Hwang}},
  \bibinfo {author} {\bibfnamefont {J.~C.}\ \bibnamefont {Lemyre}}, \bibinfo
  {author} {\bibfnamefont {T.}~\bibnamefont {Tanttu}}, \bibinfo {author}
  {\bibfnamefont {W.}~\bibnamefont {Huang}}, \bibinfo {author} {\bibfnamefont
  {K.~W.}\ \bibnamefont {Chan}}, \bibinfo {author} {\bibfnamefont {K.~Y.}\
  \bibnamefont {Tan}}, \bibinfo {author} {\bibfnamefont {F.~E.}\ \bibnamefont
  {Hudson}}, \bibinfo {author} {\bibfnamefont {K.~M.}\ \bibnamefont {Itoh}},
  \bibinfo {author} {\bibfnamefont {A.}~\bibnamefont {Morello}}, \bibinfo
  {author} {\bibfnamefont {A.}~\bibnamefont {Laucht}}, \bibinfo {author}
  {\bibfnamefont {M.}~\bibnamefont {Pioro-Ladri{\`{e}}re}}, \bibinfo {author}
  {\bibfnamefont {A.}~\bibnamefont {Saraiva}},\ and\ \bibinfo {author}
  {\bibfnamefont {A.~S.}\ \bibnamefont {Dzurak}},\ }\bibfield  {title}
  {\bibinfo {title} {{Coherent spin control of s-, p-, d- and f-electrons in a
  silicon quantum dot}},\ }\href {https://doi.org/10.1038/s41467-019-14053-w}
  {\bibfield  {journal} {\bibinfo  {journal} {Nat. Commun.}\ }\textbf {\bibinfo
  {volume} {11}},\ \bibinfo {pages} {797} (\bibinfo {year} {2020})}\BibitemShut
  {NoStop}%
\bibitem [{\citenamefont {Van~Dijk}\ \emph
  {et~al.}(2020{\natexlab{a}})\citenamefont {Van~Dijk}, \citenamefont {Patra},
  \citenamefont {Subramanian}, \citenamefont {Xue}, \citenamefont
  {Samkharadze}, \citenamefont {Corna}, \citenamefont {Jeon}, \citenamefont
  {Sheikh}, \citenamefont {Juarez-Hernandez}, \citenamefont {Esparza},
  \citenamefont {Rampurawala}, \citenamefont {Carlton}, \citenamefont
  {Ravikumar}, \citenamefont {Nieva}, \citenamefont {Kim}, \citenamefont {Lee},
  \citenamefont {Sammak}, \citenamefont {Scappucci}, \citenamefont {Veldhorst},
  \citenamefont {Vandersypen}, \citenamefont {Charbon}, \citenamefont
  {Pellerano}, \citenamefont {Babaie},\ and\ \citenamefont
  {Sebastiano}}]{vanDijk2020}%
  \BibitemOpen
  \bibfield  {author} {\bibinfo {author} {\bibfnamefont {J.~P.~G.}\
  \bibnamefont {Van~Dijk}}, \bibinfo {author} {\bibfnamefont {B.}~\bibnamefont
  {Patra}}, \bibinfo {author} {\bibfnamefont {S.}~\bibnamefont {Subramanian}},
  \bibinfo {author} {\bibfnamefont {X.}~\bibnamefont {Xue}}, \bibinfo {author}
  {\bibfnamefont {N.}~\bibnamefont {Samkharadze}}, \bibinfo {author}
  {\bibfnamefont {A.}~\bibnamefont {Corna}}, \bibinfo {author} {\bibfnamefont
  {C.}~\bibnamefont {Jeon}}, \bibinfo {author} {\bibfnamefont {F.}~\bibnamefont
  {Sheikh}}, \bibinfo {author} {\bibfnamefont {E.}~\bibnamefont
  {Juarez-Hernandez}}, \bibinfo {author} {\bibfnamefont {B.~P.}\ \bibnamefont
  {Esparza}}, \bibinfo {author} {\bibfnamefont {H.}~\bibnamefont
  {Rampurawala}}, \bibinfo {author} {\bibfnamefont {B.~R.}\ \bibnamefont
  {Carlton}}, \bibinfo {author} {\bibfnamefont {S.}~\bibnamefont {Ravikumar}},
  \bibinfo {author} {\bibfnamefont {C.}~\bibnamefont {Nieva}}, \bibinfo
  {author} {\bibfnamefont {S.}~\bibnamefont {Kim}}, \bibinfo {author}
  {\bibfnamefont {H.-J.}\ \bibnamefont {Lee}}, \bibinfo {author} {\bibfnamefont
  {A.}~\bibnamefont {Sammak}}, \bibinfo {author} {\bibfnamefont
  {G.}~\bibnamefont {Scappucci}}, \bibinfo {author} {\bibfnamefont
  {M.}~\bibnamefont {Veldhorst}}, \bibinfo {author} {\bibfnamefont {L.~M.~K.}\
  \bibnamefont {Vandersypen}}, \bibinfo {author} {\bibfnamefont
  {E.}~\bibnamefont {Charbon}}, \bibinfo {author} {\bibfnamefont
  {S.}~\bibnamefont {Pellerano}}, \bibinfo {author} {\bibfnamefont
  {M.}~\bibnamefont {Babaie}},\ and\ \bibinfo {author} {\bibfnamefont
  {F.}~\bibnamefont {Sebastiano}},\ }\bibfield  {title} {\bibinfo {title} {{A
  Scalable Cryo-CMOS Controller for the Wideband Frequency-Multiplexed Control
  of Spin Qubits and Transmons}},\ }\href
  {https://doi.org/10.1109/JSSC.2020.3024678} {\bibfield  {journal} {\bibinfo
  {journal} {IEEE J. Solid-State Circuits}\ }\textbf {\bibinfo {volume} {55}},\
  \bibinfo {pages} {2930} (\bibinfo {year} {2020}{\natexlab{a}})}\BibitemShut
  {NoStop}%
\bibitem [{\citenamefont {Van~Dijk}\ \emph
  {et~al.}(2020{\natexlab{b}})\citenamefont {Van~Dijk}, \citenamefont {Hart},
  \citenamefont {Kiene}, \citenamefont {Overwater}, \citenamefont {Padalia},
  \citenamefont {Van~Staveren}, \citenamefont {Babaie}, \citenamefont
  {Vladimirescu}, \citenamefont {Charbon},\ and\ \citenamefont
  {Sebastiano}}]{vanDijk2020a}%
  \BibitemOpen
  \bibfield  {author} {\bibinfo {author} {\bibfnamefont {J.}~\bibnamefont
  {Van~Dijk}}, \bibinfo {author} {\bibfnamefont {P.}~\bibnamefont {Hart}},
  \bibinfo {author} {\bibfnamefont {G.}~\bibnamefont {Kiene}}, \bibinfo
  {author} {\bibfnamefont {R.}~\bibnamefont {Overwater}}, \bibinfo {author}
  {\bibfnamefont {P.}~\bibnamefont {Padalia}}, \bibinfo {author} {\bibfnamefont
  {J.}~\bibnamefont {Van~Staveren}}, \bibinfo {author} {\bibfnamefont
  {M.}~\bibnamefont {Babaie}}, \bibinfo {author} {\bibfnamefont
  {A.}~\bibnamefont {Vladimirescu}}, \bibinfo {author} {\bibfnamefont
  {E.}~\bibnamefont {Charbon}},\ and\ \bibinfo {author} {\bibfnamefont
  {F.}~\bibnamefont {Sebastiano}},\ }\bibfield  {title} {\bibinfo {title}
  {{Cryo-CMOS for Analog/Mixed-Signal Circuits and Systems}},\ }in\ \href
  {https://doi.org/10.1109/CICC48029.2020.9075882} {\emph {\bibinfo {booktitle}
  {Proceedings of the Custom Integrated Circuits Conference}}}\ (\bibinfo
  {publisher} {IEEE},\ \bibinfo {year} {2020})\ pp.\ \bibinfo {pages}
  {1--8}\BibitemShut {NoStop}%
\bibitem [{\citenamefont {Paul}(2008)}]{Paul2008}%
  \BibitemOpen
  \bibfield  {author} {\bibinfo {author} {\bibfnamefont {C.~R.}\ \bibnamefont
  {Paul}},\ }\href@noop {} {\emph {\bibinfo {title} {{Analysis of
  Multiconductor Transmission Lines}}}}\ (\bibinfo  {publisher} {Wiley-IEEE
  Press},\ \bibinfo {year} {2008})\BibitemShut {NoStop}%
\bibitem [{\citenamefont {Wong}\ \emph {et~al.}(1998)\citenamefont {Wong},
  \citenamefont {Liu}, \citenamefont {Ru},\ and\ \citenamefont
  {Lin}}]{Wong1998}%
  \BibitemOpen
  \bibfield  {author} {\bibinfo {author} {\bibfnamefont {S.-C.}\ \bibnamefont
  {Wong}}, \bibinfo {author} {\bibfnamefont {P.~S.}\ \bibnamefont {Liu}},
  \bibinfo {author} {\bibfnamefont {J.-W.}\ \bibnamefont {Ru}},\ and\ \bibinfo
  {author} {\bibfnamefont {S.-T.}\ \bibnamefont {Lin}},\ }\bibfield  {title}
  {\bibinfo {title} {{Interconnection capacitance models for VLSI circuits}},\
  }\href {https://doi.org/10.1016/S0038-1101(98)00088-4} {\bibfield  {journal}
  {\bibinfo  {journal} {Solid-State Electron.}\ }\textbf {\bibinfo {volume}
  {42}},\ \bibinfo {pages} {969} (\bibinfo {year} {1998})}\BibitemShut
  {NoStop}%
\bibitem [{\citenamefont {Razavi}(2016)}]{Razavi2016}%
  \BibitemOpen
  \bibfield  {author} {\bibinfo {author} {\bibfnamefont {B.}~\bibnamefont
  {Razavi}},\ }\href@noop {} {\emph {\bibinfo {title} {Design of Analog CMOS
  Integrated Circuits}}},\ \bibinfo {edition} {2nd}\ ed.\ (\bibinfo
  {publisher} {McGraw-Hill Education},\ \bibinfo {year} {2016})\BibitemShut
  {NoStop}%
\bibitem [{\citenamefont {Patra}\ \emph {et~al.}(2020)\citenamefont {Patra},
  \citenamefont {Mehrpoo}, \citenamefont {Ruffino}, \citenamefont {Sebastiano},
  \citenamefont {Charbon},\ and\ \citenamefont {Babaie}}]{Patra2020}%
  \BibitemOpen
  \bibfield  {author} {\bibinfo {author} {\bibfnamefont {B.}~\bibnamefont
  {Patra}}, \bibinfo {author} {\bibfnamefont {M.}~\bibnamefont {Mehrpoo}},
  \bibinfo {author} {\bibfnamefont {A.}~\bibnamefont {Ruffino}}, \bibinfo
  {author} {\bibfnamefont {F.}~\bibnamefont {Sebastiano}}, \bibinfo {author}
  {\bibfnamefont {E.}~\bibnamefont {Charbon}},\ and\ \bibinfo {author}
  {\bibfnamefont {M.}~\bibnamefont {Babaie}},\ }\bibfield  {title} {\bibinfo
  {title} {Characterization and analysis of on-chip microwave passive
  components at cryogenic temperatures},\ }\href
  {https://doi.org/10.1109/jeds.2020.2986722} {\bibfield  {journal} {\bibinfo
  {journal} {{IEEE} Journal of the Electron Devices Society}\ }\textbf
  {\bibinfo {volume} {8}},\ \bibinfo {pages} {448} (\bibinfo {year}
  {2020})}\BibitemShut {NoStop}%
\bibitem [{\citenamefont {Schuch}\ and\ \citenamefont
  {Siewert}(2003)}]{Schuch2003}%
  \BibitemOpen
  \bibfield  {author} {\bibinfo {author} {\bibfnamefont {N.}~\bibnamefont
  {Schuch}}\ and\ \bibinfo {author} {\bibfnamefont {J.}~\bibnamefont
  {Siewert}},\ }\bibfield  {title} {\bibinfo {title} {{Natural two-qubit gate
  for quantum computation using the {\$}XY{\$} interaction}},\ }\href
  {https://doi.org/10.1103/PhysRevA.67.032301} {\bibfield  {journal} {\bibinfo
  {journal} {Phys. Rev. A}\ }\textbf {\bibinfo {volume} {67}},\ \bibinfo
  {pages} {032301} (\bibinfo {year} {2003})}\BibitemShut {NoStop}%
\end{thebibliography}%
\end{document}